\documentclass[aps,rmp,twocolumn,floats,floatfix,longbibliography]{revtex4-1}
\usepackage{amsmath,amssymb}
\usepackage{graphicx}
\newcommand{\shortcite}[1]{\cite{#1}}
\newcommand{\shortciteN}[1]{\textcite{#1}}
\newcommand{\shortciteNP}[1]{\textcite{#1}}

\begin{document}

\title{Random matrix approaches to open quantum systems}
\author{Henning Schomerus}
\affiliation{Department of Physics, Lancaster University, Lancaster, LA1 4YB, UK}

\date{July 2015 (lectures), May 2016 (v1), December 2016 (v2)}

\begin{abstract}
Over the past decades, a great body of theoretical and mathematical work has been devoted to random-matrix descriptions of open quantum systems. In these notes, based on lectures delivered at the Les Houches Summer School ``Stochastic Processes and Random Matrices'' in July 2015, we review the physical origins and mathematical structures of the underlying models, and collect key predictions which give insight into the typical system behaviour. In particular, we aim to give an idea how the different features are interlinked. The notes mainly focus on elastic scattering but also include a short detour to interacting systems, which we motivate by the overarching question of ergodicity.
The first chapters introduce general notions from random matrix theory, such as the ten universality classes and ensembles of hermitian, unitary, positive-definite and non-hermitian matrices. We then review microscopic scattering models that form the basis for statistical descriptions, and consider signatures of random scattering in decay, dynamics and transport. The last chapter briefly touches on Anderson localization and localization in interacting systems.
\end{abstract}

\maketitle

\tableofcontents

%\maintext

\section{Introduction}

\subsection{Welcome}
Open quantum systems come in two variants. The first variant (on which we will focus more) are scattering systems in which the dynamics allow particles to enter and leave  \shortcite{newton_scattering_2002,messiah2014quantum}. One then normally defines a scattering region, outside of which particles move free of any external forces or interactions. This situation is realised (at least to some level of approximation) in many decay or radiation processes \shortcite{RevModPhys.81.539}, but is also useful to describe phase-coherent  transport in mesoscopic devices \shortcite{datta_electronic_1997,RevModPhys.69.731,blanter_shot_2000,nazarov2009quantum} or photonic structures \shortcite{RevModPhys.87.61}.
The second variant (which we will encounter only briefly) are interacting systems in which the studied dynamical degrees of freedom are influenced by other degrees of freedom in the environment \shortcite{breuer_theory_2002}. This situation spans from the quantum-statistical foundations of thermodynamics \shortcite{gemmer_quantum_2010} to the description of  decoherence \shortcite{weiss_quantum_2008}, with ample applications to quantum optics \shortcite{carmichael2009open}, quantum-critical phenomena
\shortcite{sachdev_quantum_1999} and quantum information processing \shortcite{nielsen2010quantum}.

While these two scenarios of openness are in many ways quite distinct, they have some important features in common---in particular, in both scenarios we are led to restrict our attention to a subsystem, while the processes that are involved often are very complex (meaning that we have no realistic handles to describe them in detail), be it due to underlying classical chaos, disorder, or uncontrolled interactions. Taken together, these features lay the foundations for a statistical description where individual systems are replaced by an appropriate ensemble. These ensembles are typically formulated in terms of effective models, e.g., for the Hamiltonian, the scattering matrix, or the density matrix, in which only the fundamental symmetries and the most essential time and energy scales are retained.
Quantitative predictions then follow  from explicit calculations and often turn out to be universal, i.e., applicable to generic representatives of the ensemble.

Over the past decades, a great body of theoretical and mathematical work has been devoted to these random-matrix descriptions
\shortcite{RevModPhys.69.731,Guhr1998189,mehta2004random,stockmann_quantum_2006,haake_quantum_2010,forrester2010log,RMTHandbook,pastur2011eigenvalue,RevModPhys.87.1037}. In these notes we review the physical origins and mathematical structures of the underlying models,
and collect key predictions which give insight into the typical system behaviour.
In particular, we aim to give an idea how the different features are interlinked. This includes a detour to interacting systems, which we motivate by the overarching question of ergodicity.
With this selection of topics, we hope to provide a useful bridge  to the many excellent advanced sources, including the monographs and reviews mentioned above, which contain detailed expositions of the random-matrix calculations and further applications not covered here. In the remainder of this introduction, we provide some basic background.

\subsection{Primer}
These lectures were delivered to a mixed audience of mathematicians and physicists. To establish some common language, let us first  review some basic notions of quantum mechanics \shortcite{peres_quantum_2002}. This also gives us the opportunity to pinpoint the fundamental origins of the mathematical concepts and physical phenomena that we will encounter throughout these notes---and further explain what these notes are really about.

Let us recall, then, that quantum mechanics describes the physical states of a system in terms of vectors $|\psi\rangle$, $|\phi\rangle,\ldots$ in a complex Hilbert space $\mathcal{H}$. The superposition principle means that the vectors can be freely combined to yield new physical states $\alpha|\psi\rangle+\beta|\phi\rangle$, $\alpha, \beta\in \mathbb{C}$.
All vectors $\alpha|\psi\rangle$ that differ only by a multiplicative factor $\alpha\neq 0$ describe the same physical state, which is often
exploited to impose the convenient normalization $\langle\psi|\psi\rangle=1$. Following physics convention, we here use (what we term) the scalar product with $\langle\phi|(\alpha\psi+\beta\chi)\rangle=\alpha\langle\phi|\psi\rangle+\beta\langle\phi|\chi\rangle$, $\langle\psi|\phi\rangle=\langle\phi|\psi\rangle^*$, $\langle\psi|\psi\rangle>0$ unless $|\psi\rangle=0$,
where ${}^*$ denotes complex conjugation. Two states with $\langle\phi|\psi\rangle=0$ are called orthogonal, and a discrete basis with $\langle n|m\rangle=\delta_{nm}$ is called orthonormal. For a continuous basis, this is replaced by $\langle x|x'\rangle=\delta(x-x')$ with Dirac's delta function. In a given basis, states can be expanded as $|\psi\rangle=\sum_n
\psi_n|n\rangle$ where $\psi_n=\langle n |\psi\rangle$, with the sum replaced by an integral when the basis is continuous.

Observables are represented by  hermitian linear operators $\hat A$, with $\hat A|\psi\rangle\equiv|\hat A\psi\rangle\in \mathcal{H}$ such that $\langle\phi|\hat A\psi\rangle=\langle\hat A\phi|\psi\rangle$.
According to the measurement axiom, these operators predict physical observations
via the expectation values $\mathcal{E}_\psi(A)=\langle\psi|\hat A\psi\rangle/\langle\psi|\psi\rangle$, which in reality are obtained by averaging the outcomes of experiments
on systems in the same quantum state.
The associated uncertainty (variance) is obtained from $\Delta A =[\mathcal{E}_\psi(A^2)-\mathcal{E}^2_\psi(A)]^{1/2}$, which in general is finite.
Denoting by $E_{a}=\sum_n|\psi_{a,n}\rangle\langle\psi_{a,n}|$ the projector onto states that guarantee an outcome $a$ with vanishing uncertainty $\Delta A=0$, one finds that these are eigenstates with $\hat A|\psi_{a,n}\rangle=a|\psi_{a,n}\rangle$.
In a general state, the probability of these outcomes  $|\psi\rangle$ are then $P(a)=\langle\psi|E_{a}|\psi\rangle/\langle\psi|\psi\rangle$;
no outcomes other than the associated eigenvalues are allowed.
 Beyond this probabilistic description,
outcomes of individual experiments are unpredictable. Finally, the measurement axiom stipulates that right after the measurement with an outcome $a$, the quantum system acquires the state $E_{a}|\psi\rangle$.

Adopting the conventional Schr{\"o}dinger picture, the time dependence of the quantum state arises from the Schr{\"o}dinger equation \begin{equation}
\label{eq:schr}
i\hbar\frac{d}{dt}|\psi(t)\rangle=\hat H(t)|\psi(t)\rangle.
\end{equation}
Here $\hat H$ is the Hamiltonian, a hermitian operator which represents energy, while $\hbar=h/2\pi$ is the reduced Planck's constant. The general solution can be written as $|\psi(t)\rangle=\hat U(t,t')|\psi(t')\rangle$, where $\hat U(t,t')$ is a unitary operator ($\hat U$ is unitary if always $\langle \hat U\phi|\hat U\psi\rangle=
\langle \phi|\psi\rangle$).
If $\hat H$ is independent of time, we can separate variables as $|\psi(t)\rangle=\exp(-iEt/\hbar)|\phi\rangle$ and arrive at the stationary Schr{\"o}dinger equation $E|\phi\rangle=\hat H|\phi\rangle$. In this case,
$\hat U(t,t')=\exp(-i\hat H(t-t')/\hbar)$.

In order to describe the incoherent mixture of normalised quantum states $|\psi_n\rangle$ one introduces the density matrix (statistical operator)  $\hat\rho=\sum_n p_n|\psi_n\rangle\langle\psi_n|$ with positive weights $p_n$ summing to $\sum p_n=1$, so that $\mathrm{tr}\,\hat \rho=1$. The expectation values $\mathcal{E}_\rho(A)=\mathrm{tr}\,(\hat A\hat \rho)=\sum_np_n\mathcal{E}_{\psi_n}(A)$ are a combination of the quantum-mechanical average in each quantum state and the classical average over the weights $p_n$. The density operator is hermitian and positive semidefinite, and for a pure state (with only one finite $p_n=1$) becomes a projector, $\hat \rho^2=\hat\rho$. To capture the departure from this situation one can consider the purity $\mathcal{P}=\mathrm{tr }\,\hat \rho^2$, which equals unity only for a pure state, as well as the von Neumann entropy $\mathcal{S}=-\mathrm{tr}\, \hat\rho \ln \hat\rho$, which vanishes for a pure state.

\subsection{Open systems}
The superposition principle mentioned above is the origin of wave-like interference effects, the complexity of which we will aim to capture in a statistical description. To provide the states with some structure, we can often think of the state space being divided into sectors (which we here call regions), $\mathcal{H}=\mathcal{H}_1\oplus\mathcal{H}_2$. We then can start to talk about local and non-local processes, within or between the regions, and introduce basic notions such as the exchange of particles or energy.
An additional layer of complexity is added when we  can view the system as being composed of separate degrees of freedom (which we here call  parts). The Hilbert space then takes the form of a tensor product
$\mathcal{H}=\mathcal{H}_1\otimes\mathcal{H}_2$, with proper symmetrization or antisymmetrization if the parts are, in a physical sense,  indistinguishable (e.g., when they describe identical bosonic or fermionic particles). Separable states are of the form $|\phi\rangle\otimes|\chi\rangle$, while superpositions of such states lead to quantum  correlations (entanglement) that deeply enrich the behaviour of interacting systems.
Based on these elements of structure, let us now agree, within the confines of these notes, on two notions of open quantum systems. These are systems in which we can naturally focus on some region or part $\mathcal{H}_1$, which is either locally confined (as in $\mathcal{H}=\mathcal{H}_1\oplus\mathcal{H}_2$) or constrained to some of the degrees of freedoms (as in $\mathcal{H}=\mathcal{H}_1\otimes\mathcal{H}_2$). We are then naturally led down two roads: Scattering-like scenarios, which describe the exchange of particles between a confined region and its surrounding environment, and scenarios dominated by the interactions, which often concern the exchange of energy and creation of entanglement. In order for this division to make some sense, the environment  must be sufficiently structureless---either because the dynamics are simple and predictable (typically, the point of view taken in the case of scattering), or because they are so complex that they can be described in a simple statistical picture (typically, the point of view taken in the case of interactions).  Physically, this requires that the rest of the system is large, and of a nature where incoming and outgoing particles are only simply correlated, and so is the energy flowing in or out of the system.

\begin{figure}[t]
\includegraphics[width=\columnwidth]{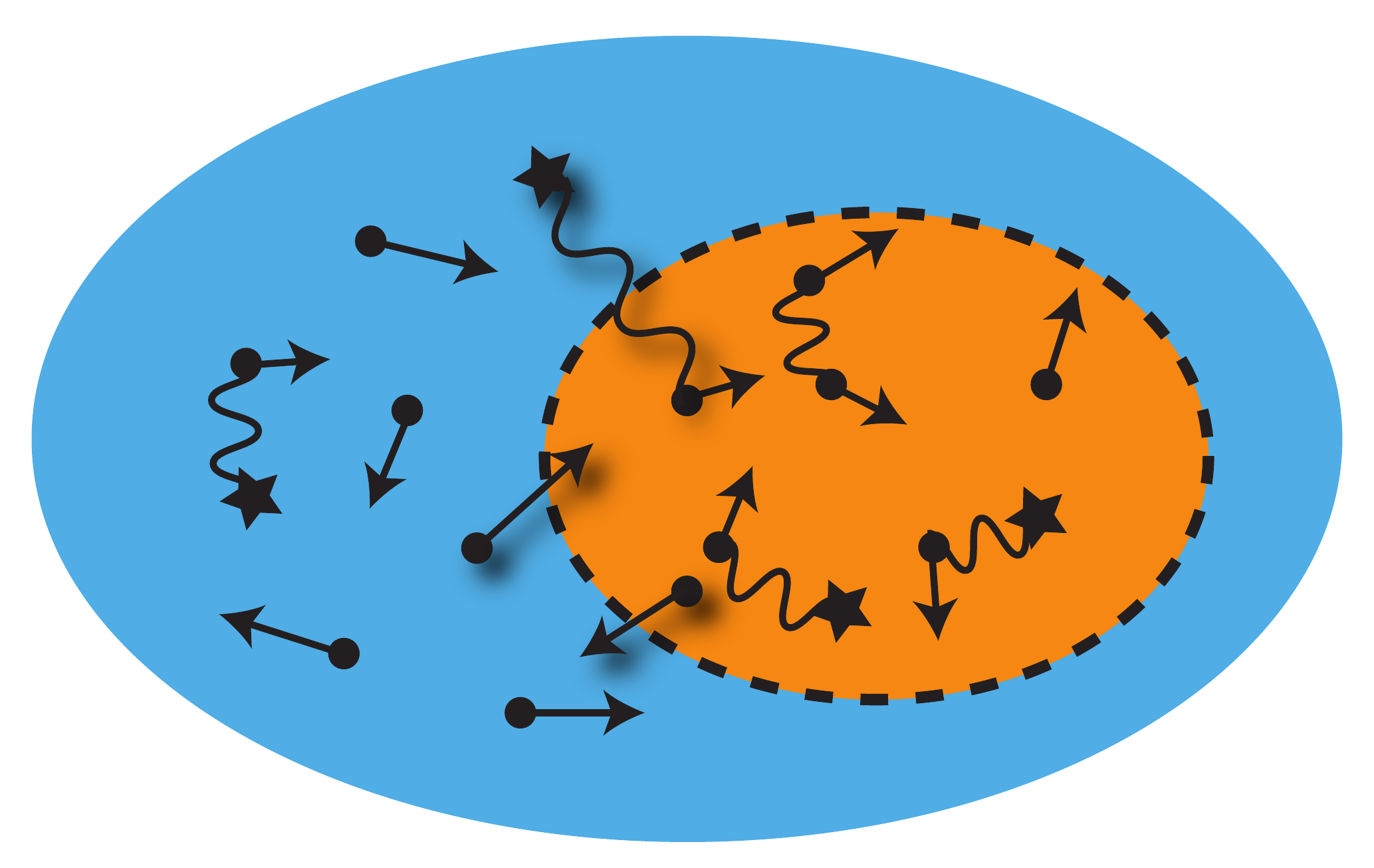}
\caption{Quantum systems couple to their environment by the exchange of particles and energy, and thereby by processes connected to the kinetic freedom of motion and the interactions of the various components.}
\label{fig1}
\end{figure}

\subsection{Preview}
With these concepts at hand, we can now define our mission---to provide a statistical description of open quantum systems in terms of random matrices. This succeeds in situations where we can apply statistical considerations also to the complex dynamics in the region or part of interest, with constraints only arising from fundamental symmetries. We describe both settings in their purest incarnation.

(i) Our main focus is the elastic scattering of a non-interacting particle, which can undergo complex dynamics in the region of interest but enters and leaves in predictable ways. This is quantified in terms of the amplitudes of the incoming and outgoing waves, which are linearly related by a unitary scattering matrix $S$.
As this pure setting is stationary, we can work in the energy domain, while time scales follow when we consider variations in energy.
This setting also covers decay processes, where we initially  confine the particle  within the region of interest---effectively, this is described by a non-hermitian Hamiltonian, with eigenvalues that coincide with the poles of the scattering matrix.

(ii) In a small detour at the end of these notes, we consider purely interacting systems, with localized degrees of freedom that cannot move but evolve under the influence of their mutual environment. We quantify this in terms of a reduced density matrix, a hermitian, positive semidefinite matrix which represents the quantum state when one ignores the other degrees of freedom. We again assume complex internal dynamics,
and consider entropies that quantify entanglement.

As indicated, we will encounter, amongst others, random hermitian Hamiltonians, unitary scattering matrices, positive semidefinite density and time-delay matrices, and non-hermitian effective Hamiltonians.
These are all naturally linked to canonical random-matrix ensembles (not surprisingly, as many of these ensembles were developed with such applications in mind), which we review in Chapter \ref{chap:found}.
In Chapter \ref{chap:scatt} we formulate effective scattering models that link these ensembles to physical effects. Chapter \ref{chap:scattres} provides an overview of key results concerning the decay, dynamics and transport, where we focus on systems with fully random internal dynamics.
Chapter \ref{chap:loc} describes how localizing effects in low dimensions let systems depart from this ergodic behaviour, first  for non-interacting systems and then in the context of interacting systems. Chapter \ref{chap:concl} gives a brief outlook, while the Appendix collects some simple derivations of relevant eigenvalue densities.

\section{Foundations of random-matrix theory}
In this chapter we review a range of classical random-matrix ensembles against the backdrop of closed-system behaviour, which informs the subsequent applications to open systems.

\label{chap:found}
\subsection{Random Hamiltonians and Gaussian hermitian ensembles}

Random-matrix descriptions in quantum mechanics naturally start out with considerations of closed systems. In this setting, the main object of interest is the Hamiltonian $\hat H$, whose eigenvalues give the energy levels. The energy spectrum can be characterised very neatly if one manages to identify a number of conserved  quantities that commute with the Hamiltonian and amongst each other; considering joint eigenstates of these quantities helps to bring some order to the spectrum. In sufficiently complex systems, however, effects such as chaotic or diffractive  scattering and interactions eliminate all conserved quantities, and the energy spectrum lacks any apparent regularities. It is natural to compare the resulting features with the  case where the Hamiltonian can be considered as random. This was first proposed in the 1950's by  \shortciteN{wigner1956conf}, who sought ways to analyse resonances in heavy nuclei. The idea is to focus on a suitable  energy range,  where the local spectral properties can then be studied by replacing the full Hamiltonian with a randomly chosen $M\times M$-dimensional hermitian matrix (the limit $M\to\infty$ can be imposed later on).

The quality of this descriptions  depends on the identification of a suitable random-matrix ensemble. To achieve this task we are allowed to incorporate any general feature of the system. These are, in particular, fundamental symmetries, rough geometric constraints such as dimensionality, as well as natural  time and energy scales.

The consideration of fundamental symmetries leads to ten symmetry classes
\shortcite{Zirnbauer1996,RMTHandbookZirnbauer,RevModPhys.87.1037}.
These comprise the three Wigner-Dyson classes based on time-reversal symmetry \shortcite{dyson_statistical_1962-I,Porter1965,mehta2004random,Guhr1998189,haake_quantum_2010}, three corresponding classes with chiral symmetry \shortcite{PhysRevLett.72.2531,doi:10.1146/annurev.nucl.50.1.343}, and four classes based on a charge-conjugation symmetry \shortcite{PhysRevB.55.1142}. These classes are developed in the present section, and listed in Table \ref{table1}. We also describe the corresponding hermitian matrix ensembles for the simplest situation, geometrically featureless systems in which the only relevant energy scale is the mean level spacing $\Delta$. This reasonably applies when all system-specific information becomes indiscernible after a short time $T_{\rm erg}$, which in particularly is much shorter than the Heisenberg time $T_H=2\pi \hbar/\Delta$ (the minimal observation time at which individual energy levels can be resolved).
Examples where this is realised are sufficiently featureless disordered
\shortcite{efetov_supersymmetry_1996} or classically chaotic systems \shortcite{stockmann_quantum_2006,haake_quantum_2010}. The short-ranged level statistics then becomes universal, and can be captured by ensembles with Gaussian statistics of the matrix elements \shortcite{mehta2004random,Guhr1998189,haake_quantum_2010,forrester2010log}.

\begin{table*}[t]
%\tableparts
%{
\caption{Fundamental symmetries of hermitian random-matrix ensembles}
\label{table1}
%}
%{
\begin{tabular}{l@{\hspace{.5cm}}l@{\hspace{.5cm}}l@{\hspace{.5cm}}l@{\hspace{.5cm}}l}
\hline
class &
symmetries &  constraints & realization ($H_{mn}^*=H_{nm}$) & Gaussian ensemble\\
\hline
 & & & & \\[-6pt]
 UE &
 no symmetries  &none besides $H=H^\dagger$ & $H_{nm}\in \mathbb{C}$
 & GUE
 \\[3pt]
  OE &
 $\mathcal{T}=K$ & $H^*=H$ &  $H_{nm}\in \mathbb{R}$
 & GOE
 \\[3pt]
 SE &
  $\mathcal{T}=\Omega K$ & $H^*=\Omega H\Omega^{-1}$ & $H_{nm}\in \mathbb{H}$
 & GSE
 \\[3pt]
  RE &
$\mathcal{C}=K$ & $H^*=-H$ &  $H_{nm}\in i\mathbb{R}$
& GRE
\\[3pt]
  T-RE &
$\mathcal{C}=K$, $\mathcal{T}=\Omega K$ & $H^*=-H=\Omega H\Omega^{-1}$ &
$H_{nm}=-\sigma_yH_{nm}\sigma_y\in \mathbb{H}$
& T-GRE
\\[3pt]
  QE &
$\mathcal{C}=\Omega K$ & $H^*=-\Omega H\Omega^{-1}$ & $H_{nm}\in i \mathbb{H}$
& GQE
\\[3pt]
  T-QE &
$\mathcal{C}=\Omega K$, $\mathcal{T}= K$ & $H^*=H=-\Omega H\Omega^{-1} $& $H_{nm}=-\sigma_yH_{nm}\sigma_y\in i\mathbb{H}$
& T-GQE
\\[3pt]
  chUE &
$\mathcal{X}=\tau_z\equiv{\rm diag}\,(1_{M_1},-1_{M_2})$ & $H=-\tau_z H\tau_z$ & $H=\left(  \begin{array}{cc} 0 & A \\ A^\dagger & 0
\end{array}\right)$, $A_{nm}\in \mathbb{C}$
& chGUE
\\[3pt]
  chOE &
$\mathcal{X}=\tau_z$, $\mathcal{C}=K$ ($\mathcal{T}=\mathcal{X}\mathcal{C}$) & $H=-\tau_z H\tau_z=-H^*$ & $H=\left(  \begin{array}{cc} 0 & A \\ A^\dagger & 0
\end{array}\right)$, $A_{nm}\in i\mathbb{R}$
& chGOE
\\[3pt]
  chSE &
$\mathcal{X}=\tau_z$, $\mathcal{C}=\Omega K$ ($\mathcal{T}=\mathcal{X}\mathcal{C}$) & $H=-\tau_z H\tau_z=-\Omega H^*\Omega^{-1}$ & $H=\left(  \begin{array}{cc} 0 & A \\ A^\dagger & 0
\end{array}\right)$, $A_{nm}\in i\mathbb{H}$
& chGSE
\\[3pt]
\hline
\end{tabular}
%}
\end{table*}

\subsubsection{Time-reversal symmetry and the Wigner-Dyson ensembles}
We start by considering the role of time reversal \shortcite{dyson_statistical_1962-I,haake_quantum_2010}, instituted by an anti-unitary operator $\mathcal{T}$ fulfilling $\langle\mathcal{T}\phi|
\mathcal{T}\psi\rangle=\langle\psi|\phi\rangle=\langle\phi|\psi\rangle^*$, which consequently may square to $\mathcal{T}^2=1$ or $\mathcal{T}^2=-1$.

If the Hamiltonian obeys a time-reversal symmetry $\mathcal{T}H\mathcal{T}^{-1}=H$ with $\mathcal{T}^2=1$, we can adopt an invariant  basis $|n\rangle$ in which
$\langle\mathcal{T} n|\psi\rangle=\langle n|\psi\rangle$ for any $|\psi\rangle$. This implies
 $\langle n|\mathcal{T}\psi\rangle=\langle n|\psi\rangle^*$, so that the time-reversal operation $\mathcal{T}=K$ amounts to the complex conjugation of
the expansion coefficients $\psi_n=\langle n|\psi\rangle$ of any state.  In this basis the matrix elements $H_{lm}=\langle l|\hat H|m\rangle=\langle \mathcal{T}l|\hat H \mathcal{T}|m\rangle=H_{lm}^*$ are real, while hermiticity implies that the matrix is symmetric, $H_{ml}=H_{lm}$.  This is known as the \emph{orthogonal symmetry class} (OE), to which we associate the symmetry index $\beta=1$.

In absence of any time-reversal symmetry,  matrix elements of the Hamiltonian are in general complex, with  $H_{lm}=H_{ml}^*$ because of hermiticity, which defines the \emph{unitary symmetry class} (UE) with symmetry index $\beta=2$.

If we have a time-reversal symmetry $\mathcal{T}H\mathcal{T}^{-1}=H$ with $\mathcal{T}^2=-1$ (\emph{symplectic symmetry class} SE with symmetry index $\beta=4$), we can adopt a basis arranged in pairs $|n\rangle=\mathcal{T}|\bar n\rangle$, so that the Hilbert space dimension $2M$ must be even. In this basis, $T=\Omega K$ where $\Omega=i\sigma_y\otimes 1_M$, while the blocks
$\left(  \begin{array}{cc} H_{lm} & H_{l\bar m} \\ H_{\bar lm} & H_{\bar l\bar m}
\end{array}\right)=a_{lm} 1+ib_{lm}\sigma_x+ic_{lm}\sigma_y+id_{lm}\sigma_z \in \mathbb{H}$ can be reinterpreted as quaternions,  with real coefficients $a_{lm},b_{lm},c_{lm},d_{lm}$ and Pauli matrices $\sigma_r$.  Hermiticity requires that $a_{lm}=a_{ml}$ forms a symmetric matrix while $b_{lm}=-b_{ml}$, $c_{lm}=-c_{ml}$, $d_{lm}=-d_{ml}$ are antisymmetric.
Expressed as an $M\times M$-dimensional matrix of quaternions, $H=\overline{H}$ is then seen to be quaternion self-conjugate, where by definition $(\overline{H})_{lm}=\overline{H_{ml}}=a_{ml} 1-ib_{ml}\sigma_x-ic_{ml}\sigma_y-id_{ml}\sigma_z$. For such a matrix, all energy levels appear in degenerate pairs, a phenomenon known as Kramers degeneracy; in all  following considerations we count each  pair as a single level. In keeping with this, the quaternion trace is defined as ${\rm tr}\,H=\sum_n a_{nn}$ (so differs by a factor of two from the conventional trace), and the quaternion determinant is similarly modified to maintain the relation $\mathrm{det}\,\exp  A = \exp\,\mathrm{tr}\,A$, which makes it equivalent to a Pfaffian \shortcite{dyson1970}.

The symmetry index $\beta=1,2,4$ mentioned above counts the real degrees of freedom in the matrix elements. The corresponding notions of orthogonal, unitary and symplectic symmetry classes
refer to the transformations  $H=U\,D\,U^\dagger$, $D=\mathrm{diag} (E_n)$ that diagonalise these Hamiltonians. For $\beta=1$ the matrix $U$  is orthogonal, $UU^T=1$, and hence belongs to the group $\mathrm{O}(M)$; for $\beta=2$ $U\in\mathrm{U}(M)$ is a unitary matrix with $UU^\dagger=1$, and for $\beta=4$ the matrix is unitary symplectic, $U\in\mathrm{Sp}(2M)$  with $U\overline{U}=1$.
This `threefold way' can be further justified within representation theory \shortcite{dyson_threefold_1962}.

Within these three Wigner-Dyson classes,
the universal spectral features encountered in ergodic systems are captured by the Gaussian orthogonal, unitary, or symplectic ensemble (GOE, GUE, GSE), where the Hamiltonian obeys a probability density of the form $P(H)\propto \exp(-c_\beta \,{\rm tr}\,H^2)$ with $c_\beta=\beta\pi^2/4M\Delta^2$.
The spectral statistics can then be determined from the joint probability distribution
\begin{equation}
\label{eq:pgauss}
P(\{E_n\})\propto \prod_{n<m}|E_n-E_m|^\beta\prod_k\exp(-c_\beta E_k^2),
\end{equation}
which follows by a change of variables from the Hamiltonian to its eigenvalues and eigenvectors.
This result can be obtained by sophisticated  methods in the language of differential geometry \shortcite{forrester2010log}, but in this specific incarnation also follows from elementary means and then acquires a simple geometric meaning.
Given that $dH=dU D U^\dagger +U dD U^\dagger -U D U^\dagger U U^\dagger$, consider
the squared line element
\begin{align}
&\sum_{lm} |dH_{lm}|^2=\mathrm{tr}\,(dHdH)
\nonumber\\
&\quad=\mathrm{tr}\,(dXD-DdX)^\dagger(dXD-DdX) + \sum_m (dE_m)^2,
\end{align}
where $D$ contains $M$ real parameters (the eigenvalues) while $d X=-iU^\dagger dU$ depends on $M(M-1)/2$ real, complex or quaternion parameters in the set of eigenvectors. The latter parameters can be  associated with the rotations $R^{(nm)}$ in the $nm$ plane of the diagonalised system, spanned by the eigenvectors with eigenvalues $E_n$ and $E_m$. Each of these rotations then translates into a line element in the space of Hamiltonians of length $\propto |E_n-E_m|^\beta$, where the power arises from the fact that the rotation is parameterised by $\beta$ real variables. (In particular, if we rotate the basis in a degenerate subspace the Hamiltonian does not change.) Hence $d\mu(H) \propto \left(\prod_{n<m}|E_n-E_m|^\beta\right) \left( \prod_k dE_k\right) d\mu(U)$, where $\mu(U)$ is the Haar measure arising from the form $dX$ in the corresponding group of transformations. This measure is uniquely defined by the requirement that it is invariant under $U\to V' U V$ for any fixed $V$, $V'$ form the same group.

The main characteristics of  \eqref{eq:pgauss}
is a universal degree of level repulsion $P(s)\sim s^\beta$ for small level spacings $s=|E_n-E_m|$ in the bulk of the spectrum. This feature was first realised by  Wigner, who put forward the famous surmise $P(s)\sim s^\beta \exp(-c s^2)$ with a suitable scale factor $c$
\shortcite{Porter1965}.
As it turned out,
this surmise is exact only for $M=2$, but provides a very accurate estimate for any $M$. The exact result can be established by  the method of orthogonal polynomials (here based on Hermite polynomials), which provides the complete set of correlation functions \shortcite{mehta2004random}.  When applied to a particular system, these correlations describe the short-ranged statistics in the bulk, i.e., over sufficiently small spectral ranges where the mean level spacing $\Delta$ is well defined  (possibly, after some unfolding of the spectrum).  In particular, the amount of level repulsion is considered as a prime indicator of whether a system displays the required ergodic dynamics, as further discussed in Chapter \ref{chap:loc}.

The mean level spacing itself is not universal; in real systems it varies systematically with energy, but for any comparison we wish to have it well defined in any given ensemble. In the Gaussian ensembles, this is guaranteed by the form of the eigenvalue density, which for large matrix dimensions $M\to\infty$ approaches the famous Wigner semicircle law \shortcite{wigner_distribution_1958}
\begin{equation}
\label{eq:semi}
\rho(E)
=\frac{1}{ \Delta }\sqrt{1-E^2/E_0^2}\quad \mbox{for }|E|<E_0=2M\Delta /\pi.
\end{equation}
A derivation of this classical result is given in the Appendix.
It reveals that $\Delta=1/\rho(0)$ is the mean level spacing at $E=0$,
defining the middle of the bulk around which we then determine the universal spectral features. Universal level statistics are also encountered around the spectral edges $\pm E_0$,
%%% (see the accompanying notes by \shortciteANP{ThisIssueVirag}),
whose actual positions are again  system specific.

\subsubsection{Chiral symmetry}

Additional positions within the spectrum deserve dedicated attention when further symmetries come into play. In particular, this is encoutered when the Hamiltonian is antisymmetric under a suitable unitary or antiunitary   transformation, an effect which often occurs in single-particle descriptions of fermions. Energy levels then appear in pairs $E_n$, $E_{\tilde n}=-E_n$, with the possible exception of levels pinned to the spectral symmetry point $E=0$.

If $\mathcal{X}H\mathcal{X}=-H$ with a unitary  involution $\mathcal{X}$ (such that $\mathcal{X}^2=1$), we
talk of a chiral symmetry \shortcite{PhysRevLett.72.2531,doi:10.1146/annurev.nucl.50.1.343}.
For a finite system of dimension $M=M_1+ M_2$, we can choose $\mathcal{X}={\rm diag}\,(1_{M_1},-1_{M_2})\equiv \tau_z$, so that the Hamiltonian takes a block form
\begin{equation}
\label{eq:chiralh}
H=\left(  \begin{array}{cc} 0 & A \\ A^\dagger & 0
\end{array}\right)
\end{equation}
where $A$ is an $M_1\times M_2$-dimensional rectangular matrix.

The chiral symmetry arises in elementary particle physics  \shortcite{doi:10.1146/annurev.nucl.50.1.343,ThisIssueAkemannPreprint}, but can also be realised as an effective symmetry in electronic
\shortcite{brouwer_zero_2002}, superconducting \shortcite{PhysRevLett.100.096407}
and photonic systems \shortcite{schomerus_parity_2013,lu_topological_2014,poli_selective_2015}.
Given the structure \eqref{eq:chiralh}, the symmetry generally applies to systems with two sublattices, termed A and B, when the couplings within each isolated sublattice vanish \shortcite{PhysRevB.34.5208}.
The mentioned electronic and photonic implementations  naturally extend this idea to suitably coupled subsystems.

An interesting aspect of these classes is the appearance of topological invariants, associated with the number of eigenenergies pinned to the symmetry point
\shortcite{PhysRevLett.62.1201,PhysRevLett.72.2531,brouwer_zero_2002}.
For a Hamiltonian of the form
\eqref{eq:chiralh} with some finite $\nu=M_2-M_1$ (so that $A$ is not square), there are at least $|\nu|$ such zero modes. If  $\nu<0$ the associated eigenstates are of the form $\psi=(\psi_A,0)^T$ with $A^\dagger\psi_A=0$, while for $\nu>0$  we have
$\psi=(0, \psi_B)^T$ with $A\psi_B=0$.
The remaining paired levels with finite energy can be determined from the positive definite matrix  $A^\dagger A$ or $AA^\dagger$, whose eigenvalues are given by $E_n^2$.

In combination with considerations of time-reversal symmetry one can now define \emph{chiral orthogonal, unitary or symplectic symmetry classes} (chOE, chUE, chSE) \shortcite{PhysRevLett.72.2531,doi:10.1146/annurev.nucl.50.1.343,ThisIssueAkemannPreprint}, which are again associated with a symmetry index $\beta=1,2,4$.
Taking $A$ as a random matrix with real, complex, or quaternion entries and $P(A)\propto \exp(-c_\beta \,{\rm tr}\,A^\dagger A)$  then leads to the
Gaussian chiral ensembles (chGOE, chGUE, and chGSE), for which the positive energy levels in each pair follow the joint distribution
\begin{align}
 \label{eq:pchiral}
P(\{E_n\})\propto \prod_{\substack{n<m,\\ E_{n,m}>0}}|E_n^2-E_m^2|^\beta\prod_{k, E_{k}>0} E_k^{(|\nu|+1)\beta-1}
e^{-c_\beta E_k^2}.
%\exp(-c_\beta E_k^2).
\end{align}
The terms $E_n^2-E_m^2=(E_n-E_m)(E_n+E_m)$ include the repulsion from the negative-energy levels,
while $E_k^{(|\nu|+1)\beta-1}$ includes the repulsion from the mirror level at $E_{\tilde k}=-E_k$ and from the zero modes. This modified repulsion follows again from the geometric argument above, where the subspace to be explored by the rotations $R^{(nm)}$ corresponds to the case   $M_1=|\nu|+1$, $M_2=1$. In this space, $A$ becomes a vector and the eigenvalues and the squared eigenvalues $E_n^2=E_{\tilde n}^2=|A|^2$ obey a $\chi^2$ distribution.

These modifications affect the eigenvalue density around $E=0$ over a range of a few level spacings,
\begin{equation}
\label{eq:rhoEd}
\rho(E)-|\nu|\delta(E)\propto |E|^{(|\nu|+1)\beta-1}\quad\mbox{for small }|E|,
\end{equation}
which now becomes a universal spectral characteristic of the system.
For a macroscopic number of zero modes with $M_2\gg M_1\gg1$, the repulsion yields a  hard gap around the symmetry point, corresponding to the mean density
\begin{align}
\rho(E)&=\frac{\pi}{M_1\Delta^2 E}\sqrt{(E^2-E_-^2)(E_+^2-E^2)},
\nonumber \\
 E_\pm&=\frac{M_1\Delta}{\pi}(\sqrt{M_2/M_1}\pm 1)
\label{eq:rhochiral}
\end{align}
for the $M_1$ positive eigenvalues (this expression  follows from the Marchenko-Pastur law derived in the Appendix). For $M_1=M_2\gg 1$ this eigenvalue density reverts to a Wigner semicircle law \eqref{eq:semi}, normalised to $2M_1$ eigenvalues in the whole energy range (the level repulsion \eqref{eq:rhoEd} is not resolved as in this limit $\Delta\to 0$).

\subsubsection{Charge-conjugation symmetry}
If we admit for an antisymmetry $\mathcal{C}\mathcal{H}\mathcal{C}^{-1}=-\mathcal{H}$ with an antiunitary operator $\mathcal{C}$ we encounter four additional cases
\shortcite{PhysRevB.55.1142}. Two of these arise from the choices $\mathcal{C}^2=\pm 1$, while the other two arise from
an additional time-reversal symmetry with $\mathcal{T}^2=-\mathcal{C}^2$.

If the antisymmetry is $\mathcal{C}=K$ ($\beta=\beta'=2$), the Hamiltonian is imaginary and antisymmetric, $H=-H^*=-H^T$, and can be written in terms of matrix elements $H_{nm}\in i\mathbb{R}$. It is useful to denote this as the \emph{real symmetry class} (RE) \shortcite{RevModPhys.87.1037}.
If we have in addition a time-reversal symmetry $\mathcal{T}=\Omega K$  ($\beta=4$, $\beta'=3$) we can write
the Hamiltonian in the block form
\begin{equation}
\label{eq:block}
H=\left(  \begin{array}{cc} A & B \\ B & -A
\end{array}\right),
\end{equation}
where $A=-A^T$, $B=-B^T$ are antisymmetric and  $A_{nm}, B_{nm}\in i \mathbb{R}$. This can be usefully denoted as the \emph{time-invariant real symmetry class} (T-RE).

For the antisymmetry  $\mathcal{C}=\Omega K$ ($\beta=2$, $\beta'=0$) the Hamiltonian $\overline{H}=-H$ is anti-selfconjugate,
and thus can be written in terms of matrix elements $H_{nm}\in i\mathbb{H}$. If in addition we also have the time-reversal symmetry
$T=K$ ($\beta=1$, $\beta'=0$), the Hamiltonian takes the block form
\eqref{eq:block}
with symmetric matrices $A=A^T$, $B=B^T$ and elements $A_{nm}, B_{nm}\in \mathbb{R}$. The two cases define the \emph{quaternion symmetry class} (QE) and the \emph{time-invariant quaternion symmetry class} (T-QE)

In the two classes with $\mathcal{C}^2=1$, where the Hamiltonian can be made anti-symmetric by an appropriate basis choice, a topologically protected zero mode exists if $M$ is odd (when we have an additional time-reversal symmetry with $\mathcal{T}^2=-1$ this mode is Kramers-degenerate). The topological invariant counting such modes is then set to $\nu=1$, while for even  $M$ we set $\nu=0$. No such symmetry-protected zero modes exist in the two classes with $\mathcal{C}^2=-1$.

Adopting again a Gaussian distribution $P(H)\propto \exp[-(c_\beta/2) \,{\rm tr}\,H^2]$
of matrix elements, these symmetry classes provide the joint probability density
\begin{equation}
P(\{E_n\})\propto \prod_{\substack{n<m,\\ E_{n,m}>0}}|E_n^2-E_m^2|^\beta\prod_{k, E_k>0} E_k^{(|\nu|+1)\beta-\beta'}
e^{-c_\beta E_k^2},
%\exp(-c_\beta E_k^2),
\end{equation}
where $\beta'$ modifies the repulsion from the mirror level as specified above (this follows again from the geometric argument in the small subspaces spanned by a level pair and any zero modes).
As in the chiral classes, the spectral symmetry and the zero mode thus directly affect the level statistics in the closed system.

The symmetry associated with $\mathcal{C}$ is known as a charge-conjugation or particle-hole symmetry, and arises naturally in the context of superconducting systems. In a mean-field description, excitations are described as quasi-particles that obey the Boguliubov-de Gennes Hamiltonian
\begin{equation}
\mathcal{H}=\left(  \begin{array}{cc} H_0-E_F & -i\sigma_y\otimes\Delta \\ i\sigma_y\otimes\Delta^* & E_F-H_0^*
\end{array}\right),
\end{equation}
where the blocks refer to the electron-like and hole-like degrees of freedom (addressed by Pauli matrices $\tau_i$), the Pauli matrix $\sigma_y$ acts in spin space, and $\Delta=\Delta^T$ is the s-wave pair potential. The charge-conjugation is of the form $\mathcal{C}=\tau_x K$ and squares to $\mathcal{C}^2=1$. If $H_0=H_+\oplus H_-$ and $\Delta=\Delta_+\oplus \Delta_-$ preserve the spin we can rearrange the Hamiltonian into two systems with
$\mathcal{H}_\pm=\left(  \begin{array}{cc} H_\pm-E_F & \mp\Delta_\pm \\ \mp\Delta_\pm^* & E_F-H_\pm^*
\end{array}\right)$, for which the charge-conjugation symmetry $\mathcal{C}=\Omega K$ with $\Omega=i\tau_y$ squares to $\mathcal{C}^2=-1$.

In this setting,  the zero modes
in the classes with $\mathcal{C}^2=1$ are associated with Majorana fermions \shortcite{0034-4885-75-7-076501,0268-1242-27-12-124003,doi:10.1146/annurev-conmatphys-030212-184337}, previously elusive quasi-particles with possible applications for topological quantum computation \shortcite{RevModPhys.80.1083}.
These concepts can be generalised to surface and interface states in systems of specified spatial dimensions \shortcite{:/content/aip/proceeding/aipcp/10.1063/1.3149495,PhysRevB.82.115120,1367-2630-12-6-065010}, which are encountered in topological insulators and superconductors
\shortcite{RevModPhys.82.3045,RevModPhys.83.1057}.

\subsection{Random time-evolution operators and circular ensembles}
To prepare how these considerations about the Hamiltonian translate to open systems, it is useful to turn to the dynamics and identify the corresponding symmetry classes of unitary matrices  that exemplify  the time evolution in the system. Of particular interest is the time evolution over a fixed time interval $T_0$, which also admits situations in which the Hamiltonian is itself time-dependent with that period. With a nod to the notion of a Floquet-operator in the latter setting, we denote this stroboscopic time-evolution operator over a fixed time interval as $F$. Its eigenvalues $z_n=\exp(-i\varepsilon_n)$ lie on the unit circle, where the phases
$\varepsilon_n$ can be interpreted as quasi-energies.
Similar considerations apply to quantum maps \shortcite{haake_quantum_2010} and quantum walks \shortcite{2010PhRvB..82w5114K}.

As the time evolution is generated by the Schr{\"o}dinger equation \eqref{eq:schr}, we can symbolically write  $F=\exp(-iH T_0/\hbar)$ with a suitable effective Hamiltonian $H$. The symmetries of $F$ then follow from the symmetries of $H$, and thus comply with the ten symmetry classes described above \shortcite{Zirnbauer1996}.
In the resulting spaces of unitary matrices,
some segments are smoothly connected to the identity,
while others form disconnected pieces. This once more provides scope for topological invariants \shortcite{PhysRevB.83.155429,RevModPhys.87.1037}, which we specify in the following explicit constructions.

\begin{table*}[t]
%\tableparts
%{
\caption{Classification of unitary matrix ensembles}
\label{table2}
%}
%{
\begin{tabular}{l@{\hspace{.5cm}}l@{\hspace{.5cm}}l@{\hspace{.5cm}}l@{\hspace{.5cm}}l@{\hspace{.5cm}}l}
\hline\\[-.7cm]
class &
symmetries & unitary  matrices & space& \parbox[b]{1cm}{\flushleft Cartan label} & \parbox[b]{1.28cm}{\flushleft{}circular ensemble}\\
\hline
 & & &  \\[-6pt]
 UE &
no symmetries        & $F^{-1}=F^\dagger$  &$\mathrm{U}(M)$ & A
& CUE
 \\[3pt]
 OE &
 $\mathcal{T}=K$ &   $F^{-1}=F^*$  &$\mathrm{U}(M)/\mathrm{O}(M)$ & AI
 & COE
 \\[3pt]
  SE &
 $\mathcal{T}=\Omega K$ &  $F^{-1}=\Omega F^*\Omega^{-1}$  &$\mathrm{U}(2M)/\mathrm{Sp}(2M)$ & AII
  & CSE
 \\[3pt]
  RE &
 $\mathcal{C}= K$ & $F^{-1}=F^T$  &$\mathrm{O}(M)$ & D
   & CRE
 \\[3pt]
   T-RE &
 $\mathcal{C}=K$,  $\mathcal{T}=\Omega K$& $F^{-1}=F^T=\Omega F \Omega^{-1}$  &$\mathrm{O}(2M)/\mathrm{U}(M)$ & DIII
    & T-CRE
\\[3pt]
    QE &
 $\mathcal{C}=\Omega K$& $F^{-1}=\Omega F^T\Omega^{-1}$  &$\mathrm{Sp}(2M)$ & C
 & CQE
  \\[3pt]
    T-QE &
 $\mathcal{C}=\Omega K$, $\mathcal{T}=K$& $F^{-1}=F^*=\Omega F \Omega^{-1}$  &$\mathrm{Sp}(2M)/\mathrm{U}(M)$ & CI
  & T-CQE
 \\[3pt]
     chUE &
 $\mathcal{X}=\tau_z$ & $(\mathcal{X}F)=(\mathcal{X}F)^\dagger$ &$\mathrm{U}(M_1+M_2)/\mathrm{U}(M_1)\otimes \mathrm{U}(M_2)$ & AIII
   & chCUE
 \\[3pt]
     chOE &
$\mathcal{X}=\tau_z$, $\mathcal{C}=K$
& $(\mathcal{X}F)=(\mathcal{X}F)^T=(\mathcal{X}F)^*$  &$\mathrm{O}(M_1+M_2)/\mathrm{O}(M_1)\otimes \mathrm{O}(M_2)$ & BDI
   & chCOE
 \\[3pt]
     chSE &
 $\mathcal{X}=\tau_z$, $\mathcal{C}=\Omega K$
& $(\mathcal{X}F)=(\mathcal{X}F)^\dagger=\Omega(\mathcal{X}F)^*\Omega^{-1}$  &$\mathrm{Sp}(2M_1+2M_2)/\mathrm{Sp}(2M_1)\otimes \mathrm{Sp}(2M_2)$ & CII
   & chCSE
 \\[3pt]
\hline
\end{tabular}
%}
\end{table*}

For the time-evolution operator, time-reversal symmetry implies $\mathcal{T}F\mathcal{T}^{-1}=F^{-1}$.
Given a time-reversal symmetry with $\mathcal{T}^2=1$
(orthogonal symmetry class with $\beta=1$) and adopting  a canonical basis where this is represented by $\mathcal{T}=K$, we find that $F$ is symmetric under transposition, $F=F^T$.
In absence of any symmetries (unitary symmetry class with $\beta=2$), $F$ is only constrained by $F^{-1}=F^\dagger$, so a member of the unitary group $\mathrm{U}(M)$.
For time-reversal symmetry with $\mathcal{T}^2=-1$ (symplectic symmetry class with $\beta=4$), the choice $\mathcal{T}=\Omega K$ implies that $F=\overline{F}$ is quaternion self-conjugate.
The matrix $F_\Omega=\Omega F$ with elements $F_{\Omega,nm}=i\sigma_yF_{nm}$, written as a normal $2M\times2M$ matrix, is then antisymmetric,
$F_\Omega^T=-F_\Omega$.
Notably, in the two classes arising from time-reversal symmetry, even though denoted as orthogonal and symplectic, the spaces of matrices differ from the groups of orthogonal and symplectic matrices encountered in the diagonalisation of the corresponding Hamiltonians. Only in the case of broken time reversal symmetry the space remains associated with the unitary group.

In each of these three spaces we can again determine a Haar measure $\mu(F)$. This is uniquely defined by the requirement that the measure is invariant under transformations $F\to U' F U$, but now with unitary matrices $U$, $U'$ that are subject to the constraints $U'=U^T$ in the orthogonal symmetry class, and $U'=\overline{U}$ in the symplectic symmetry class.
Equipped with this measure, the corresponding ensembles are known as the circular ensembles (COE, CUE and CSE) \shortcite{dyson_statistical_1962-I}. The joint distributions of phases $\varphi_n$ in the unimodular eigenvalues $z_n=e^{i\varphi_n}$ is given by
\begin{equation}
\label{eq:pphi}
P(\{\varphi_n\})\propto \prod_{n<m}|e^{i\varphi_n}-e^{i\varphi_m}|^{\beta},
\end{equation}
and their density is uniform.

Chiral symmetry implies $\mathcal{X}F\mathcal{X}=F^\dagger$, so that $F_X=\mathcal{X}F$ is hermitian and only has eigenvalues $\pm1$.
A topological invariant can then be defined as $\nu'=\frac{1}{2}{\rm tr}\,(F_X)=(M_+-M_-)/2$, where $M_\pm$ counts the eigenvalues of either sign. One can again introduce a Haar measure, which in combination with the possible constraints from time-reversal symmetry
leads to three chiral circular  ensembles (chCOE, chCUE and chCSE).

A charge-conjugation symmetry implies  $\mathcal{C}F\mathcal{C}^{-1}=F$. When we express $\mathcal{C}=K$ with $\mathcal{C}^2=1$ this implies that $F=F^*$ is real, and thus an element of the orthogonal group  $\mathrm{O}(M)$ (as the label OE is already taken this justifies the notion of the real symmetry class RE). We then have the invariant $\nu'={\rm det}\,F$,  where $\nu'=1$ accounts for matrices from $\mathrm{SO}(M)$. If in addition we have a time-reversal symmetry with $\mathcal{T}=K\Omega$ (class T-RE), such an invariant can be formulated with help of the Pfaffian $\nu'={\rm pf}F_\Omega$ of the real antisymmetric matrix $F_\Omega=\Omega F$.
For $\mathcal{C}=\Omega K$ with $\mathcal{C}^2=-1$, the constraint can be written as $F^T \Omega F=\Omega$, which identifies $F$ as symplectic (in quaternion language, $F\overline{F}=1$, which justifies the notion of the quanternion universality class QE). If in addition we have a time-reversal symmetry with $\mathcal{T}=K$ (class T-QE), the matrix is furthermore constrained to be symmetric. Equipped with a Haar measure, the corresponding real and quaternion circular ensembles are denoted as CRE, T-CRE, CQE and T-CQE \shortcite{RevModPhys.87.1037}.

In a specific mathematical sense, it can now be argued that these ten classes provide a complete classification of random-matrix ensembles \shortcite{Zirnbauer1996,Caselle200441,RMTHandbookZirnbauer}---they arise from the  groups of unitary, orthogonal and symplectic matrices and the associated  compact symmetric Riemannian spaces, as classified by Cartan and summarised in Table \ref{table2}.
The three Wigner-Dyson classes with unitary, orthogonal and sympletic symmetry (UE, OE and SE) are the labelled A, AI, AII; the corresponding chiral classes (chUE, chOE and chSE) are labelled AIII, BDI, CII; the classes with charge-conjugation symmetry $\mathcal{C}^2=1$ and topological invariants (RE and T-RE) are labelled D and DIII, while the remaining to classes with $\mathcal{C}^2=-1$  (QE and T-QE)  are labelled C and CI.

\subsection{Positive-definite matrices and Wishart-Laguerre ensembles}

As we have seen in the construction of the ten Hamiltonian ensembles, it is often useful to study the blocks of a matrix, and compose new matrices out  from them. This leads to natural extensions of the ensembles encountered so far, which can be justified via their connection to orthogonal polynomials  \shortcite{mehta2004random,forrester2010log}.  From this perspective, the Gaussian hermitian matrix ensembles in the Wigner-Dyson classes are related to Hermite polynomials, while the other ensembles are related to Laguerre polynomials.
As mentioned for the chiral symmetry classes, these ensembles are naturally related to positive semidefinite matrices $W=X^\dagger X$, where $X$ is an $M'\times M$-dimensional matrix. It suffices to consider the case $M\leq M'$, as otherwise we can simply study $W=X X^\dagger$.

We again use the symmetry index $\beta=1,2,4$ to distinguish settings where the matrix elements $X_{lm}$ are real, complex or quaternion. A Gaussian distribution
\begin{equation}
\label{eq:px}
P(X)\propto \exp(-c'_{\beta}\,{\rm tr}\,X^\dagger X)
\end{equation}
then defines the Wishart-Laguerre ensemble for $W$, where we set $c'_{\beta}=\beta /2\sigma^2$. This ensemble was first introduced by \shortciteN{wishart_generalised_1928} in the context of multivariate statistics, which marks the historical beginnings of random-matrix applications.
The joint probability density of the eigenvalues $\lambda$ of $W$ is given by
\begin{equation}
\label{eq:pwish}
P(\{\lambda_n\})\propto \prod_{n<m}|\lambda_n-\lambda_m|^\beta
\prod_k \lambda_k^{\beta(1+M'-M)/2 -1}
e^{-c'_\beta\lambda_k},
%\exp(-c'_\beta\lambda_k),
\end{equation}
which relates to the previously encountered eigenvalue distributions by the substitution $\lambda_n=E_n^2$.
As mentioned above, the resulting eigenvalue correlations can be expressed in terms of Laguerre polynomials.

For large matrix dimensions the eigenvalue density approaches the Marchenko-Pastur law \shortcite{0025-5734-1-4-A01}
\begin{equation}
\label{eq:mp}
\rho(\lambda)=\frac{MT_0}{2\pi\lambda}\sqrt{(\lambda-\lambda_-)(\lambda_+-\lambda)}
\quad\mbox{for }\lambda_-<\lambda<\lambda_+,
\end{equation}
where $\lambda_\pm=(\sqrt{M'}\pm \sqrt{M})^2/\sigma^2$ defines the range where this density is finite.
This expression is derived in the Appendix.

\subsection{Jacobi ensembles}
A third class of classical orthogonal polynomials appearing in random-matrix problems are the Jacobi polynomials. These are associated with joint probability distributions of the form
\shortcite{forrester2010log}
\begin{equation}
\label{eq:jac0}
P(\{\mu_n\})\propto\prod_{n<m}|\mu_n-\mu_m|^\beta\prod_k(1-\mu_k)^{a\beta/2}(1+\mu_k)^{b\beta/2},
\end{equation}
where $\mu_m\in[-1,1]$, $m=1,2,3,\ldots,M$.

Such distributions arise, for instance, when one considers the singular values of an $M'\times M$ dimensional off-diagonal block $t$ of a suitable $N\times N$ dimensional unitary matrix $F$ \shortcite{RevModPhys.69.731,RevModPhys.87.1037}. In particular, setting $N=M+M'$ with $M'\geq M$  and generating $F$ from the three standard circular ensembles (COE, CUE or CSE), the eigenvalues $T_n=(1-\mu_n)/2\in[0,1]$ of $t^\dagger t$ obey a Jacobi ensemble with $a=M'-M+1-2/\beta$, $b=0$; similarly, if $F$ is taken from $\mathrm{O}(M+M')$ or $\mathrm{Sp}(2M+2M')$ (symmetry class D or C) one finds the same $a$ but $b=1-2/\beta$; the complete picture is presented in Section  \ref{sec:transport}.

Alternatively \shortcite{forrester2010log}, the quantities $T_n$ can be interpreted as the eigenvalues of a
so-called MANOVA matrix $(X^\dagger X+Y^\dagger Y)^{-1}X^{\dagger} X$,
where $X$ and $Y$ are matrices of dimensions $M_x\times M$ and  $M_y\times M$, distributed as Gaussians with equal variance $\sigma$ according to Eq.~\eqref{eq:px}. In this case,
$a=M_x-M+1-2/\beta$, $b=M_y-M+1-2/\beta$.
As shown based on this realization in the Appendix, in the limit of a large matrix dimension $M$ with fixed  $c_x=M_x/M$, $c_y=M_y/M$ the eigenvalue density approaches
\begin{equation}
\label{eq:rhot2}
\rho(T)=\frac{M(c_x+c_y)\sqrt{(T-T_-)(T_+-T)}}{2\pi T(1-T)},
\end{equation}
where
\begin{equation}
T_\pm=\frac{1}{1+\lambda_\mp},\quad \lambda_\pm=\left(\frac{\sqrt{c_xc_y}\pm\sqrt{c_x+c_y-1}}{c_x-1}\right)^2
\end{equation}
determines the range where the density is finite. In terms of the variables $\mu_n$,
this takes the form
\begin{equation}
\rho(\mu)=\frac{M(c_x+c_y)}{2\pi}\frac{\sqrt{(\mu-\mu_-)(\mu_+-\mu)}}{1-\mu^2},
\end{equation}
within the boundaries given by $\mu_\pm=(\lambda_\pm-1)/(\lambda_\pm+1)$.

\subsection{Non-hermitian matrices}\label{sec:nonherm}
The eigenvalues $\lambda_n$ in the Wishart matrix $W=X^\dagger X$ are the squared singular values of the matrix $X$.  For a square matrix of dimensions $M\times M$ we can also study the complex eigenvalues $z_n$ of $X$, obtained from $X\mathbf{v}_n=z_n \mathbf{v}_n$ with eigenvectors $\mathbf{v}_n$.
This leads to entirely different classes of random matrices \shortcite{ginibre_statistical_1965,RMTHandbookKhoruzhenko}.
Since $X$ is in general not normal (in particular neither hermitian nor unitary), there is no direct relation between the real singular values  and the complex eigenvalues $z_n$. This key difference is intimately related to the fact that the eigenvectors $\mathbf{v}_n$ are not orthogonal to each other, so that the spectral decomposition $X=VDV^{-1}$ with $D=\mathrm{diag}(z_n)$ involves a non-unitary matrix $V$. We therefore need to distinguish the right eigenvectors $\mathbf{v}_n$, which form the columns of $V$, from the left eigenvectors $\mathbf{w}_n$, which are obtained from $\mathbf{w}_nX=s_n \mathbf{w}_n$.
Imposing the biorthogonality condition $\mathbf{w}_m\mathbf{v}_n=\delta_{nm}$, the left eigenvectors form the rows of $V^{-1}$.

This biorthogonal set of eigenvectors is in general no longer normalised.
The extent of mode non-orthogonality can thus be quantified by the condition numbers
\shortcite{PhysRevLett.81.3367,janik_correlations_1999,schomerus_quantum_2000}
\begin{equation}
O_{mn}=\frac{(\mathbf{v}_m^\dagger \mathbf{v}_n)(\mathbf{w}_n\mathbf{w}_m^\dagger)}{(\mathbf{v}_m^\dagger \mathbf{w}_m^\dagger) (\mathbf{w}_n\mathbf{v}_n)},
\end{equation}
which we have written in a way that does not rely on the chosen normalisation condition.
In terms of the matrix $V$,
\begin{equation}
\label{eq:condnum}
O_{mn}=(V^\dagger V)_{mn}(V^{-1} V^{-1\dagger})_{nm}.
\end{equation}

The diagonal elements $K_m=O_{mm}$ are real and obey $K_m\geq 1$, with $K_m=1$ for all $m$ only if $V$ is unitary. These quantities  become large in particular when two eigenvalues approach each other closely, and indeed diverge at eigenvalue degeneracies, so-called exceptional points \shortcite{berry_physics_2004,heiss_physics_2012}. Close to such a degeneracy with a coalescing pair $z_{n+1}=z_n$, $X$ cannot be diagonalised but only be brought into a form involving Jordan blocks
\begin{equation}
\left(\begin{array}{cc} z_n & 1 \\ 0  & z_n \end{array} \right).
\end{equation}
This means that the eigenvectors of the modes become identical, in sharp contrast to hermitian systems where the eigenvectors remain orthogonal as one approaches a degeneracy.

The probability distribution \eqref{eq:px} for $M\times M$-dimensional square matrices $X$ defines the Ginibre ensemble \shortcite{ginibre_statistical_1965,RMTHandbookKhoruzhenko}. For the complex Ginibre ensemble ($\beta=2$), the joint distribution of eigenvalues is
\begin{equation}
P(\{z_n\})\propto\prod_{n<m}|z_n-z_m|^2\prod_k \exp(-c'_\beta z_k^2).
\end{equation}
In the quaternion case $\beta=4$ eigenvalues come in conjugate pairs, and the joint distribution of eigenvalues in the upper half of the complex plane
\begin{equation}
P(\{z_n\})\propto\prod_{n<m}|z_n-z_m|^2|z_n-z_m^*|^2\prod_k |z_k-z_k^*|^2
e^{-c'_\beta z_k^2}
%\exp(-c'_\beta z_k^2)
\end{equation}
contains the expected self-repulsion terms. For the real case
$\beta=1$, much more complicated expressions arise due to  the accumulation of $O(\sqrt{M})$ eigenvalues on the real axis \shortcite{Lehmann1991,forrester_eigenvalue_2007}. What is common to all three cases are the local spectral correlations of  eigenvalues well inside the complex support (away from the boundaries and spectral symmetry lines), which irrespective of $\beta$ are determined by the factors $|z_n-z_m|^2$. This yields a cubic level repulsion $P(s)\propto s^3$ for small spacings $s=|z_n-z_m|$, where one power of $s$ arises from the area element in the complex plane.

As shown in the Appendix for the complex Ginibre ensemble, for a variance scaled to $\sigma^2=1/M$ and $M\to\infty$ the eigenvalue density in the complex plane approaches Ginibre's circular law $\rho(z)=\frac{M}{\pi}\Theta(1-|z|)$, where $\Theta$ denotes the unit step function. As a side product of the calculation  presented there \shortcite{janik_correlations_1999}, the condition number $\overline {K_{m}}|_{z_m=z}\sim M(1-|z|^2)$ turns out to be large, unless one approaches the boundaries of the eigenvalue support.

From the general perspective of commutation and anticommutation with unitary and anti-unitary symmetries, non-hermitian matrices admit a very large number of symmetry classes \shortcite{magnea_random_2008}.
For a physical setting that illustrates this richness, we can consider photonic systems with absorption and amplification \shortcite{RevModPhys.87.61}. Without further constraints we may model these as a complex Ginibre ensemble ($\beta=2$) with different weights of the hermitian and non-hermitian contributions, where the eigenvalue support becomes elliptic \shortcite{girko_elliptic_1986}.
Time-reversal symmetry in optics (reciprocity) makes the matrix complex symmetric, $H=H^T\neq H^*$, which modifies the statistics but does not entail any spectral constraints.  As a template for the real Ginibre ensemble  ($\beta=1$), we can take a system with balanced amplification and absorption, situated in regions that are mapped onto each other by a reflection or inversion  $P$ \shortcite{PhysRevLett.100.103904,ruter_observation_2010}. We then obtain a non-hermitian PT-symmetric system  with $PHP=H^*\neq H^T$ \shortcite{bender_making_2007}, which in a suitable basis  is represented by a real asymmetric matrix.
In combination with magneto-optical effects, we can similarly construct PTT$'$-symmetric systems with  $PHP=H^\dagger\neq H$ \shortcite{schomerus_scattering_2013}. The spectrum
remains symmetric about the real axis,  and a random-matrix analysis reveals a close connection  to the real Ginibre ensemble, including the same accumulation of $O(\sqrt{M})$ eigenvalues on the real axis \shortcite{birchall_random-matrix_2012}.
Further examples can be constructed by modifying the role of $P$.
In an optical system where $P$ represents  a chiral symmetry, we can realize the case $H=-PH^*P$ in which eigenvalues are symmetric
with respect to the imaginary axis \shortcite{schomerus_parity_2013,schomerus_topologically_2013,poli_selective_2015}, as well as the case
$H=H^*=-P H P$ in which eigenvalues are symmetric with respect to  both  the real and the imaginary axis \shortcite{PhysRevLett.115.200402}. For a symmetry with $P^2=-1$ (hence $P=-P^T$, assuming $P$ is real), two interesting cases are the so-called Hamiltonian ensembles with $PHP= H^T$, as well as the skew-Hamiltonian ensembles with $PHP=-H^T$ (these notions  relate to the symplectic structure of classical Hamiltonians,  generated by an antisymmetric involution such as $P$; see \shortciteNP{PhysRevLett.111.037001}). For a real Hamiltonian matrix with Gaussian statistics,  $O(\sqrt{M})$ eigenvalues accumulate both on the real and on the imaginary axis; for a real skew-Hamiltonian matrix, all eigenvalues are twofold degenerate and $O(\sqrt{M})$ of these pairs accumulate on the real axis.

In the next Chapter we will see that non-hermitian matrices play a crucial role in the description of open scattering systems, where additional constraints arise from the physical constraints of unitarity and causality.

\section{The scattering matrix}
\label{chap:scatt}
\begin{figure}[b]
\includegraphics[width=\columnwidth]{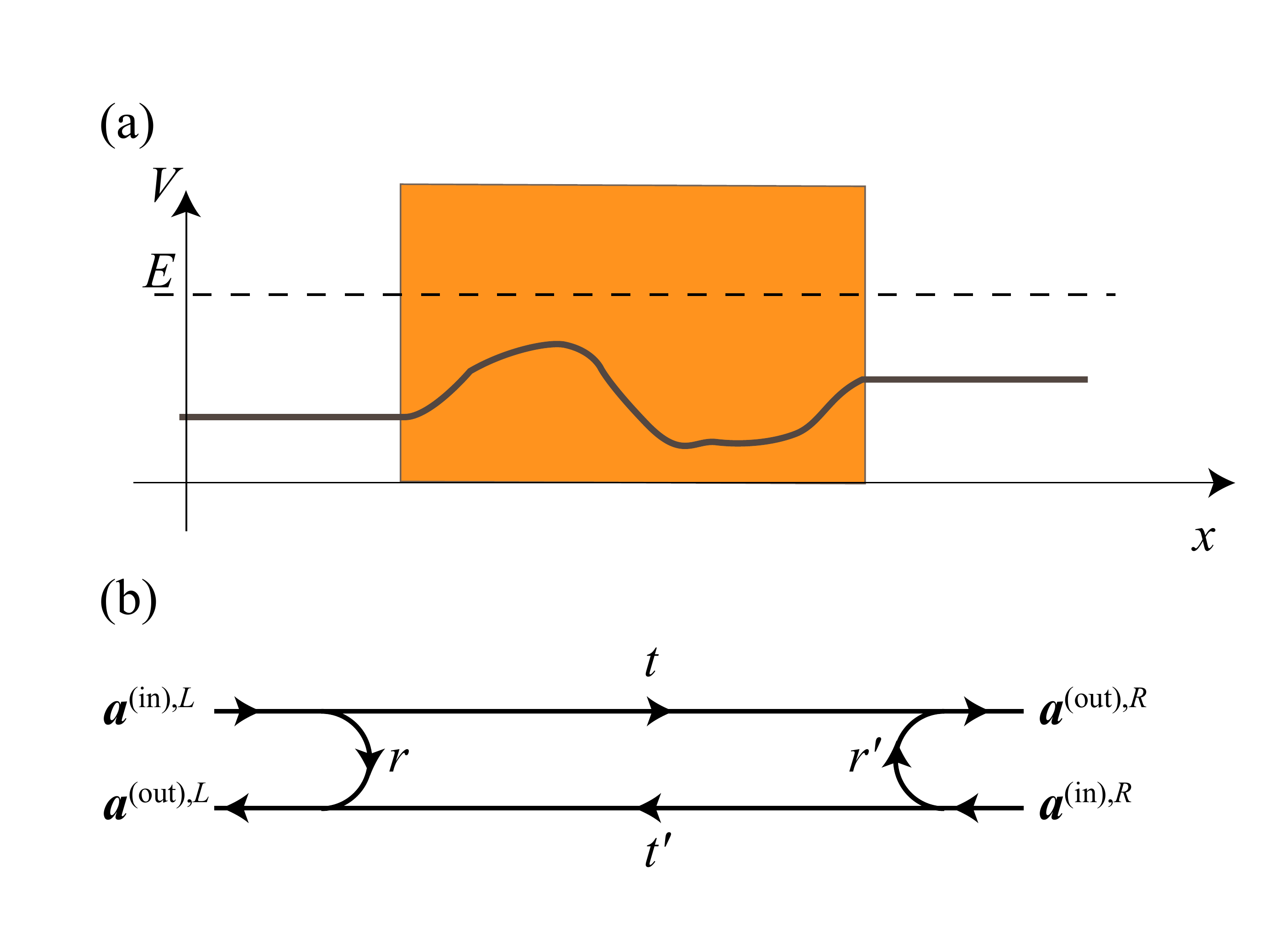}
\caption{(a) Sketch of a scattering region with a varying potential $V(x)$ in a one-dimensional system, with ideal leads attached to either side. Note that the potential does not need to be identical in both leads. (b) Scattering processes relating the amplitudes of propagating waves in the leads.}
\label{fig2}
\end{figure}

In this chapter we develop effective models for the scattering matrix and use these to identify the associated random-matrix ensembles.

\subsection{Points of interest}
Consider a particle moving through a scattering region with a spatially varying potential energy $V$, as sketched for a simple one-dimensional setting in Fig.~\ref{fig2}. The corresponding Hamiltonian is $\hat H= \hat T+ \hat V$, where $\hat T$ represents the kinetic energy.
Here are some natural phenomena that we may wish to consider:
Decay, where we address the escape rate of a particle inserted into the scattering region;
transport, where we address the probability for an incident  particle to be transmitted or reflected; dynamics, where we ask how long the particle engages with the scattering region and how many internal states it explores. We may also wish to identify system-specific details beyond the fundamental symmetries, such as regarding the role of different scattering subregions or the role of the contacts.
All of these questions (and many more) can be addressed with the help of a single object, the scattering matrix $S(E)$.

\subsection{Definition of the scattering matrix}
To define the scattering matrix \shortcite{newton_scattering_2002,messiah2014quantum}
we stipulate that the motion outside the scattering region is ballistic. At any energy $E$, we then have access to a complete set of propagating scattering states $|\psi_n^{({\rm in})}\rangle$ in which the particle is approaching the scattering region (incoming channels), and a corresponding set of propagating  states $|\psi_n^{({\rm out})}\rangle$   where the particle is moving away from the region (outgoing channels). These states are taken to be normalised to a unit probability flux through any closed surface surrounding the scattering region.
We may also encounter a set of non-propagating (evanescent) states $|\psi_m^{({\rm ev})}\rangle$  which decay away from the scattering region and do not carry any flux.
Outside the scattering region, we then can write a state with a given energy as
\begin{equation}
|\psi\rangle=\sum_{n=1}^N  a_n^{(\mathrm{in})}|\psi_n^{(\mathrm{in})}\rangle+
\sum_{n=1}^N  a_n^{(\mathrm{out})}|\psi_n^{(\mathrm{out})}\rangle
+\sum_l a_l^{(\mathrm{ev})}|\psi_m^{(\mathrm{ev})}\rangle,
\end{equation}
where $N$ fixes the number of scattering channels. We collect the expansion coefficients into vectors
$\mathbf{a}^{(\mathrm{in})}$, $\mathbf{a}^{(\mathrm{out})}$ and $\mathbf{a}^{(\mathrm{ev})}$.

Inside the scattering region, we may expand the state in terms of any suitable complete set of modes, $|\psi\rangle=\sum_m b_m|\chi_m\rangle$ with a coefficient vector $\mathbf{b}$. With help of the stationary Schr{\"o}dinger equation \eqref{eq:schr}, the states inside and outside the scattering region are uniquely related. In particular, if we fix $\mathbf{a}^{(\mathrm{in})}$ then the solution of the Schr{\"o}dinger equation uniquely fixes $\mathbf{a}^{(\mathrm{out})}$, $\mathbf{a}^{(\mathrm{ev})}$, and $\mathbf{b}$,  up to effectively decoupled parts that can be treated as a separate system. These relations must be linear, so that
\begin{equation}
\mathbf{a}^{(\mathrm{out})}=S(E)\mathbf{a}^{(\mathrm{in})}.
\end{equation}
This defines the scattering matrix. Flux normalization ensures that for real energies $S(E)$ is unitary, hence $S(E)\in \mathrm{U}(N)$. Causality ensures that the poles $E_l$ of $S$ at complex energies are all confined to the lower half of the complex plane, $\mathrm{Im}\,E_l<0$. The number of propagating scattering channels $N$ may change at certain energies, which gives rise to branch cuts in the complex-energy plane.

\subsection{Preliminary answers}\label{sec:prelim}
The scattering matrix addresses the phenomena listed at the beginning of this chapter in the following ways.

\emph{Decay.}---The complex poles $E_l=E_l'-i\hbar \gamma_l/2$ of the scattering matrix provide solutions where
$\mathbf{a}^{(\rm out)}$ is finite while $\mathbf{a}^{(\rm in)}=0$. These quasi-bound states  provide a fundamental description of decay and resonant scattering \shortcite{Guhr1998189,RevModPhys.81.539,moiseyev_non-hermitian_2011}.
The time dependence of the quasibound states
follows from the amplitude factor $A(t)=\exp(-itE_l/\hbar)=\exp(-itE_l'/\hbar)\exp(-t\gamma_l/2)$, so that the corresponding intensity $|A(t)|^2=\exp(-t\gamma_l)$ decays with rate $\gamma_l$.
For a particle prepared in this state at $t=0$, the Fourier signal
\begin{equation}
A(\omega)=\int_0^\infty A(t)e^{i\omega t}\,dt=i[(\omega-E_l'/\hbar)+i\gamma_l/2]^{-1}
\end{equation}
delivers the resonance-like frequency-resolved signal
\begin{equation}
|A(\omega)|^2=\frac{1}{(\omega-E_l'/\hbar)^2+\gamma_l^2/4},
\end{equation}
a Lorentzian centred at $E_l'/\hbar$ with full width at half maximum $\gamma_l$.
When the particle is prepared in a superposition of quasi-bound states, the resulting decay for long times depends on the characteristic decay rate $\gamma_0=\mathrm{inf}\,\gamma_l$, defined such that $\gamma_l\geq \gamma_0$ for all contributing states. If $\gamma_0>0$ the decay becomes exponential, while for $\gamma_0=0$ one typically encounters a power-law.

\emph{Transport}.---For a particle incoming in channel $n$, the probability to scatter into the outgoing channel $n'$ is given by $|S_{n'n}|^2$. The unitarity of the scattering matrix guarantees that the sums of probabilities $\sum_n|S_{n'n}|^2=\sum_{n'}|S_{n'n}|^2=1$ are normalised.
This normalization also holds for an incident particle in any superposition of incoming modes, $|\mathbf{a}^{(\mathrm{out})}|^2=|\mathbf{a}^{(\mathrm{in})}|^2$.
These features are at the heart of the scattering approach to transport
\shortcite{RevModPhys.69.731,blanter_shot_2000,nazarov2009quantum}.

In many settings, we are let to group the scattering amplitudes into subcomponents $\mathbf{a}^{(\rm in),s}$, $\mathbf{a}^{(\rm out),s}$, where $s$ labels different asymptotic regions (leads). The scattering matrix is then formed of blocks  $S_{s's}$ describing transmission from lead $s$ to lead $s'$, and reflections back into lead $s$ if $s'=s$. The associated transmission probability is quantified by the dimensionless conductance $g_{s's}=\mathrm{tr}\, (S_{s's}^\dagger S_{s's})$.
In the case of two leads, designated as a left lead $s=L$ with $N_L$ channels and a right lead $s=R$ with $N_R$ channels, we write the blocks as
\begin{equation}
\label{eq:sblock}
S=\left(  \begin{array}{cc} r & t' \\ t & r'
\end{array}\right),
\end{equation}
where $r$ and $t$ describe the reflection and transmission of particles arriving from the left,
while $r'$ and $t'$ describe these processes for particles arriving from the right.
This designation is illustrated in Fig.~\ref{fig2}(b).
The dimensionless conductance is then given by $g=\mathrm{tr}\, t^\dagger t=\mathrm{tr}\, {t'}^\dagger t'=N_L-\mathrm{tr}\, r^\dagger r=N_R-\mathrm{tr}\, {r'}^\dagger r'$, where the stated identities follow from unitarity.

The eigenvalues $T_n\in[0,1]$ of the hermitian matrix $t^\dagger t$ are known as the transmission eigenvalues, and determine the dimensionless conductance via
$g=\sum_nT_n$.
The quantities $\sqrt{T_n}$ can be interpreted as the singular values of $t$, which generalises to the polar decomposition of the scattering matrix,
\begin{align}
S&=\left(\begin{array}{cc}   V & 0 \\ 0 & V'
\end{array}\right)\left(  \begin{array}{cc} \sqrt{1-T} & \sqrt{T} \\ \sqrt{T} & -\sqrt{1-T}
\end{array}\right)\left(  \begin{array}{cc} V'' & 0 \\ 0 & V'''
\end{array}\right),
\nonumber\\
 T&=\,{\rm diag}\,(T_n)
\label{eq:polar}
\end{align}
with unitary matrices $V$, $V'$, $V''$ and $V'''$.

The transmission eigenvalues determine many other transport properties,
including the full counting statistics of electrons at low temperatures
\shortcite{1993JETPL..58..230L},
with the shot noise characterised by the second binomial cumulant \shortcite{PhysRevLett.65.2901,blanter_shot_2000}
\begin{equation}
\label{eq:snoise}
\sum_nT_n(1-T_n).
\end{equation}
Another example is the charge transport through a normal conductor into a conventional superconducting lead \shortcite{PhysRevB.46.12841,RevModPhys.69.731}, for which the dimensionless conductance at vanishing magnetic fields is given by
\begin{equation}
\label{eq:gns}
g_{NS}=\sum_nT_n^2/(2-T_n)^2.
\end{equation}

\emph{Dynamics}.---Complementing the information about the scattering probabilities, the phase $\varphi$ of a scattering amplitude $S_{n'n}=|S_{n'n}|e^{i\varphi}$ provides insight into the dynamics \shortcite{de_carvalho_time_2002,texier_wigner_2016}. For instance,
for ballistic propagation through a region of length $L$ at a constant momentum $p(E)$, the particle picks up the dynamical phase $\varphi=pL/\hbar$. The energy sensitivity $\hbar d\varphi/dE=L/v=\tau$ of the phase therefore gives an indication of the travel time. In a semiclassical description of scattering from a slowly varying potential, we have $\varphi=S_{\mathrm{cl}}/\hbar$, where the classical action $S_{\mathrm{cl}}$ again obeys $dS_{\mathrm{cl}}/dE=\tau$.

These observations lead to the formal definition of the delay time of a particle that passes through the scattering region. For injection and extraction in individual channels, the delay time can be isolated by the logarithmic derivative ${\rm Im}\,S_{n'n}^{-1}dS_{n'n}/dE$.
For multi-channel scattering this is generalised by the Wigner-Smith time-delay matrix \shortcite{PhysRev.98.145,PhysRev.118.349}
\begin{equation}
Q=-i\hbar S^\dagger dS/dE.
\end{equation}
The unitarity of $S$ at any energy ensures that $Q=Q^\dagger$ is hermitian, while causality ensures that $Q$ is positive semidefinite. Therefore, the eigenvalues $\tau_n$ of $Q$ are real and positive. These eigenvalues are known as the proper delay times.

Noting that $v^{-1}=dp/dE$ also appears in semiclassical estimates of the accessible phase-space volume, the delay times are intimately related to the density of states. Indeed, the Wigner-Smith matrix directly quantifies the global density of states in the system, in terms of the Birman-Krein formula \shortcite{birman_theory_1962}
\begin{equation}
\label{eq:krein}
\rho(E)=\frac{1}{2\pi\hbar} {\rm tr}\,Q.
\end{equation}
Replacing the derivative $d/dE$ by a local variation of the potential $\partial /\partial V(x)$, this approach can be extended to obtain the local density of states \shortcite{Gasparian1996}. Analogous variations with respect to other parameters deliver a wide range of response functions, which can for instance be used to study adiabatic transport and quantum pumping \shortcite{Buttiker1994,brouwer_scattering_1998}.

\emph{System-specific details.}---When we separate the scattering region into subregions, we can  build up the total scattering matrix from the scattering problems of the subregions  \shortcite{datta_electronic_1997,RevModPhys.69.731,nazarov2009quantum}. This can be done exactly if we extend the scattering matrix to include evanescent states, and often still very reliably if we only account for the propagating states. The simple idea is to inspect each interface and identify the amplitudes of outgoing states from one region with the amplitudes of incoming states into the adjacent region.

For the case of two adjacent regions with scattering matrices $S_1$, $S_2$ of the form
\eqref{eq:sblock}, the wave-matching of propagating states leads to the composition law
\begin{equation}
\label{eq:scomp}
S_{1\oplus 2}=
\left(  \begin{array}{cc} r_1+t_1'r_2\frac{1}{1-r_1'r_2}t_1 &t_1'\frac{1}{1-r_2r_1'}t_2'
\\ t_2\frac{1}{1-r_1'r_2}t_1  & r_2'+t_2r_1'\frac{1}{1-r_2r_1'}t_2'
\end{array}\right).
%\left(  \begin{array}{cc} r_1+t_1'r_2(1-r_1'r_2)^{-1}t_1 &t_1'(1-r_2r_1')^{-1}t_2'
%\\ t_2(1-r_1'r_2)^{-1}t_1  & r_2'+t_2r_1'(1-r_2r_1')^{-1}t_2'
%\end{array}\right).
\end{equation}
This rule can be reformulated as a simple matrix multiplication $M=M_2M_1$ for the transfer matrix
\begin{equation}
\label{eq:transfer}
M=\left(\begin{array}{cc} t^{\dagger-1} &  r^{\prime}t^{\prime-1}
\\ r^{\prime\dagger}t^{\dagger-1}  &  t^{\prime-1} \end{array}\right),
\end{equation}
which relates modes on the left and right according to
\begin{equation}
\left(\begin{array}{c}\mathbf{a}^{\mathrm{out},R}\\ \mathbf{a}^{\mathrm{in},R}\end{array}\right)
=M
\left(\begin{array}{c}\mathbf{a}^{\mathrm{in},L}\\ \mathbf{a}^{\mathrm{out},L}\end{array}\right).
\end{equation}
Flux conservation translates to the property
$M^\dagger\sigma_z M=\sigma_z$,
so that $M$ is complex symplectic. The eigenvalues of $M^\dagger M$ and $(M^\dagger M)^{-1}=\sigma_z M^\dagger M \sigma_z$ are thus identical and appear in reciprocal pairs, which are given by $(\sqrt{1/T_n}\pm\sqrt{-1+1/T_n})^2$.

We note that in the composed system, according to Eq.~\eqref{eq:scomp}
poles from the multiple scattering across the interface arise from
\begin{equation}
\label{eq:sctquant1}
{\rm det} [1-r_2(E)r_1'(E)]=0.
\end{equation}

Similarly, the role of a contact can be studied by inserting a static tunnel barrier at the corresponding boundary of the scattering region \shortcite{Brouwer1995,RevModPhys.69.731}. For example, the scattering matrix
\begin{equation}
S_B=
\left(  \begin{array}{cc} \sqrt{1-\Gamma^2} &\sqrt{\Gamma}
\\ \sqrt{\Gamma}  & -\sqrt{1-\Gamma^2}
\end{array}\right)
\end{equation}
describes a barrier with uniform transparency $\Gamma\in[0,1]$ in all channels.
If we send $\Gamma\to 0$ for all contacts the system becomes closed. Poles approaching the real axis become the energy levels
of the closed system, while poles moving deep into the complex plane are associated with direct reflection processes from the outside.

We can also artificially separate a closed system into two open systems joined by an interface. For a left and a right region, this is described by scattering matrices $S_1=r_1'$ and  $S_2=r_2$, both  only composed of a reflection block back to the interface. The quantization condition \eqref{eq:sctquant1} can then be rewritten as
\begin{equation}
{\rm det}\, (S_1(E)S_2(E)-1)=0,
\end{equation}
which determines the energies of the closed systems. This scattering quantization approach becomes exact when one includes the evanescent modes into the scattering description \shortcite{0951-7715-5-5-003,a.baecker2003}, and can be extended, e.g., to superconducting systems \shortcite{beenakker_andreev_2005} and non-hermitian photonic systems \shortcite{schomerus_scattering_2013}.

\subsection{Effective scattering models}
In practice, many methods are available to calculate the scattering matrix in specific settings. This includes wave matching, Green function methods and the boundary integral method, as well as iterative procedures based on the composition rule \eqref{eq:scomp} of scattering matrices, and analogous rules for the Green function \shortcite{datta_electronic_1997}.
For the purpose of  a statistical description, however, we require a generic model that captures the essential features of the internal dynamics and the coupling to the leads. This is delivered by the Mahaux-Weidenm{\"u}ller formula
\shortcite{mahaux_shell-model_1969,Livsic1973,Verbaarschot1985367,Guhr1998189} % ,ThisIssueWeidenmueller}
\begin{equation}
\label{eq:weidenprev}
S(E)=\frac{i\pi W^\dagger(E-H)^{-1}W-1}{i\pi W^\dagger(E-H)^{-1}W+1},
\end{equation}
where $H$ is an effective internal Hamiltonian of dimension $M\times M$ while $W$ is a suitable $M\times N$-dimensional coupling matrix, specified fully in Eq.~\eqref{eq:wnn}.

We provide a motivation of this formula via a detour to the stroboscopic scattering problem \shortcite{fyodorov_spectra_2000,tworzydlo_dynamical_2003}, which leads to its close cousin
\begin{equation}
\label{eq:strobsmatnonball2prev}
S(\varepsilon)=\frac{K\mathcal{A}K^T-1}{K\mathcal{A}K^T+1}, \quad \mathcal{A}=\frac{1+e^{i\varepsilon}F}{1-e^{i\varepsilon}F}=-\mathcal{A}^\dagger.
\end{equation}
Here $F$ is an effective internal time-evolution operator over a fixed time period $T_0$, $\varepsilon$ is the associated quasi-energy, and the coupling matrix $K$ is fully specified in \eqref{eq:strobsmatnonball2}.
The Mahaux-Weidenm{\"u}ller formula then follows in the continuum limit $T_0\to 0$.
We present this construction because it gives rather direct intuitive insight into scattering  and decay problems, and also helps to isolate and justify  the general features of the scattering matrix described in the previous section.

\subsubsection{Stroboscopic scattering approach}

\begin{figure}[t]
\includegraphics[width=\columnwidth]{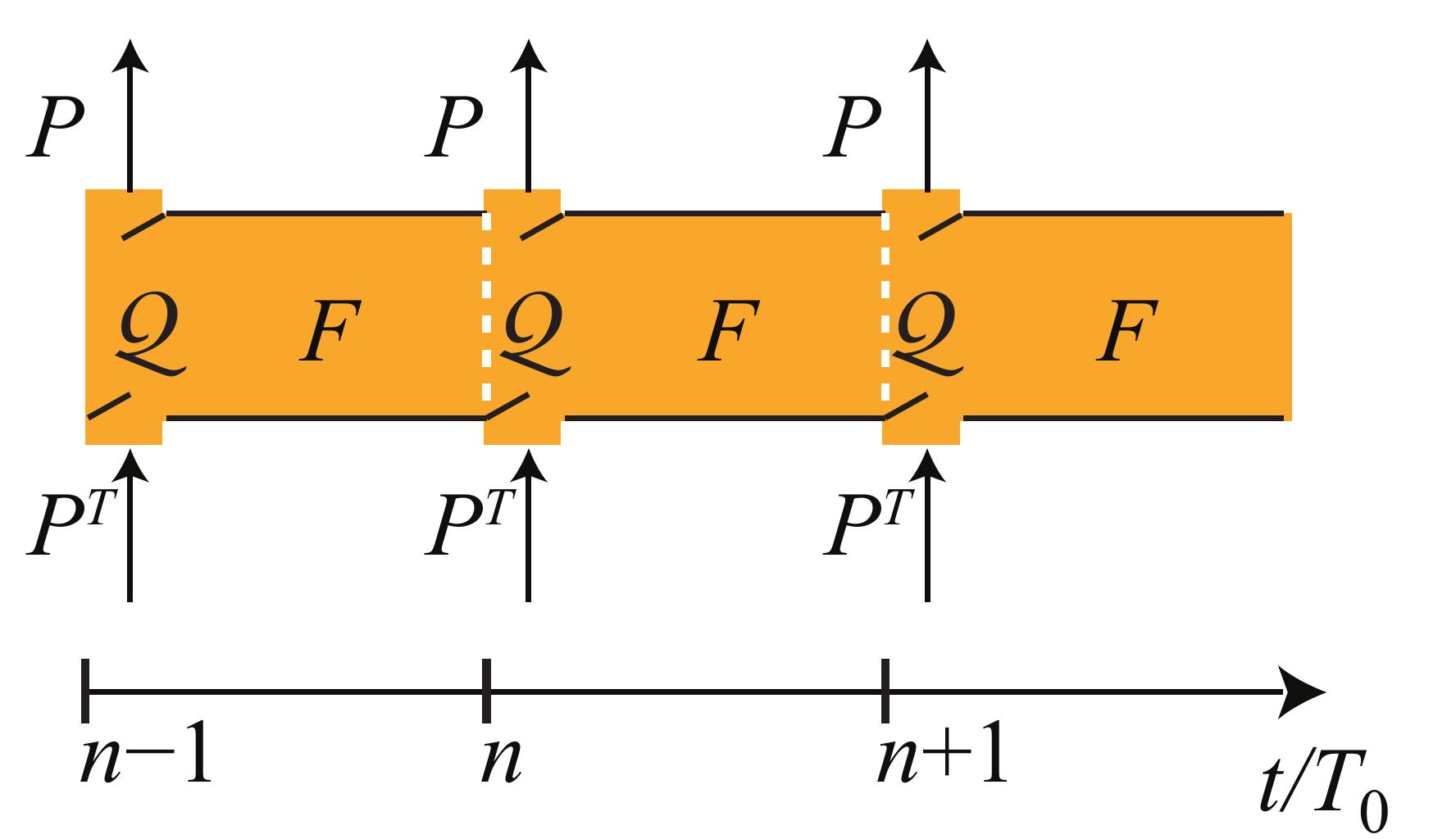}
\caption{Illustration of the stroboscopic scattering approach, in which particles are injected and collected at regular intervals.}
\label{fig3}
\end{figure}

\paragraph{Stroboscopic ballistic decay}
Our starting point is a simple, highly idealised scenario, which nonetheless can be easily extended to capture a large range of other cases. Consider a situation where the coupling of the scattering region to the outside occurs stroboscopically, at periodically spaced, discrete times $t=nT_0\equiv t_n$, $n=0,1,2,3,\ldots$ (see Fig.~\ref{fig3}).
Let us denote the state within the system just before these times as
$|\psi_n\rangle=|\psi(t_n^-)\rangle$.
This state evolves stroboscopically according to
\begin{equation}
\label{eq:strobdecayinternal}
|\psi_{n}\rangle=F\mathcal{Q}|\psi_{n-1}\rangle=(F\mathcal{Q})^{n}|\psi_{0}\rangle,
\end{equation}
where $F$ is the unitary operator that describes the time evolution when the system is closed, while $\mathcal{Q}$ is a projector that describes what remains in the system when the system is open. In other words, in each time interval, we lose some internal wave amplitude according to the complementary projector $\mathcal{P}=1-\mathcal{Q}$, while the remaining amplitude is propagated by the unitary time evolution operator $F$.
As we assume that $F$ and $\mathcal{Q}$ are independent of the time index $n$, we require that the details of the coupling are otherwise time-independent and the internal dynamics are autonomous, or at least themselves time-periodic with period $T_0$. The fact that we take $\mathcal{Q}$ as a projector means that the coupling is \emph{ballistic}---the opening is fully transparent, without any partial reflection of the passing wave. This is also called an ideal lead.

According to Eq.~\eqref{eq:strobdecayinternal}, the decay of the amplitude within this system is described by the non-unitary operator $F\mathcal{Q}$. In a basis where $\mathcal{Q}$  is diagonal this corresponds to \emph{truncating} the unitary operator $F$. Let us specify this for  a system with a finite internal Hilbert space of dimension $M$, coupled to $N$ external channels such that ${\rm rank}\,\mathcal{Q}=M-N$. In the basis where
$\mathcal{Q}={\rm diag}(0,0,0,\ldots,0,1,1,\ldots,1)$ ($N$ zeros followed by $M-N$ ones), $F\mathcal{Q}$ is then obtained from $F$ by setting the first $N$ columns to zero.

In this setting, the quasibound states $|\phi_m\rangle$ are obtained from the eigenvalue problem
\begin{equation}
\label{eq:truncatedeval}
F\mathcal{Q}|\phi_m\rangle=z_m|\phi_m\rangle,\quad=1,2,\ldots,M.
\end{equation}
Due to the projective nature of $\mathcal{Q}$, there will by $N$ vanishing eigenvalues, while the remaining eigenvalues are in general complex and finite, with $|z_m|<1$.
Each eigenvalue describes the exponential stroboscopic decay of the associated
quasibound state---if the initial state is $|\psi_0\rangle=|\phi_m\rangle$, the intensity within the system decays as
\begin{equation}
\langle \psi_n|\psi_n\rangle=|z_m|^{2n}\langle \psi_0|\psi_0\rangle.
\end{equation}
Writing $z_m=\exp[-i(\varepsilon_m-i\gamma_m/2)]$, the decay constant over a period $T_0$ is given by $\gamma_m$. As indicated, this decay constant is best viewed as arising from the imaginary part of a complex quasienergy $\varepsilon^\star_m=\varepsilon_m-i\gamma_m/2$, where the real part is defined modulo $2\pi$.

\paragraph{Stroboscopic scattering with ideal contacts}
We now turn the stroboscopic decay problem into a stroboscopic scattering problem. This requires to define how the escape from the system translates into  particles detected outside, as well as how to feed particles into the system. In other words, we need to define objects that connect the state within the system (residing in the internal Hilbert space in which $F$ and $\mathcal{Q}$ operate) to the amplitudes of the $N$  incoming modes (states $|\psi^{\rm (in)}_n\rangle$) and the $N$ outgoing modes (states $|\psi^{\rm (out)}_n\rangle$) outside the system.

In the case of ballistic coupling that we study thus far, the outgoing state can be taken of the simple form
\begin{equation}
\label{eq:strobscatout}
|\psi^{\rm (out)}_n\rangle=P|\psi_n\rangle,
\end{equation}
with $P$ such that $\mathcal{P} = P^TP=1-\mathcal{Q}$ recovers the rank-$N$ projector that complements $\mathcal{Q}$ in the internal Hilbert space. It  follows that $PP^T=1$ is the identity  in the space of the external scattering channels (the rank does not change under the reordering and the resulting object is still a projector).
Recall that the internal state refers to the instance just before we open the system. Therefore, the incoming particle injected in the previous step modifies this state
according to
\begin{eqnarray}
|\psi_{n}\rangle&=&F\mathcal{Q}|\psi_{n-1}\rangle+FP^T|\psi^{\rm (in)}_{n-1}\rangle
\nonumber
\\
&=&(F\mathcal{Q})^{n}|\psi_{0}\rangle+
\sum_{l=0}^{n-1}   (F\mathcal{Q})^l FP^T |\psi^{\rm (in)}_{n-l-1}\rangle,
\label{eq:strobscatinternal}
\end{eqnarray}
which replaces  Eq.~\eqref{eq:strobdecayinternal}.
Combining these expressions, we find
\begin{equation}
|\psi^{\rm (out)}_n\rangle=
P(F\mathcal{Q})^{n}|\psi_{0}\rangle+
P\sum_{l=0}^{n-1}   (F\mathcal{Q})^l F P^T |\psi^{\rm (in)}_{n-l-1}\rangle.
\end{equation}
The first part recovers the decay of the initial state, while the remaining part describes  the scattering. The pure decay problem is characterised by the absence of the incoming state, while the pure scattering problem is characterised by the absence of the initial state.

Both these problems now turn out to be intimately related. For this, we revert back to a continuous time variable, $|\psi^{\rm (out)}(t)\rangle=\sum_n\delta(t-nT_0) |\psi^{\rm (out)}_n\rangle$,
and perform a Fourier decomposition of the scattered signal,
\begin{align}
\label{eq:strobscatoutfinal1}
&|\psi^{\rm (out)}(\varepsilon)\rangle=\sum_{n=0}^\infty e^{i \varepsilon n } |\psi^{\rm (out)}_n\rangle
\\&
=
\sum_{l=0}^\infty\sum_{n=l+1}^\infty e^{i \varepsilon l}
P   (F\mathcal{Q})^l e^{i \varepsilon}F P^T e^{i \varepsilon (n-l-1)}|\psi^{\rm (in)}_{n-l-1}\rangle,
\end{align}
hence
\begin{equation}
\label{eq:strobscatoutfinal}
|\psi^{\rm (out)}(\varepsilon)\rangle=
S(\varepsilon)
|\psi^{\rm (in)}(\varepsilon)\rangle
\end{equation}
with the stroboscopic scattering matrix
\begin{equation}
\label{eq:strobsmat}
S(\varepsilon)=P\sum_{l=0}^\infty [e^{i \varepsilon}F\mathcal{Q}]^l e^{i \varepsilon}F P^T=
P   \frac{1}{1-e^{i \varepsilon}F\mathcal{Q}} e^{i \varepsilon}F P^T.
\end{equation}
We now observe that the  poles of the scattering matrix coincide with the complex quasienergies $\varepsilon^\star_m$, as determined by the eigenvalue problem \eqref{eq:truncatedeval}.

It is convenient to bring the scattering matrix
\eqref{eq:strobsmat} into the equivalent form
\begin{equation}
\label{eq:strobsmat2}
S(\varepsilon)=\frac{P\mathcal{A}P^T-1}{P\mathcal{A}P^T+1}, \quad \mathcal{A}=\frac{1+e^{i \varepsilon}F}{1-e^{i \varepsilon}F}=-\mathcal{A}^\dagger.
\end{equation}
We then see that the scattering matrix is indeed unitary. Furthermore, this expression nicely generalises to the case of non-ideal contacts, which we address next.

\paragraph{Stroboscopic scattering with non-ideal contacts}
To account for non-ideal coupling we insert an energy-independent scatterer at the place of the contact. The contact can be viewed as a region with $N$ channels coupled to the outside and $N$ channels coupled to the inside, and thus is described by a $2N\times 2N$-dimensional unitary scattering matrix
\begin{equation}
S_B=\left(\begin{array}{cc}r_B & t_B' \\ t_B & r_B' \end{array}\right).
\end{equation}
The blocks $r_B$ and $r_B'$ describe the partial reflection in the external and internal channels,
while $t_B$ and $t_B'$ describe the transmission  into and out of the system. This matrix is assumed to be energy-independent, meaning that the reflection and transmission processes from the contact are instantaneous.  The return of the particle to the contact is described by the ballistic scattering matrix $S_0$.

We can now match the waves at the contact according to Eq. \eqref{eq:scomp}, which
results in the total scattering matrix
\begin{equation}
\label{eq:sbarrier}
S=r_B+t'_BS_0(1-r_B'S_0)^{-1}t_B.
\end{equation}
This expression has a simple interpretation:
The incident wave is either directly reflected according to $r_B$, or enters into the system according to $t_B$. Once in the system, the wave undergoes a sequence of $l$ events, each consisting of an internal scattering round trip $S_0$ followed by a partial reflection $r_B'$, until after another return $S_0$ to the contact it escapes according to $t_B'$. Equation \eqref{eq:sbarrier} follows by summing over $l$, which is of the form of a geometric series.

Inserting for $S_0$ the stroboscopic scattering matrix
\eqref{eq:strobsmat} for ideal contacts, we find that this can be written more directly as
\begin{equation}
\label{eq:strobsmatnonball}
S=r_B+t'_BP\frac{1}{1-e^{i \varepsilon}F(\mathcal{Q}+P^Tr_B'P)}e^{i \varepsilon}FP^Tt_B.
\end{equation}
To further simplify this expression we choose an appropriate basis for the internal state, as well as for the incoming and the outgoing state. This follows from the polar decomposition \eqref{eq:polar}, which we need to adopt in the slightly more general form
\begin{align}
&S_B=\left(\begin{array}{cc}V & 0 \\ 0 & V' \end{array}\right)
\left(\begin{array}{cc}-\Sigma\sqrt{1-\Gamma^2} \!\!\!\!\!& \sqrt{\Gamma} \\ \sqrt{\Gamma} & \!\!\!\!\!\Sigma\sqrt{1-\Gamma^2} \end{array}\right)
\left(\begin{array}{cc}V'' & 0 \\ 0 & V''' \end{array}\right),
\nonumber\\
&
%\left\{
%  \begin{array}{l}
    \Gamma=\,{\rm diag}\,(\Gamma_n)
    %\\
    , \quad
    \Sigma=\,{\rm diag}\,(\sigma_n)
%  \end{array}
%\right.
%\quad
.
\end{align}
Here $\Gamma_n\in[0,1]$ are the transmission eigenvalues of the contact, while $\sigma_n=\pm 1$ discriminates two distinct ways to close a channel.
The unitary matrices $V$, $V'$, $V''$ and $V'''$ can all be absorbed into the basis choice, which means that $S_B$ is block diagonal and real. Starting from  \eqref{eq:strobsmatnonball},
this basis choice results in the desired generalization of Eq.~\eqref{eq:strobsmat2},
\begin{align}
S&=\frac{K\mathcal{A}K^T-1}{K\mathcal{A}K^T+1}, \quad \mathcal{A}=\frac{1+e^{i\varepsilon}F}{1-e^{i\varepsilon}F}=-\mathcal{A}^\dagger,
\nonumber\\ K&={\rm diag}\,(\kappa_n^{\sigma_n}) P,
\label{eq:strobsmatnonball2}
\end{align}
where the contact is now characterized by  the coupling coefficients
\begin{equation}
\label{eq:kappa}
\kappa_n=\Gamma_n^{-1/2}(1-\sqrt{1-\Gamma_n^2})
.\end{equation}
These coefficients take the value $\kappa_n=1$ for $\Gamma_n=1$ and $\kappa_n\approx\sqrt{\Gamma_n}/2$ for $\Gamma_n\ll 1$.  As they enter the matrix $K$ to the power $\sigma_n$, a semitransparent contact can be achieved both by decreasing the coupling ($\sigma_n=1$) or by
increasing the coupling ($\sigma_n=-1$).
This completes the derivation of the stroboscopic scattering matrix
\eqref{eq:strobsmatnonball2prev}.

\subsubsection{Continuous-time scattering theory}\label{sec:contlim}
To realize the time-continuous limit of the
stroboscopic scattering theory, we set $\varepsilon =E T_0/\hbar$, $F=\exp(-iT_0 H/\hbar)$, and equate $T_0\equiv 2\pi\hbar/M\Delta=T_H/M$ to the dwell time in a continuous system with $M$ channels and mean level spacing $\Delta$ (this is the mean time for a round trip $F$ in the system). In the leading orders of $T_0$,
we can approximate
\begin{equation}
\label{eq:cayley}
e^{i\varepsilon}F\approx\frac{1-iT_0(H-E)/2\hbar}{1+iT_0(H-E)/2\hbar},
\end{equation}
so that
\begin{equation}
\mathcal{A}=\frac{1+e^{i\varepsilon}F}{1-e^{i\varepsilon}F}\approx\frac{2i\hbar}{T_0}G(E),\quad G(E)=\frac{1}{E-H},
\end{equation}
where $G(E)$ is the Green function (or resolvent) of the closed system.
For the ideal case with scattering matrix \eqref{eq:strobsmat2},
we then have
\begin{equation}
\label{eq:weidenball}
S=\frac{\frac{2i\hbar}{T_0}P(E-H)^{-1}P^T-1}{\frac{2i\hbar}{T_0}P(E-H)^{-1}P^T+1},
\end{equation}
while for non-ideal leads $P$ is replaced by $K$. Inserting $T_0$ completes the derivation of the Mahaux-Weidenm{\"u}ller formula
\eqref{eq:weidenprev},
\begin{equation}
\label{eq:weidenprev2}
S(E)=\frac{i\pi W^\dagger(E-H)^{-1}W-1}{i\pi W^\dagger(E-H)^{-1}W+1},
\quad W=\frac{\sqrt{M\Delta}}{\pi} K^\dagger,
\end{equation}
where the $M\times M$-dimensional hermitian matrix $H$  represents the Hamiltonian of the closed systems, while
the $M\times N$-dimensional matrix  $W$ describes the coupling
to the $N$ scattering channels. With our basis choice, $W$ is diagonal, with elements
\begin{equation}
\label{eq:wnn}
W_{nn}=\frac{\sqrt{M\Delta}}{\pi}\kappa_n^{\sigma_n}
\end{equation}
specified according to Eq.~\eqref{eq:kappa}.
The form of $W$ in the non-ideal case can also be obtained by starting with the scattering matrix \eqref{eq:weidenball} for ideal contacts
and adding barriers by the construction \eqref{eq:sbarrier}.

Equation
\eqref{eq:weidenprev2} can be rewritten in the equivalent form
\begin{equation}
\label{eq:weiden}
S(E)
=
-1+2\pi i W^\dagger(E-H+i\pi WW^\dagger)^{-1}W.
\end{equation}
According to this, the poles of the scattering matrix are given by the eigenvalues of the effective non-hermitian Hamiltonian
$H-i\pi WW^\dagger$. The poles all lie in the lower half of the complex plane, as required by causality.
Furthermore, the Wigner-Smith time-delay matrix $Q=-i\hbar S^\dagger dS/dE$ takes the form
\begin{equation}
\label{eq:wsmh}
Q=2\pi\hbar W^\dagger (E-H-i\pi WW^\dagger)^{-1} (E-H+i\pi WW^\dagger)^{-1}W,
\end{equation}
which is explicitly positive semidefinite, as again required by causality.

\subsection{Merits}
\label{sec:merits}

\begin{figure}[t]
\includegraphics[width=\columnwidth]{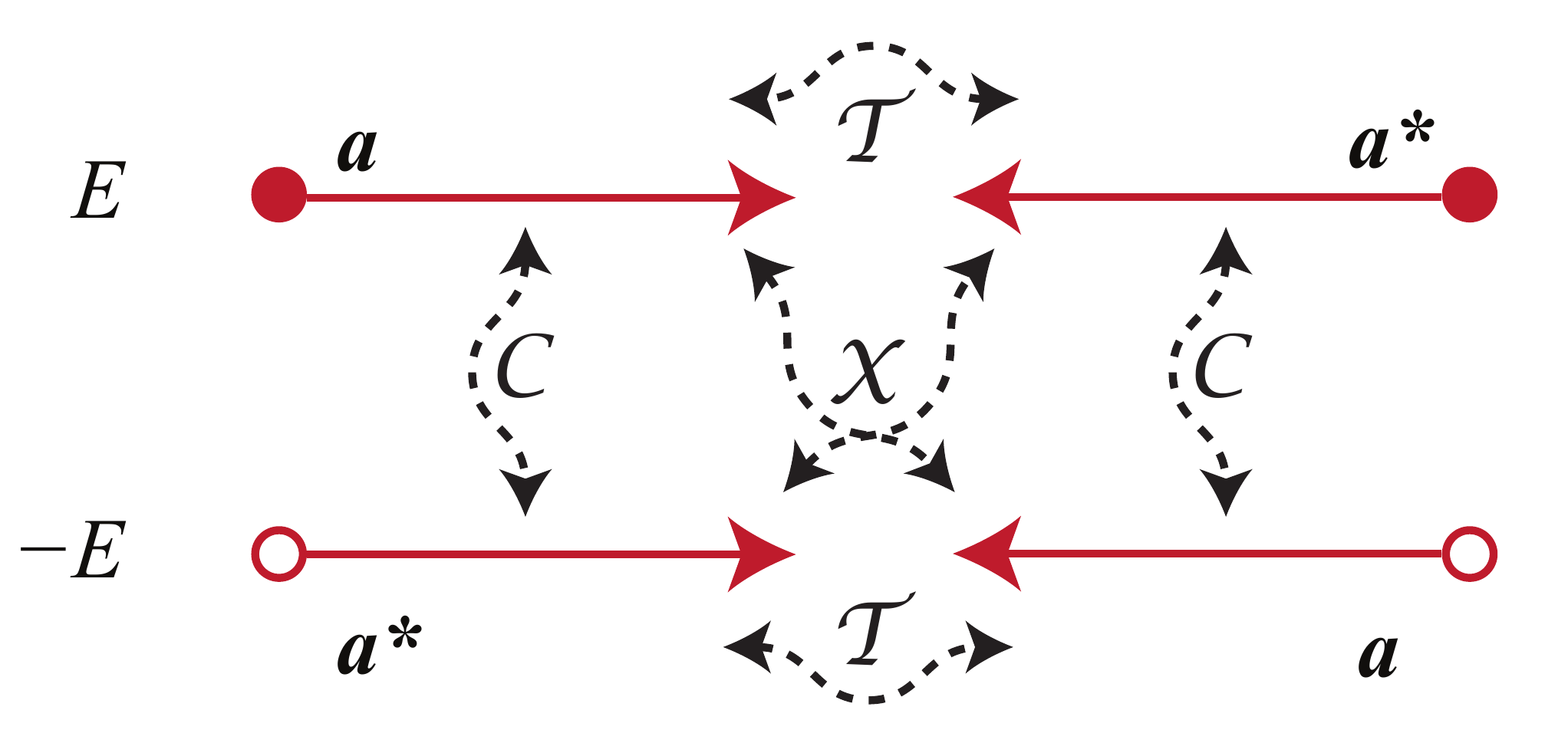}
\caption{Fundamental symmetries relate various states of motion, which constrains the scattering matrix in accordance to the ten universality classes for unitary matrix.}
\label{fig4}
\end{figure}

Via the stroboscopic model \eqref{eq:strobsmatnonball2}, the orthogonal, unitary,  or symplectic  symmetry of $F$ in the  three Wigner-Dyson classes
with different form of time-reversal symmetry
translates  directly into a corresponding
symmetry of $S$.  Via the continuous model \eqref{eq:weiden}, one finds that this also agrees with the corresponding symmetry class for $H$.
In the symmetry classes with chiral or charge-conjugation symmetry, this translation holds when the scattering matrix is evaluated at the spectral symmetry points $E=0$ or $\varepsilon=0,\pi$ (away from these points, the symmetry reduces to the three Wigner-Dyson classes).
Thus, the ten symmetry classes listed in Table \ref{table2} directly apply to the scattering matrix, with energy fixed to the symmetry point where required
\shortcite{RevModPhys.87.1037}.

It is instructive to verify these statements directly within the scattering picture (see Fig.~\ref{fig4}).
For this, consider that the time-reversal operation $\mathcal{T}$ transforms incoming modes into outgoing modes. If this is a  symmetry of the Hamiltonian then the correspondingly transformed scattering state
must be described by the original scattering matrix. For $\mathcal{T}=K$ this delivers
\begin{equation}
{\mathbf{a}^{(\mathrm{in})}}^*=S(E){\mathbf{a}^{(\mathrm{out})}}^*=S(E)S^*(E){\mathbf{a}^{(\mathrm{in})}}^*,
\end{equation}
such that $S^T(E)=S(E)$, as anticipated. Analogously, a time-reversal symmetry with $T=\Omega K$ implies
$S^T(E)=\Omega S(E)\Omega^{-1}$, hence $[\Omega S(E)]^T=-\Omega S(E)$. For a chiral symmetry $\mathcal{X}$, we transform a
solution at energy $E$ into a solution at energy $-E$. This inverts the group velocity of the propagating modes, thus again transforms incoming modes into outgoing modes. It follows that
$\mathcal{X}S(E)\mathcal{X}=S^\dagger(-E)$,
and  hence $[\mathcal{X}S(-E)]^\dagger=\mathcal{X} S(E)$.
For a charge-conjugation symmetry $\mathcal{C}$, both effects on the propagation direction cancel such that $S(-E)=S^*(E)$ if $\mathcal{C}=K$, while  $S(-E)=\Omega S^*(E) \Omega^{-1} $ if $\mathcal{C}=\Omega K$. This recovers all constraints in Table \ref{table2}.

Based on this correspondence, the effective scattering models deliver an independent view on the topological quantum numbers associated with the Hamiltonian \shortcite{PhysRevB.83.155429,RevModPhys.87.1037,PhysRevLett.114.166803}.
In systems with a chiral symmetry,
the matrix $S_{X0}=\mathcal{X} S(0)$ is unitary and hermitian, so that the trace
$\nu_0=\frac{1}{2}\mathrm{tr}\,S_X$ quantifies the difference between eigenvalues $\pm 1$.
According to Eq.~\eqref{eq:weiden} with a chiral Hamiltonian of the form \eqref{eq:chiralh}, this topological quantum number can then be expressed as $\nu_0=[\nu+(N_A-N_B)/2]_{|\nu_0|\leq N/2}$, where $N_A$ and $N_B$ count the number of channels coupled to the two different chiral sectors; as indicated by the brackets this saturates at $|\nu_0|=N/2$ where $N=N_A+N_B$. In systems with a charge-conjugation symmetry, where the Hamiltonian can be made anti-symmetric by an appropriate basis choice and displays a zero mode if $M$ is odd (modulo possible Kramers degeneracy), $\nu_0=\mathrm{det}\,S(0)=\nu$ (class D)
and $\nu_0=\mathrm{pf}\,\Omega S(0)=\nu$ (class DIII) remain directly related to the internal topological quantum number.

Beyond the pure symmetry classification, and perhaps even more importantly, the effective scattering models also determine the appropriate statistical ensembles for the scattering matrix for ergodic internal wave propagation \shortcite{Brouwer1995}.
For ideal contacts,
the circular ensembles for $F$ translate via Eq.~\eqref{eq:strobsmat} into the corresponding circular ensembles for the ballistic scattering matrix $S$, with energy again
fixed to the symmetry point where required.
In the presence of a tunnel barrier, the Haar measure is deformed according to Eq.~\eqref{eq:strobsmatnonball2}. In the three Wigner-Dyson classes this takes the form of a Poisson-kernel
\begin{equation}
\label{eq:poisson}
P(S)\propto |\mathrm{det}\,(1-\overline{S}^\dagger S)|^{-\beta N-2+\beta},
\end{equation}
where the
non-ideal contacts are encoded in the average scattering matrix $\overline{S}=(1-KK^T)/(1+KK^T)$.
In the additional symmetry classes with chiral or charge-conjugation symmetry, the analogue of the Poisson kernel can be constructed based on Eq.~\eqref{eq:sbarrier}
\shortcite{PhysRevB.79.214506,marciani_effect_2016}, which we briefly illustrate in Section \ref{sec:delay}.

By carrying out the continuum limit for large $M$, one furthermore finds that the internal Hamiltonian $H$ in the Mahaux-Weidenm{\"u}ller formula \eqref{eq:weidenprev2} complies with the corresponding Gaussian ensemble
(see again \shortciteNP{Brouwer1995}). In the three Wigner-Dyson ensembles, the Cayley transform  \eqref{eq:cayley}  implies at $E=0$
\begin{equation}
F^\dagger dF =-i\Sigma dH \Sigma^\dagger,\quad \Sigma=\frac{1}{1+iHT_0/2\hbar},
\end{equation}
which allows to calculate the Jacobian for the transformation from $F$ to $H$. This leads to a Cauchy distribution
\begin{equation}
P(H)\propto \mathrm{det}\,(1+H^2T_0^2/4\hbar^2)^{-(\beta M+2-\beta)/2},
\end{equation}
which for large $M$ shares all leading $p$-point correlations functions
with the corresponding Gaussian ensemble.

These considerations provide a solid link between the random-matrix models for closed and open systems with ergodic internal dynamics. For ideal leads, the stationary scattering at fixed energy is described by a unitary scattering matrix from a circular ensemble, while the related Poisson kernel applies when the contacts are non-ideal.
Based on the appropriate Gaussian ensemble for $H$, the effective scattering model can also be employed to study the energy-dependence, including the crossover between symmetry classes as the energy is steered away from a spectral symmetry point. Guided by the list of questions posed at the beginning of this chapter, we can now set out to describe scattering and decay from a random-matrix perspective.

\section{Decay, Dynamics and Transport}
We now turn to the random-matrix description of the physical phenomena outlined in Section \ref{sec:prelim}.

\label{chap:scattres}
\subsection{Scattering poles}
According to the Mahaux-Weidenm{\"u}ller formula \eqref{eq:weiden}, the complex energies of the quasibound states (poles of the scattering matrix) are obtained from the  eigenvalue problem
\begin{equation}
\label{eq:effH}
E_m|\phi_m\rangle  =
H_{\mathrm{eff}}|\phi_m\rangle,
\end{equation}
where  the $M\times M$ dimensional effective non-hermitian Hamiltonian is of the form $H_{\mathrm{eff}}=H-i\pi WW^\dagger$
\shortcite{fyodorov_statistics_1997,fyodorov_random_2003,RMTHandbookFyodorov}. This consists of a hermitian part $H$ which represents the dynamics in the closed system, and an anti-hermitian part involving a positive semidefinite matrix $WW^\dagger$ of rank $N$.
The eigenvalues are therefore confined to the lower half of the complex plane, where ${\rm Im}\, E_m=-\hbar\gamma_m/2$ encodes the positive decay rates $\gamma_m$. Analogously, the poles of the stroboscopic scattering matrix can be read off Eq.~\eqref{eq:strobsmatnonball}, according to which they are obtained from the eigenvalue problem
\begin{equation}
z_m|\phi_m\rangle=F(\mathcal{Q}+P^Tr_B'P)|\phi_m\rangle,
\end{equation}
with $z_m=\exp(-i \varepsilon_m)$ confined by $|z_m|\leq 1$. The two problems are then related by identifying $\varepsilon_m=  E_m T_0/\hbar$ with $T_0=2\pi \hbar/M\Delta$; see our discussion in Section~\ref{sec:contlim}.

In a random-matrix description with large matrix dimension $M$, one typically finds that the eigenvalues populate a well-defined region, with universal statistics in the bulk \shortcite{PhysRevLett.83.65,forrester2010log,RMTHandbookKhoruzhenko}.
In particular, well inside the eigenvalue support the level repulsion is typically captured by a factor $\prod_{n<m}|E_n-E_m|^2$, as already encountered for the Ginibre ensembles, which then yields cubic level repulsion.
For many physical applications, however, we are mainly interested in the properties of the longest-living modes in a given energy range, which approach the real axis closest from below, and are automatically situated at the boundary of the spectral support. These modes determine the noticeable resonance patterns that one observes, e.g., in the scattering and decay of nuclei  \shortcite{RevModPhys.81.539} or in the emission properties of optical microresonators \shortcite{RevModPhys.87.61}. To determine their properties we need to work directly with the effective scattering models.

\begin{figure*}[t]
\includegraphics[width=\textwidth]{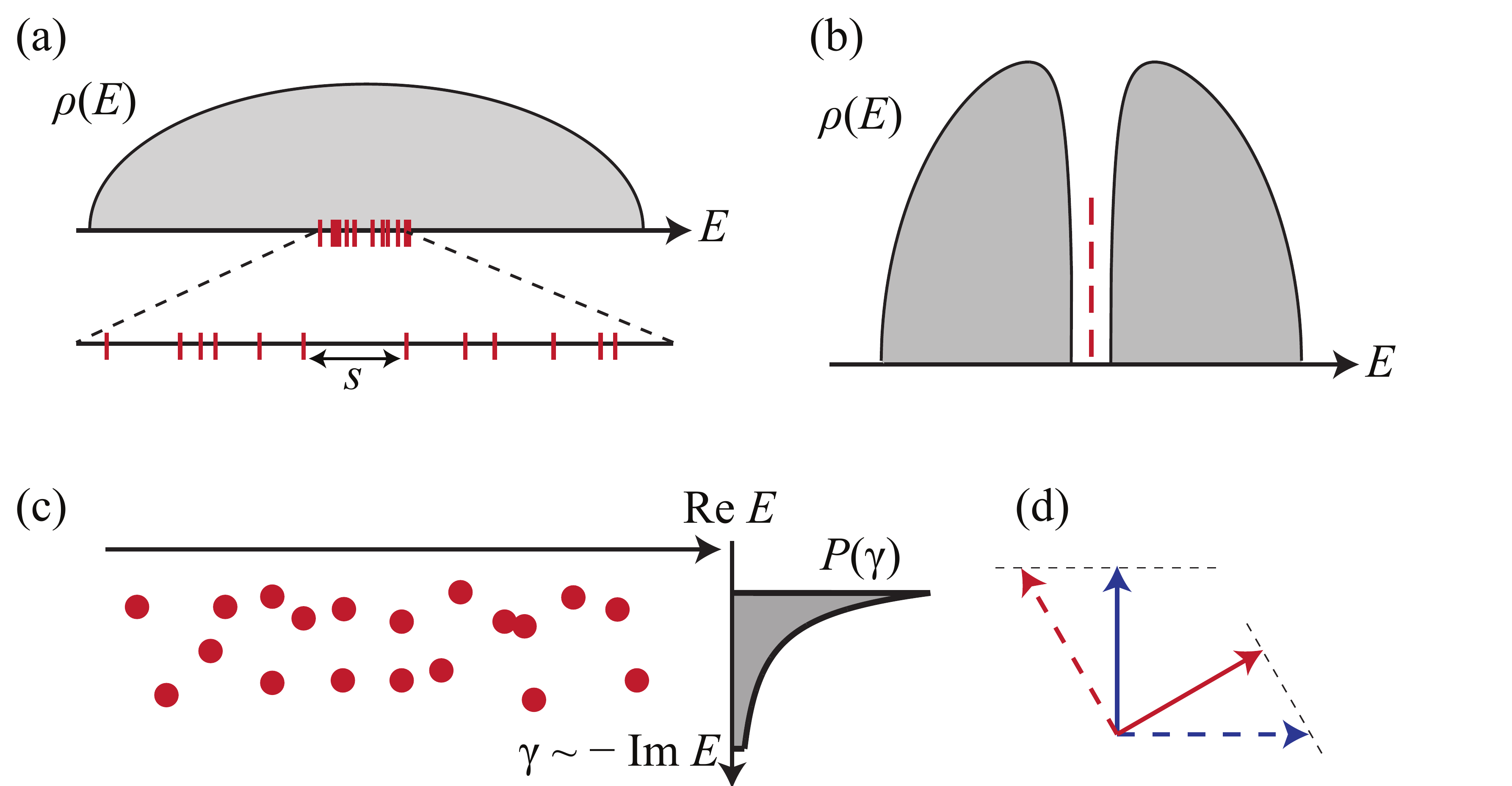}
\caption{(a) In a closed system, energy levels are constrained to be real, and random-matrix theory focusses on the spectral fluctuations, e.g. of the level spacings $s$. These occur against the non-universal backdrop of the mean density of states $\rho(E)$, here illustrated as the Wigner semicircle law \eqref{eq:semi}. (b) Fundamental symmetries can introduce spectral symmetries which induce universal aspects into the mean density of states. At the symmetry point, topologically protected zero modes can appear. This is here illustrated for the case of the chiral symmetry, with the mean density of states given by Eq.~\eqref{eq:rhochiral}.
(c) In an open system, the corresponding energies are complex and attention shifts to the decay rates $\gamma$ of the states, here given in accordance to Eq.~\eqref{eq:pycl2}. (d) The states become non-orthogonal, which requires to introduce a bi-orthogonal system as here illustrated for a pair of states.}
\label{fig5}
\end{figure*}

Particularly compact expression for the distribution of decay rates can be obtained
for the stroboscopic model \eqref{eq:strobsmat} with ideal leads \shortcite{Zyczkowski2000}.
The quasi-bound states are then obtained from the eigenvalue problem \eqref{eq:truncatedeval} for the truncated time-evolution operator  $F\mathcal{Q}$. We assume that $F\in \mathrm{U}(M)$ is a random unitary matrix of dimension $M\times M$, distributed according to the Haar measure $\mu(F)$, which places us into the circular unitary ensemble (CUE) for systems without any further symmetries.
Averaging over this ensemble, it is then possible to determine the density of eigenvalues $z_m$ in the complex plane. In  a first step, one finds the joint distribution of the nontrivial eigenvalues $z_m\neq 0$,
to which we assign the indices $m=1,2,\ldots,M-N$. This joint distribution is given by
\begin{equation}
P(\{z_m\}) \propto \prod_{i<j}^{M-N}|z_i-z_j|^2\prod_{k=1}^{M-N}(1-|z_k|^2)^{N-1} ,
\end{equation}
where the first term signifies the expected level repulsion.  The density of the eigenvalues in the complex plane follows by integrating out all but one eigenvalue, which gives
\begin{equation}
\rho(z)\propto(1-|z|^2)^{N-1}\!\sum_{l=1}^{M-N}\!\frac{(N+l-1)!}{(l-1)!}|z|^{2l-2}
\quad\mbox{for }|z|<1.
\end{equation}

This density has several interesting limits.
For $M,N\to \infty$ at fixed $N/M=1-\mu$, the modulus $r=|z|$ obeys
\begin{equation}
\label{eq:pymf}
P(r)=(\mu^{-1}-1)\frac{2r}{(1-r^2)^2}\Theta(\mu-r^2),
\end{equation}
while for $M\to\infty$ at fixed $N$
we have, setting  $(1-r)/T_0\to \gamma/2$,
\begin{equation}
\label{eq:py}
P(\gamma)=\frac{\gamma^{N-1}}{(N-1)!}\left(\frac{-d}{d\gamma}\right)^N\frac{1-e^{-\gamma T_H}}{\gamma T_H},
\end{equation}
where $T_H=2\pi\hbar/\Delta$ is the Heisenberg time.

According to Eq.~\eqref{eq:pymf}, in the considered limit all poles are confined to the region $r<\sqrt{\mu}$, thus do not approach the unit circle closely. Such a hard gap is also obtained from large-$N$ limit of
equation \eqref{eq:py}   (thus $1\ll N\ll M$), in which
\begin{equation}
\label{eq:pycl}
P(\gamma)=\frac{\gamma_0}{\gamma^2} \quad \mathrm{if}\,\,\gamma>\gamma_0, \,\quad 0\mbox{ otherwise}.
\end{equation}
Here $\gamma_0=N\Delta/2\pi\hbar=1/T_D$ coincides with the classical decay rate out of a system with dwell time $T_D=T_H/N$. The corresponding energy scale $E_{\mathrm{Th}}=\hbar/T_D=N\Delta/2\pi$ is known as the Thouless energy.

These results recover the main features earlier obtained by a direct analysis of the
non-hermitian eigenvalue problem \eqref{eq:effH}. The most comprehensive insight is obtained using supersymmetric integration techniques, which predict Eq.~\eqref{eq:py} for ideal coupling and extend it to non-ideal leads  \shortcite{Fyodorov1996,fyodorov_statistics_1997,fyodorov_random_2003}.
The result is
\begin{eqnarray}
P(\gamma)&=&\frac{\hbar\pi}{\Delta}
\mathcal{F}_1\left(\frac{\hbar\pi}{\Delta}\gamma\right)
\mathcal{F}_2\left(\frac{\hbar\pi}{\Delta}\gamma\right)
,
\nonumber
\\
\mathcal{F}_1(y)&=&
\frac{1}{2\pi}\int_{-\infty}^\infty dx\, e^{-ixy}
\prod_{n=1}^N\frac{1}{x_n-ix},
\nonumber
\\
\mathcal{F}_2(y)&=&
\frac{1}{2}\int_{-1}^1 dx\, e^{-xy}\prod_{n=1}^N(x_n+x),
\label{eq:pgue}
\end{eqnarray}
where $x_n=-1+2/\Gamma_n$ encodes the transparency of the contact.
For a barrier with uniform transparency $\Gamma$ (hence dimensionless conductance $g_c=\Gamma N$),
the distribution function can be written compactly as
\begin{equation}
P(\gamma)=\frac{\Delta}{2\pi\hbar \gamma^2 (N-1)!}
\int_{N(1-\Gamma)\gamma/\gamma_0}^{N\gamma/\gamma_0}dx\,x^Ne^{-x}
,
\label{eq:pgcon}
\end{equation}
where now $\gamma_0=\Gamma N\Delta/2\pi\hbar$.
The large-$N$ limit \eqref{eq:pycl} is then replaced by
\begin{equation}
\label{eq:pycl2}
P(\gamma)=\frac{\gamma_0}{\Gamma \gamma^2} \quad \mathrm{if}\,\,\gamma_0<\gamma<\gamma_0/(1-\Gamma),
\end{equation}
so that the decay rates are reduced according to the increased classical dwell time
$T_D=T_H/(\Gamma N)$.

The random-matrix results for the unitary symmetry class can be extended to the other symmetry classes. As with the Ginibre ensembles, many of the common characteristics remain unchanged, with the main modifications arising from
spectral symmetries. In particular, in systems with time-reversal symmetry (orthogonal and symplectic symmetry class) no further spectral symmetries arise
(these cases are therefore quite  distinct from the real and symplectic Ginibre ensemble, which lends further justification to their careful construction). The main modifications arise from the altered level repulsion in the closed limit, which is felt by the longest-living states \shortcite{Sommers1999,RMTHandbookFyodorov}. At large matrix dimensions $N$ and $M$, these modifications do not matter and a hard gap of order $\gamma_0$ again emerges for the decay rates \shortcite{haake_statistics_1992,LEHMANN1995223,janik_non-hermitian_1997}. This induces the emergence of classical exponential decay in the time domain \shortcite{PhysRevE.56.R4911}.

In the classes with chiral or charge-conjugation symmetries, all poles come in pairs $E_l$, $-E_l^*$ which are symmetrically arranged with respect to the imaginary axis ${\rm Re}\,E=0$. The exception are unpaired modes pinned to the imaginary axis, $\mathrm{Re}\,E_l=0$, that arise from the zero modes in the closed setting,
and add a topological feature to the complex spectrum
\shortcite{Pikulin2012a,PhysRevB.87.235421}.
These symmetry-respecting poles can only depart from the imaginary axis in pairs, involving an exceptional point where two poles meet as described in Section~\ref{sec:nonherm}.
Thus, for an odd number of zero modes at least one such pole is always confined to the imaginary axis. For a superconducting system these poles describe Majorana zero modes that seep out of the system
\shortcite{Pikulin2012a,PhysRevB.87.235421,sanjose2016},
while in a photonic setting they can be employed for selective amplification \shortcite{schomerus_parity_2013,schomerus_topologically_2013,poli_selective_2015}.
Within random-matrix theory, we describe the consequences for the density of states in Section \ref{sec:delay}.

In the construction of the effective scattering models we noted that channels can also be closed by increasing the coupling beyond a certain threshold
($\sigma_n=-1$ in  Eq.~\eqref{eq:strobsmatnonball2} or Eq.~\eqref{eq:wnn}). Physically this should again result in a reduced decay rate $\gamma_0$ of the longest-living modes.
The spectral decomposition of the effective Hamiltonian, on the other hand, implies the sum rule
\begin{equation}
\mathrm{Im}\, \mathrm{tr}\, (H-i\pi WW^\dagger) =-\pi \mathrm{tr}\,WW^\dagger=\sum_m  {\rm Im}\, E_m,
\end{equation}
so that the sum of all decay rates must grow.
These two expectations can be reconciled in a careful analysis which shows that $N'$ strongly coupled channels result in a corresponding number of poles with very short life time \shortcite{haake_statistics_1992}. These poles are then well-separated from the poles describing the long-living states, which retain a typical decay rate $\gamma_0=\Gamma N\Delta/2\pi\hbar$.
This  nontrivial reorganisation
of the complex spectrum is known as resonance trapping
\shortcite{Rotter2009}. In the symmetry classes with charge-conjugation symmetry, it can affect the Majorana pole pinned to the imaginary axis, which justifies to identify the case of ideal coupling as a topological phase transition \shortcite{PhysRevLett.106.057001,marciani_effect_2016}.

The appearance of the classical decay rate in these considerations indicates that random-matrix theory is only applicable if the system-specific details become indiscernible before the classical dwell time $T_D=T_H/(\Gamma N)$. For a contact with dimensionless conductance $g_c=\Gamma N\gg 1$, this condition is  more stringent than the requirement in the closed system, where $T_D$ is replaced by $T_H$. A common occurrence where this condition is mildly violated are systems with ballistic decay routes, which result in additional short-living states that often form interweaving bands deep in the complex plane
\shortcite{Weich2014}. In a classically chaotic systems, these routes apply to trajectories that escape before the Ehrenfest time $T_{\mathrm{Ehr}}\approx \lambda^{-1}\ln N$, where $\lambda$ is the Lyapunov exponent
\shortcite{Berman1978450,PhysRevB.54.14423,SchomerusJacquod2005}.
In the limit of large $N$ and $M$,  the fraction of long-living modes is then reduced by a factor $\exp(-T_{\mathrm{Ehr}}/T_{D})=N^{-1/(\lambda T_{D})}$ \shortcite{PhysRevLett.93.154102}, a  power-law which agrees with a picture where these states are confined to the classical repeller \shortcite{PhysRevLett.91.154101,PhysRevLett.97.150406}. This modification due to ballistic chaotic decay is known as the fractal Weyl law \shortcite{0305-4470-38-49-014}.  In practice,
random-matrix theory still provides a good description of the remaining long-living modes
\shortcite{schomerus_lifetime_2009}. Furthermore, partial reflections at the contacts and disorder are very effective mechanisms to remove the ballistic decay routes.

\subsection{Mode non-orthogonality}

Since the effective Hamiltonian $H_{\mathrm{eff}}=H-i\pi WW^\dagger$ is non-hermitian, the quasibound states $|\phi_m\rangle$ from the eigenvalue problem \eqref{eq:effH} do not form an orthonormal basis. In a given basis, we thus have a spectral decompositions
$H_{\mathrm{eff}}=VDV^{-1}$, $D=\mathrm{diag}\,(E_m)$ where the matrix $V$ is not unitary.
The extent of mode non-orthogonality is then quantified by the condition numbers $O_{mn}$ introduced in Eq.~\eqref{eq:condnum}.

In order to get insight into the significance of these objects we consider the
divergent part
\begin{align}
\mathrm{tr}\,S^\dagger S
\approx & \mathrm{tr}\,[ 2\pi W^\dagger (E-H-i\pi WW^\dagger)^{-1}
\nonumber\\ &{}\times 2\pi W^\dagger W  (E-H+i\pi WW^\dagger)^{-1}W]\equiv \sigma(E)
\label{eq:sigmaE}
\end{align}
of the scattering strength for a complex energy close to a pole, $E\to E_n$  \shortcite{schomerus_quantum_2000}.
Using the spectral decomposition for the effective Hamiltonian we find
\begin{eqnarray}
\sigma(E)
&=&
\sum_{nm}
\frac{-(E_n-E_m^*)^2}{(E-E_n)(E-E_m^*)}O_{mn},
\end{eqnarray}
where we used
$2\pi WW^\dagger =i H_{\mathrm{eff}}-i H_{\mathrm{eff}}^\dagger =iVDV^{-1}-iV^{-1\dagger}D^*V^{\dagger}$.
Very close to the pole, $\sigma(E)\approx
\frac{(\hbar\gamma_n)^2}{|E-E_n|^2}K_n$ describes a Breit-Wigner resonance
with peak height proportional to $K_n=O_{nn}$.
Thus, the factors  $K_n$ are directly related to the scattering strengths of the quasibound states.

Energies in the complex plane
are effectively probed in amplifying photonic systems, which can be described in a scattering approach that is amended to account for radiation created within the medium
\shortcite{PhysRevLett.81.1829,schomerus_quantum_2000,schomerus_excess_2009}.
Under ideal conditions, an active medium with amplification rate $\gamma_a$ can generate spontaneously amplified radiation with frequency-resolved intensity
\begin{equation}
\label{eq:lorentzian}
I(\omega)\approx(2\pi)^{-1}\,\mathrm{tr}\,(S^\dagger S-1)|_{E=\hbar\omega-i\hbar\gamma_a/2}.
\end{equation}
Close to the laser threshold, a single pole $E_m=\hbar \omega_m$ lies close to the real axis, producing a well-isolated Lorentzian emission line
\begin{equation}
I(\omega)\approx
\frac{K_m}{2\pi}
\frac{\gamma_n^2}{(\omega-\omega_m)^2+(\gamma_m-\gamma_a)^2/4}.
\end{equation}
In this context,  $K_m$ is know as the Petermann factor
and signifies excess noise \shortcite{petermann_calculated_1979}.

For lasers we can ignore magneto-optical effects,
and thus are concerned with the orthogonal symmetry class where the effective Hamiltonian inherits the symmetry $H_{\mathrm{eff}}=H_{\mathrm{eff}}^T$. In this case we can normalise the right and left eigenstates so that $V^{-1}=V^T$
and find
\begin{equation}
K_m=|(V^\dagger V)_{mm}|^2.
\end{equation}

As described in Section~\ref{sec:nonherm} for the Ginibre ensemble, the Petermann factor of modes in the bulk of the complex spectrum should be large. For the effective Hamiltonian $H_{\rm eff}$ with $N, M\gg 1$, this can be verified in  the free-probability approach
\shortcite{janik_non-hermitian_1997}, according to which
\begin{equation}
\overline{K_m}\,|_{\gamma_m=\gamma}\approx N \left(\frac{\gamma}{\gamma_0}-1\right)
\left(1-\frac{(1-\Gamma)\gamma}{\gamma_0}\right)
\end{equation}
for decay rates well within the range $\gamma_0<\gamma<\gamma_0/(1-\Gamma)$.
However, this result breaks down close to the edges of the spectrum, where it violates the constraint $K_m\geq 1$, and hence does not apply to the long-living states that become the lasing modes.

These restrictions can be circumvented by the same supersymmetric techniques that address the poles \shortcite{schomerus_quantum_2000}.
Equation \eqref{eq:pgue} is then supplemented by
\begin{align}
\overline{K_m}\,|_{\gamma_m=\gamma}&=1+
\frac{2\pi\hbar }{\Delta}
\frac{S(\pi\hbar\gamma/\Delta)
}{ P(\gamma)}
,
\nonumber \\
S(y)&= - \int_0^ydy'\, \mathcal{F}_1(y')\frac{\partial}{\partial y'}
\mathcal{F}_2(y'),
\label{eqs:meankfinal}
\end{align}
which for identical transparencies $\Gamma_n=\Gamma$ can be brought into a compact form using
\begin{align}
&S(\pi\gamma/\hbar\Delta)=\frac{\Delta^2}{(2\pi\hbar\gamma)^2(N-1)!}
\nonumber
\\ &{}\times
\int_{N(1-\Gamma)\gamma/\gamma_0}^{N\gamma/\gamma_0}
\!\!\!\!\!\!\!\!\!dx\,
x^{N-1}e^{-x}
\left(\frac{N(1-\Gamma)\gamma}{\gamma_0}-x\right) \left(x-\frac{N\gamma}{\gamma_0}\right)
.
\label{eq:scon}
\end{align}
For large $N$, where we can apply a saddle-point approximation, it follows that the  Petermann factor
$\overline{K_m}\,|_{\gamma_m=\gamma_0}\sim \Gamma(\sqrt{2N/\pi}+4\pi/3)$ of the long-living modes can still be parametrically large in $N$. When a large number $L$ of such modes compete for the gain, the large-deviation tail of the decay-rate distribution \eqref{eq:py} is probed, which reduces $K_m$ by a factor $\sim 1/\sqrt{\ln L}$.
For $\Gamma N\ll 1$, on the other hand, the system is almost closed, and $K_m\sim 1$ as mode-orthogonality is restored. Similarly,
for $N=1$ the typical Petermann factor $K_m\sim 1+\Gamma\hbar\gamma_m/\Delta$ is also close to unity.

In all these cases, the Petermann factors of individual states can be much larger than the typical values quoted above. This is the case because $K_m$  diverges if two complex eigenvalues become degenerate, thus, as one approaches an exceptional point. The cubic level repulsion makes such approaches rare, but long power tails still emerge in the probability distribution of $K_m$.

We mentioned that the Petermann factor signifies an enhanced sensitivity to noise generated by spontaneous emission.
Similar considerations apply when external parameters are changed.
A perturbative treatment then reveals an enhanced response compared to systems with orthogonal modes, which is again quantified by the mode non-orthogonal matrix
\shortcite{PhysRevLett.108.184101}.
Close to an exceptional point, where the eigenvectors become degenerate and $K_m$ diverges, the significantly enhanced response can be exploited for sensors \shortcite{PhysRevLett.112.203901}. This enhanced sensitivity also applies to the topological spectral transitions in non-hermitian systems with a chiral or charge-conjugation symmetry (where they occur on the imaginary axis), or non-hermitian systems with a parity-time symmetry (where they occur on the real axis). The radiation emitted from a parity-time symmetric photonic system indeed diverges when one closes the system \shortcite{schomerus_quantum_2010}. For an open system close to an exceptional point, on the other hand, the formal divergence of the Petermann factor signifies a change of the line shape from the Lorentzian \eqref{eq:lorentzian} to a squared Lorentzian \shortcite{yoo_quantum_2011}.

\subsection{Delay times}
\label{sec:delay}

We now turn to the Wigner-Smith time-delay matrix $Q=-i\hbar S^\dagger dS/dE$, which according to the
Mahaux-Weidenm{\"u}ller formula \eqref{eq:weiden} can be written in the form \eqref{eq:wsmh},
\begin{equation}
Q=2\pi\hbar W^\dagger (E-H-i\pi WW^\dagger)^{-1} (E-H+i\pi WW^\dagger)^{-1}W.
\end{equation}
This matrix is manifestly hermitian and positive-definite, as required by causality.
According to the Birman-Krein formula \eqref{eq:krein}, the density of states
is then given by
\begin{equation}
\rho(E)=\mathrm{tr}\,W^\dagger (E-H-i\pi WW^\dagger)^{-1} (E-H+i\pi WW^\dagger)^{-1}W,
\end{equation}
which is of a similar form as the scattering strength $\sigma(E)$ in Eq.~\eqref{eq:sigmaE}.
Using the spectral decomposition for the effective Hamiltonian we find
\begin{eqnarray}
\rho(E)
&=&
\frac{1}{2\pi}\sum_{nm}
i\frac{(E_n-E_m^*)}{(E-E_n)(E-E_m^*)}O_{mn}
\nonumber\\
&=&
-\frac{1}{\pi}\mathrm{Im}\,\sum_{n}
\frac{1}{(E-E_n)},
\end{eqnarray}
where we used $\sum_n O_{nm}=\sum_m  O_{nm}=1$.
Close to an isolated resonance $E\approx E_n$ this approaches
 $\rho(E)\approx
\frac{1}{\pi}
\frac{\mathrm{Im}\,E_n}{|E-E_n|^2}$,
which is a Lorentzian normalised to 1.

\begin{figure}[t]
\includegraphics[width=\columnwidth]{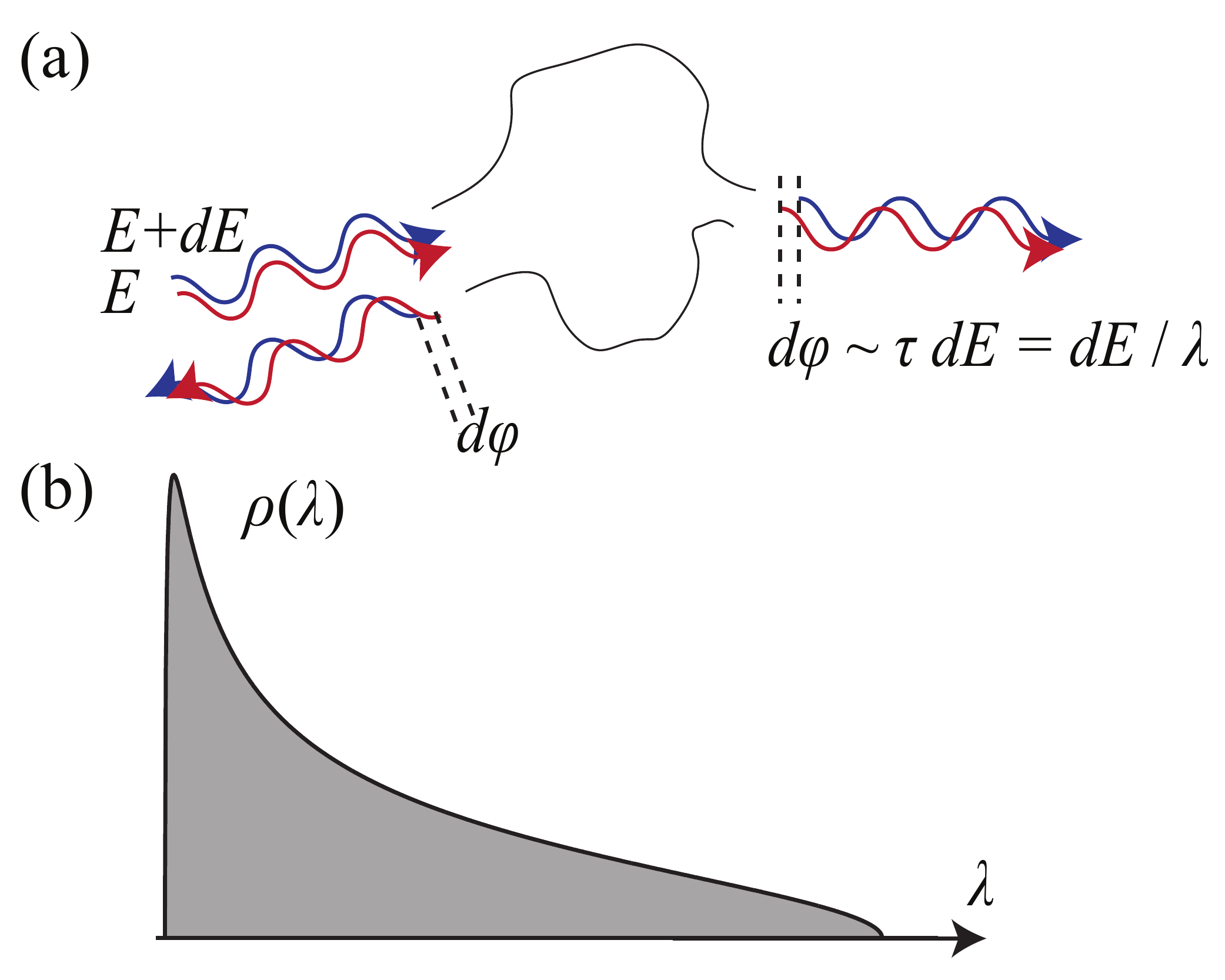}
\caption{(a) The Wigner-Smith delay times $\tau$ extract dynamical information by considering the energy sensitivity of the scattering phase in stationary scattering states. (b) Distribution of rates $\lambda=\tau^{-1}$ from random scattering, as predicted by the Marchenko-Pastur law \eqref{eq:mp} for the Wishart-Laguerre ensemble.}
\label{fig6}
\end{figure}

More direct insight into this problem is obtained from the proper delay times $\tau_n$, defined as the eigenvalues of $Q$, which are all real and nonnegative.
We first consider the case of ballistic coupling. In the three standard classes \shortcite{PhysRevLett.78.4737}, it is useful to consider the matrix $Q_S=S^{1/2}QS^{-1/2}$, which has the same eigenvalues but whose statistical distribution is the independent of $S$ itself, so that $P(S,Q_S)=P(S)P(Q_S)$. Perturbation theory around the point where $S=-1$ then shows that
the positive-definite rate matrix $Q_S^{-1}$ follows the distribution
\begin{equation}
P(Q_S^{-1})\propto (\mathrm{det}\,Q_S^{-1})^{N\beta/2}
\exp[-(\beta T_H/2)\mathrm{tr}\,Q_S^{-1}],
\end{equation}
with the Heisenberg time $T_H=2\pi \hbar/\Delta$.
This resembles a Wishart-Laguerre ensemble \eqref{eq:px}, but is directly expressed for  $Q_S$ and supplemented with a determinantal factor.
The joint distribution of rates $\lambda_n=1/\tau_n$ is given by
\begin{equation}
P(\{\lambda_n\})\propto\prod_{n<m}|\lambda_n-\lambda_m|^{\beta}\prod_k\lambda_k^{N\beta/2}
\exp(-\beta \lambda_k T_H/2),
\end{equation}
which indeed looks formally identical to the eigenvalue distribution \eqref{eq:pwish} of
a Wishart matrix, albeit with half-integer dimensions if $\beta=4$.
This still constitutes a Wishart-Laguerre ensemble.

The same independence of $S$ and $Q_S$ also
occurs in the four classes with charge-conjugation symmetry at the symmetry point $E=0$ \shortcite{PhysRevB.90.045403}, where
\begin{equation}
\label{eq:plambda}
P(\{\lambda_n\})\propto\prod_{n<m}|\lambda_n-\lambda_m|^{\beta_T}
\prod_k\lambda^{\beta_T'+N\beta_T/2}
e^{-\beta_T''\lambda_k T_H/2}
%\exp(-\beta_T''\lambda_k T_H/2)
\end{equation}
with $\beta_T=1,2,4,2$,
$\beta_T'=-1,-1,2,1$, $\beta_T''=1,2,2,1$ in the symmetry classes D, DIII, C, CI. In the classes C and CI all delay times occur in degenerate pairs, which in Eq.~\eqref{eq:plambda} are only accounted for once.

In contrast, the chiral symmetry condition $S(E)=\mathcal{X}S^\dagger(-E)S^\dagger$ implies that the hermitian unitary matrix $S_{X0}=\mathcal{X}S(0)$ commutes with $Q(0)$, so that both matrices share a common structure \shortcite{PhysRevLett.114.166803}. Recall that $S_{X0}$ has eigenvalues $\pm 1$, whose frequency is captured by the topological quantum number
$\nu_0=\frac{1}{2}\mathrm{tr}\,S_{X0}=[\nu+(N_A-N_B)/2]_{|\nu_0|\leq N/2}$ (see Section \ref{sec:merits}).
Correspondingly, the delay times can be grouped into two sets, made of $N_+=N/2+\nu_0$ delay times $\tau_n^+=1/\lambda_n^+$ associated with the subspace where the eigenvalues of $S_{X0}$ are $1$, and $N_-=N/2-\nu_0$ delay times $\tau_n^-=1/\lambda_n^-$  associated with the subspace where the eigenvalues of $S_{X0}$ are $-1$. These two sectors can be made manifest by considering the reordered matrix
\begin{equation}
\tilde Q(E)=2\pi\hbar \frac{1}{E-H+i\pi WW^\dagger}WW^\dagger \frac{1}{E-H-i\pi WW^\dagger},
%\tilde Q(E)=2\pi\hbar (E-H+i\pi WW^\dagger)^{-1}WW^\dagger (E-H-i\pi WW^\dagger)^{-1},
\end{equation}
which has the same non-vanishing eigenvalues as $Q$.  Inserting here the chiral Hamiltonian \eqref{eq:chiralh} and splitting the coupling matrix analogously into blocks $W=\mathrm{diag}\,(W_A,W_B)$ describing $N_A$  and $N_B$ open channels, respectively,
this reordered matrix becomes block diagonal,
\begin{equation}
\tilde Q(0)=2\pi\hbar\,\mathrm{diag}\,(\Lambda_-^{-1},\Lambda_+^{-1})
\end{equation}
where
\begin{eqnarray}
\Lambda_-&=&\pi^2W_AW_A^\dagger+A(W_BW_B^\dagger+0^+)^{-1}A^\dagger,\\
\Lambda_+&=&\pi^2W_BW_B^\dagger+A^\dagger(W_AW_A^\dagger+0^+)^{-1}A.
\end{eqnarray}
In the subspaces where these two matrices are finite, we can write $\Lambda_\pm=X_\pm ^\dagger X_\pm$ with an $N\times N_{\pm}$ dimensional matrix $X$. For large $M$, the matrix $X$ tends to a random Gaussian matrix, so that the two sets of decay rates are both obtained from a Wishart-Laguerre ensemble,
\begin{align}
P_\pm(\{\lambda_n^{\pm}\})&=
\prod_k \lambda_k^{\beta/2-1+(\beta/4)|N\mp2\nu\pm N_B\mp N_A|}e^{-\beta \lambda_k T_H/4}
\nonumber \\  &{}\times
\prod_{n<m}|\lambda_n^{\pm}-\lambda_m^{\pm}|^{\beta}.
\label{eq:lambdac}
\end{align}
The two sets are independent of each other, whereby the full joint distribution factorises according to
$P(\{\lambda_n^+,\lambda_n^{-}\})=P_+(\{\lambda_n^{+}\})P_-(\{\lambda_n^{-}\})$.

We note that the joint distribution \eqref{eq:plambda}
does not involve the topological quantum number $\nu$ defined in classes D and DIII.
In the chiral ensembles, on the other hand, the topological zero modes directly affect the joint distribution \eqref{eq:lambdac}.
This dependence also transfers to the mean density of states, which is given by
\begin{align}
\overline{\rho}=&\frac{1}{\Delta}\frac{N/2(N/2+1-2/\beta)+\nu_0^2}{(N/2+1-2/\beta)^2-\nu_0^2} \quad \mbox{for
}|\nu_0|<N/2,
\\
\overline{\rho}=&\frac{1}{\Delta}\frac{N/2}{|\nu-\nu_0+(N_A-N_B)/2|+1-2/\beta}
\nonumber\\ & \qquad\qquad\qquad\qquad\qquad\qquad\hfill\mbox{for }|\nu_0|=N/2,
\end{align}
with the exceptions $|\nu-N_B|\leq 1$ or $|\nu+N_A|\leq 1$ (for  $\beta=1$) and
$\nu=N_B$ or $\nu=-N_A$ (for  $\beta=2$) where the ensemble-average diverges.

These considerations can be extended to non-ideal leads \shortcite{marciani_effect_2016}, where one relates the scattering matrix $S$ via Eq.~\eqref{eq:sbarrier} to the scattering matrix $S_0$ for ballistic coupling. The time-delay matrix then changes from $Q_{S0}$ to $Q_S=\Sigma Q_{S0}\Sigma^\dagger$, where $\Sigma=(1-S^\dagger r_B)^{-1}t_B$. The transformation of the probability measure follows from the analogous relation $S^\dagger dS =\Sigma (S_0^\dagger dS_0)\Sigma^\dagger$.
For the standard symmetry classes, the factorised distribution  $P(S_0,Q_{S0}^{-1})=P(S_0)P(Q_{S0}^{-1})$ transforms into
\begin{align}
P(S,Q_S^{-1})=&(\mathrm{det}\,\Sigma \Sigma^\dagger)^{N\beta/2}(\mathrm{det} Q_S^{-1})^{N\beta/2}
\nonumber\\ & {}\times
\exp[-(\beta T_H/2)\,\mathrm{tr}\,\Sigma^\dagger Q_S^{-1}\Sigma],
\end{align}
while in the classes with charge-conjugation symmetry this takes the form
\begin{align}
P(S,Q_S^{-1})=&(\mathrm{det}\,\Sigma \Sigma^\dagger)^{N\beta_T/2}(\mathrm{det} Q_S^{-1})^{N\beta_T/2+\beta_T'}
\nonumber\\ & {}\times
\exp[-(\beta_T'' T_H/2)\,\mathrm{tr}\,\Sigma^\dagger Q_S^{-1}\Sigma].
\end{align}
The density of states
$\rho={2\pi \hbar}^{-1}\,\mathrm{tr}\,\Sigma Q_{S0}\Sigma^\dagger$
can then be analysed directly using the independence of $S$ (appearing in $\Sigma$) and $Q_{S0}$. By definition, $\mathrm{tr}\,\overline{Q_{S0}}=2\pi\hbar \rho_0$ is given by the density of states for ideal coupling, while the scattering matrix itself follows the Poisson kernel distribution
$P(S)\propto|\mathrm{det}\,(1-r_B^\dagger S)|^{-\beta_T N-2+\beta_T-2\beta_T'}$ (recovering the result of \shortciteNP{PhysRevB.79.214506}).
In class $D$, a barrier with mode-independent transparency $\Gamma$ then yields the mean density of states
\begin{equation}
\overline{\rho}=\frac{N}{(N-2)\Delta}\left(1-\frac{2}{N\Gamma}[\Gamma-1+(-1)^\nu(1-\Gamma)^{N/2}]\right),
\end{equation}
which now depends on $\nu$.
More generally, in classes D and DIII
a topological zero mode remains visible as long as none of the couplings are fully ballistic.

\subsection{Transport}
\label{sec:transport}

\begin{figure}[t]
\includegraphics[width=\columnwidth]{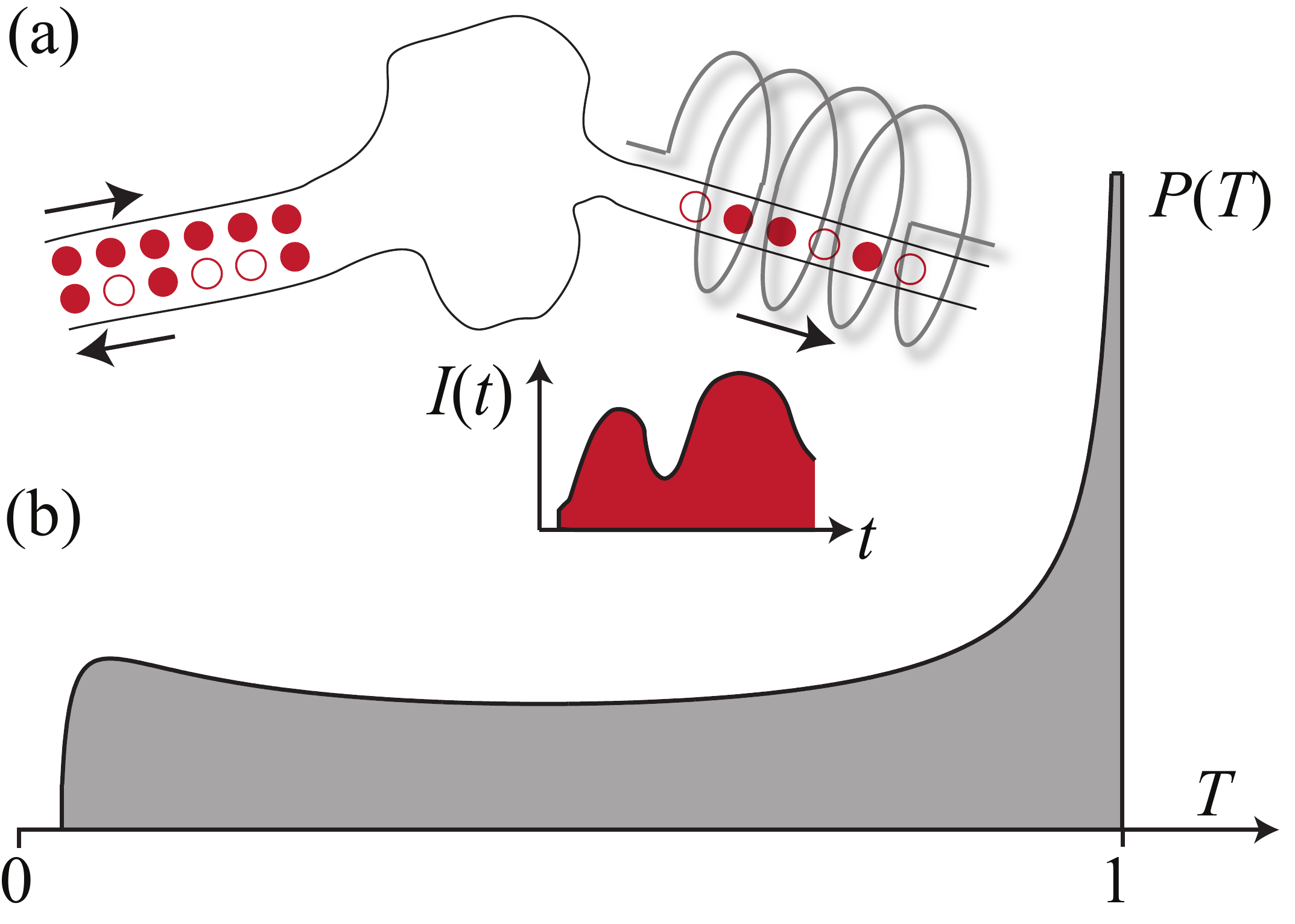}
\caption{(a) Phase-coherent electronic transport is characterised by partition noise, generated by the transmission of charge carriers with probability $T$. This noise can be detected in the current fluctuations $I(t)$. (b) Mean density of transmission probabilities $T$ from Eq.~\eqref{eq:rhot}, in accordance to the Jacobi ensemble of random-matrix theory.}
\label{fig7}
\end{figure}

Some of the best tested applications of random scattering matrices arise when one considers the low-temperature transport of electrons through a mesoscopic device in response to a small bias voltage $V_b$. These applications have been covered in two comprehensive reviews considering the standard ensembles \shortcite{RevModPhys.69.731}
and the additional ensembles with chiral or charge-conjugation symmetry \shortcite{RevModPhys.87.1037}, supplemented by a detailed review on shot noise \shortcite{blanter_shot_2000}, and we refer to these sources throughout the section.
In keeping with the rest of these notes we remain focussed on situations where the details of the geometry do not matter (this
ignores the effects of Anderson localization, which we briefly pick up in the next Chapter \ref{chap:loc}).
For the scattering at a fixed energy we are then directly led to the circular ensembles. This was first utilised by
\shortciteN{PhysRevLett.64.241}, who found that the statistics of phase shifts from chaotic scattering agree with
Eq.~\eqref{eq:pphi}, while the applications to transport were pioneered by
\shortciteN{baranger_mesoscopic_1994} and \shortciteN{jalabert_universal_1994}.

In the scattering approach to transport, the device is modelled as a scattering region attached to a left and a right lead, so that
the scattering matrix is of the form \eqref{eq:sblock}. The quantities of interest are the transmission eigenvalues $T_n\in[0,1]$ of  $t^\dagger t$, with the dimensionless conductance given by $g=\sum_n T_n$. We shall assume that the number of channels $N_R\geq N_L$ so as to avoid $N_L-N_R$ vanishing eigenvalues (otherwise we can simply study the eigenvalues of $tt^\dagger$).
In the three standard circular ensembles (COE, CUE and CSE), the joint distribution of the transmission eigenvalues is then given by
\begin{equation}
\label{eq:pt}
P(\{T_n\})\propto\sum_{n<m}|T_n-T_m|^\beta\prod_kT_k^{-1+\beta(1+|N_L-N_R|)/2},
\end{equation}
which can be interpreted as a Jacobi ensemble for variables $\mu_n=1-2T_n$.

For large number of channels $N_L,N_R\gg 1$, the mean density of eigenvalues converges to
\begin{equation}
\label{eq:rhot}
\rho(T)=\frac{N_L+N_R}{2\pi T}\left(\frac{T-T_c}{1-T}\right)^{1/2}
\end{equation}
for $1>T>T_c=(N_L-N_R)^2/(N_L+N_R)^2$;
for $N_L=N_R$ this takes the form
\begin{equation}
\rho(T)=\frac{N_L}{\pi}\frac{1}{\sqrt{T(1-T)}}.
\end{equation}

In leading order of $N_L,N_R\gg 1$, Eq.~\eqref{eq:rhot} gives the ensemble-averaged dimensionless conductance
$\overline{g}=N_LN_R/(N_L+N_R)$, so that ${\overline{g}}^{-1}=N_L^{-1}+ N_R^{-1}$ resembles the series addition of two resistances.
The exact result for finite $N_L$ and $N_R$ is
\begin{equation}
\overline{g}=
\frac{N_LN_R}{N_L+N_R-1+2/\beta},
\end{equation}
so that the next-to-leading reads
\begin{equation}
\overline{g}-\frac{N_LN_R}{N_L+N_R}
\approx (1-2/\beta)\frac{N_LN_R}{(N_L+N_R)^2} \quad\mbox{for }N_L,N_R\gg 1.
\end{equation}
This ensemble-dependent correction, known as weak localization (for $\beta=1$) and as weak anti-localization (for $\beta=4$), can be related to the
factors $T_k^{-1+\beta/2(1+|N_L-N_R|)}$ in the joint distribution \eqref{eq:pt}, which induce a bias of the transmission eigenvalues to small or large values.
The joint distribution  also determines the variance of the conductance within the ensemble,
\begin{equation}
\mathrm{var}\, g\approx \frac{(N_LN_R)^2}{\beta(N_L+N_R)^4} \quad\mbox{for }N_L,N_R\gg 1.
\end{equation}
Due to the repulsion $\sim|T_n-T_m|^\beta$ of the eigenvalues this variance is small, but depends on the symmetry class already in leading non-vanishing order.

In a more general picture, the transmission eigenvalues determine the full counting statistics of the electrons that pass through the  system. Let $Q(s)$ be the accumulated charge over a time interval $s$, via the arrival of electrons with elementary charge $\textbf{e}$. In each eigenchannel, an incoming electron is transmitted with probability $T_n$, so that the counting statistics are given by a Bernoulli process. Noting that these transmission events occur with an attempt rate $\textbf{e}V_b/h$ (with $h=2\pi\hbar$),
this process is described by the cumulant-generating function
\shortcite{1993JETPL..58..230L}
\begin{align}
\ln \langle \exp(p Q(s)/\textbf{e})\rangle&=\sum_{k=1}^\infty \langle\langle Q(s)\rangle\rangle \frac{p^k}{\textbf{e}^kk!}
\nonumber \\ &
 = s (\textbf{e}V_b/h)\sum_n\ln [1+T_n(e^p-1)].
\end{align}
The average current follows  from  $I=\lim_{s\to\infty} s^{-1}\textbf{e}\langle N(s)\rangle= (\textbf{e}^2V_b/h) g$, while the shot-noise power is $P=\lim_{s\to\infty}2s^{-1}\textbf{e}^2\langle\langle N(s)^2\rangle\rangle=(2\textbf{e}^3V_b/h)\sum_nT_n(1-T_n)$. If all transmission eigenvalues are small, the shot-noise power is  $P=2\textbf{e}I\equiv P_0$, while in  general  $P=f P_0$ with the so-called Fano factor $f=\sum_nT_n(1-T_n)/\sum_nT_n\in [0,1]$.

For $N_L=N_R\gg 1$ we can calculate the cumulant-generating function exactly
\shortcite{blanter_effect_2001},
\begin{align}
\overline{
\ln \langle \exp(p Q(s)/\textbf{e})\rangle}&=
s \frac{\textbf{e}V}{h}\int dT \rho(T)\ln [1+T(e^p-1)]
\nonumber \\ &
=
4 s g \frac{\textbf{e}V}{h} \ln[\frac{1+e^{p/2}}{2}].
\end{align}
The Fano factor is then given by $\overline {f}=1/4$. For $N_L,N_R\gg 1$ not necessarily equal, one finds
\begin{equation}
\overline{f}\approx \frac{N_LN_R}{(N_L+N_R)^2}-(1-2/\beta)\frac{(N_L-N_R)^2}{(N_L+N_R)^3},
\end{equation}
where the weak-localization correction is seen to vanish if $N_L=N_R$.

As for the decay problem, these transport properties are modified by ballistic transport routes.
A wavepacket injected into the opening can leave without any noticeable diffraction until the transport Ehrenfest time $T_{\rm Ehr}'=\lambda^{-1}\ln N^2/M$
\shortcite{PhysRevB.67.241301}, which results in transmission eigenvalues $T_n$ close to 0 and 1. In particular, these processes can yield a noticeable suppression of shot noise \shortcite{tworzydlo_dynamical_2003}.

In the chiral symmetry classes, the statistics of the transmission eigenvalues is most conveniently expressed via $T_n=\sqrt{1-r_n^2}$, where $r_n$ are the eigenvalues of the hermitian matrix $R_z=\tau_z r$ \shortcite{PhysRevB.66.041307}.
We only consider  the case $N_L=N_R$ with balanced coupling to both chiral subspaces ($N_A=N_B$), so that $\nu_0=\mathrm{tr}\,R_z$ determines the number of zero modes in the closed system.
In this case one encounters $|\nu_0|$ closed transmission channels  with $r_n^2=1$, while the remaining eigenvalues obey the joined distribution
\begin{equation}
\label{eq:ptchiral}
P(\{r_n\})\propto\prod_{n<m}|r_n-r_m|^\beta \prod_k(1-r_k^2)^{-1+(|\nu_0|+1)\beta/2}.
\end{equation}
In the symmetry classes with a charge-conjugation symmetry \shortcite{PhysRevB.82.014536},
\begin{align}
P(\{T_n\})\propto &
\prod_{n<m}|T_n-T_m|^{\beta_T}
\nonumber \\ & {}\times
\prod_k T_k^{-1+\beta_T(1+|N_L-N_R|)/2}
(1-T_k)^{\beta_T'/2},
\label{eq:ptcconj}
\end{align}
where the parameters
$\beta_T=1,2,4,2$,
$\beta_T'=-1,-1,2,1$ (classes D, DIII, C, CI) are the same as those encountered for the delay times.
In the large-$N$ limit, the eigenvalue density becomes again ensemble-independent and approaches \eqref{eq:rhot}.

Note that the topological quantum number $\nu_0$ only appears in the joint distribution \eqref{eq:ptchiral} for chiral symmetry, but not in the joint distribution \eqref{eq:ptcconj}, so that
any zero modes due to charge-conjugation symmetry cannot be detected in the transport with ideal leads---the same situation that we encountered for the density of states.
This provides an incentive to consider the role of superconductivity and tunnel barriers in such systems, which we here will discuss for the classes D and BDI  \shortcite{Pikulin2012}. Instead of applying the Poisson kernel, we consider the experimentally relevant situation \shortcite{mourik_signatures_2012} where the tunnel barrier is placed into a normal-conducting  region, which is then interfaced with a superconductor.

In the context of such superconducting systems, the dimensionless conductance $g$ relates to the particle (or heat) transport, while the charge transport is modified by the fact that holes carry an opposite charge. If a normal metallic region from the orthogonal symmetry class is attached to a conventional superconductor, the dimensionless conductance for charge transport is given by Eq.~\eqref{eq:gns}, which applies to systems with no magnetic fields and no spin-orbit scattering.
The symmetry classes D arises in the presence of spin-orbit coupling and broken time-reversal symmetry, where only the charge-conjugation symmetry with $\mathcal{C}^2=1$ remains. The class BDI emerges from an additional chiral symmetry $\mathcal{X}$ that commutes with $\mathcal{C}$, which then also implies a time-reversal symmetry $\mathcal{T}=\mathcal{X}\mathcal{C}$ with $\mathcal{T}^2=1$.
In these classes, the dimensionless conductance at the Fermi level can be written as
\begin{align}
g_{NS}=&\mathrm{tr}\,[\Gamma\frac{1}{1-U^*\sqrt{1-\Gamma}U\sqrt{1-\Gamma}}
\nonumber \\ & {}\times \Gamma
\frac{1}{1-U^\dagger\sqrt{1-\Gamma}U^T\sqrt{1-\Gamma}}],
%g_{NS}=\mathrm{tr}\,\Gamma(1-U^*\sqrt{1-\Gamma}U\sqrt{1-\Gamma})^{-1}\Gamma
%(1-U^\dagger\sqrt{1-\Gamma}U^T\sqrt{1-\Gamma})^{-1},
\end{align}
where $\Gamma=\mathrm{diag}\,(T_n)$ while the $N\times N$ dimensional unitary matrix $U$ accounts for the mode-mixing from the spin-orbit scattering.
For a large tunnel barrier in the normal region, we can assume that all transmission eigenvalues are identical, $T_n\equiv T$, so that
\begin{align}
g_{NS}&=T^2\,\mathrm{tr}\,\frac{1}{1-(1-T)X} \frac{1}{1-(1-T)X^\dagger}
\nonumber \\
&=\sum_n\frac{T^2}{|1-(1-T)x_n|^2}
\end{align}
is determined by the eigenvalues $x_n$ of $X=U^*U$.

The structure of $X$ implies that all eigenvalues occur in complex conjugated pairs $x_n$, $x_{\bar n}=x_n^*$, with the exception of possible eigenvalues pinned to $1$.
In class D, where $U$ is only constrained to be unitary, such an eigenvalue occurs if $N$ is odd; we thus have a topological index $\nu=N\,\mathrm{mod}\,2$. The paired eigenvalues can be specified by the quantities $\mu_n=(x_n+x_{\bar n})/2=\mathrm{Re}\,x_n$.
Sampling $U$ from the circular unitary ensemble, these quantities
then follow the distribution
\begin{align}
\label{eq:jac1}
P(\{\mu_n\})&\propto\prod_{n<m}(\mu_n-\mu_m)^2\prod_k\frac{1+\mu_k}{\sqrt{1-\mu_k^2}}
\quad \mbox{if } \nu=0,
\\
P(\{\mu_n\})&\propto\prod_{n<m}(\mu_n-\mu_m)^2\prod_k\sqrt{1-\mu_k^2}
\quad \mbox{if } \nu=1.
\end{align}

In class BDI, $U$ is unitary and hermitian, and the number of pinned eigenvalues $|\nu|$ follows from $\nu=\mathrm{tr}\,U$. Setting $U=V^\dagger \mathrm{diag}\, (1,1,1,\ldots,-1,-1,-1,\ldots) V$ with
$V$ again following the circular unitary ensemble,
the paired eigenvalues are then described
by the distribution
\begin{equation}
\label{eq:jac2}
P(\{\mu_n\})\propto\prod_{n<m}|\mu_n-\mu_m|\prod_k(1-\mu_k)^{(|\nu|-1)/2}.
\end{equation}

The probability distributions
\eqref{eq:jac1}
and \eqref{eq:jac2}
are both of the form of a Jacobi ensemble  \eqref{eq:jac0}.
Including the next-to-leading order in the  large-$N$ limit, the density
\begin{equation}
\rho(\mu)=\frac{N}{\pi}\frac{1}{\sqrt{1-\mu^2}}+\frac{1}{2}\delta(\mu-1)-\frac{1}{2}\delta(\mu+1)
\end{equation}
becomes independent of the symmetry class and the topological indices.
In leading orders, the ensemble-averaged dimensionless conductance is then given by
\begin{equation}
\overline{g_{NS}}=\frac{NT}{2-T}+\frac{2(1-T)}{(2-T)^2},
\end{equation}
again irrespective of the symmetry class.

Note that the joint distribution Eq.~\eqref{eq:ptchiral} of reflection coefficients in the classes with chiral symmetry can also be interpreted as a Jacobi distribution, while the joint distributions \eqref{eq:pt} and \eqref{eq:ptcconj} can be brought into this form by a suitable shift $T_n=(1-\mu_n)/2$ of the transmission coefficient.
We take the appearance of this final class of classical random-matrix ensembles as our cue to wrap up the discussion of fully ergodic elastic scattering. Much more is known, both in terms of technical details as well as in terms of practical applications. This includes the full physical implications of superconductivity, such as Andreev reflection and Josephson currents \shortcite{RevModPhys.69.731,RevModPhys.87.1037},
as well as the interpretation of zero modes in terms of Majorana fermions
\shortcite{0034-4885-75-7-076501,0268-1242-27-12-124003,doi:10.1146/annurev-conmatphys-030212-184337}. Another important aspect is the role of physical dimensions, which enters the full classification of topologically protected states
\shortcite{:/content/aip/proceeding/aipcp/10.1063/1.3149495,PhysRevB.82.115120,1367-2630-12-6-065010,RevModPhys.82.3045,RevModPhys.83.1057}.
In the following chapter \ref{chap:loc}, we turn to one specific aspect of low-dimensional physics, the phenomenon of Anderson localization which prevents the full exploration of phase space. We also take this as an opportunity for a short detour into interacting systems, for which the related question of thermalization can be addressed by the density matrix.

\section{Localization, thermalization and entanglement}
\label{chap:loc}

To round off these notes we discuss a setting for random-matrix applications which has significance also for interacting systems. This brings us back to the origins of the field, which concerned the energy levels of heavy nuclei \shortcite{wigner1956conf,Porter1965,RevModPhys.81.539}. There, interactions are sufficient to effectively couple a large number of many-body states, thus resulting in a random Hamiltonian.

Quite generally, statistical methods find broad applications to interacting systems, where the dynamics becomes particularly interesting when one considers low dimensions  \shortcite{cardy1996scaling,sachdev_quantum_1999}. An interesting question is how such systems thermalize
\shortcite{PhysRev.109.1492,PhysRevA.43.2046,srednicki_chaos_1994,gemmer_quantum_2010}. Even in absence of interactions, the spread of energy and the propagation of particles can  be inhibited by the same wave-interference effects that we so far have taken as the very justification for the application of random-matrix theory. These localization effects, first recognised by \shortciteN{PhysRev.109.1492}, arise from the sparsity of the underlying matrices, be it due to a reduced coordination number on a lattice, resulting in localization in real space
\shortcite{0034-4885-56-12-001,RevModPhys.80.1355}, or due to the presence of interactions that only involve some few-body operators, resulting in localization in Fock space \shortcite{Basko20061126,altman_universal_2015,doi:10.1146/annurev-conmatphys-031214-014726}. The main question is when this sparsity can be felt, and how.

We first discuss this question briefly for non-interacting systems, where it can be addressed by the impact on the transport properties described in the Section \ref{sec:transport}. We then turn to the many-body setting, where we focus on aspects of thermalization and entanglement.

\subsection{Anderson localization}

In random-matrix theory, systems with fully chaotic wave scattering are traditionally termed zero-dimensional systems. This is because in practice these systems are often realised by shrinking the size of two or three dimensional systems, as, e.g., in a planar quantum dot or a metallic grain. From a different perspective, such systems could be termed infinite-dimensional, as their main feature is  the efficient dynamical coupling of states in the accessible Hilbert space, which is also observed in lattices or graphs with a fixed number of vertices and increasing coordination number. In properly scaled units, we can then assume that all of Hilbert space is instantly explored (ergodic time $T_{\rm erg}=0$), so that only one relevant dynamical time scale remains---the dwell time $T_{D}$, which characterizes how long a particle will reside within the scattering region.

This approach reaches its limit when the internal transport within the system becomes inefficient. Consider a system made of $L$ random scattering regions placed in series, with contacts carrying $N\gg 1$ channels \shortcite{Iida1990219}. While the dimensionless conductance $g\sim N/2$ of each individual region may be large, the overall conductance $g_L\sim N/(1+L)$ of the composed system shrinks when $L$ is increased. Once $g\lesssim 1$, one finds that the conductance decays exponentially with $L$, $\ln g_L\sim -2L/\xi$ where $\xi\sim \beta N$ is termed the localization length. The decay arises from a similar exponential decay of the wave functions in the closed system. This phenomenon is known as Anderson localization \shortcite{PhysRev.109.1492,0034-4885-56-12-001,RevModPhys.80.1355}. Among its many signatures, it results in a significant reduction of the levels repulsion, as energy levels of wave functions localized far apart can approach each other closely.
We describe the underlying mechanism in the quasi one-dimensional setting sketched above, which is realised in a long and narrow disordered quantum wire or a disordered wave guide. Anderson localization then occurs at any strength of uncorrelated disorder.

For the detailed statistical description, one composes the system from slices of length $L_0$ that efficiently scramble all the modes according to a mean free path $l$, but are small enough so that the effect of each slice can be obtained in perturbation theory
\shortcite{dorokhov1982transmission,mello_macroscopic_1988,RevModPhys.69.731,nazarov2009quantum}.
For $\beta=2$, one obtains for each step
\begin{align}
c_n&=\frac{l}{L_0}\overline{\delta T_n}=-T_n^2+2T_n(1-T_n)\sum_{m\neq n}\frac{T_m}{T_n-T_m},
\nonumber \\
d_n&=\frac{l}{L_0}\overline{(\delta T_n)^2}=2T_n^2(1-T_n).
\end{align}
The result can be fed into a Fokker-Planck equation
\begin{equation}
Nl\frac{\partial }{\partial L}P(\{T_n\})=\sum_n\frac{\partial}{\partial T_n}
\left(-c_n +\frac{1}{2}\frac{\partial}{\partial T_n}d_n\right)P(\{T_n\})
\end{equation}
for the joint distribution of transmission eigenvalues, then known (up to a change of variables) as the Dorokhov-Mello-Pereira-Kumar (DMPK) equation.
For this particular symmetry class, the joint distribution can be found exactly  by a mapping to a Schr{\"o}dinger equation describing free fermions
\shortcite{PhysRevLett.71.3689}.  DMPK equations can also be formulated for the other symmetry classes, where they reveal delocalizing effects near the spectral symmetry points \shortcite{brouwer_zero_2002,2005cond.mat.11622B}.
In the many-channel limit $N\gg 1$, these equations make  equivalent predictions  to non-linear sigma models \shortcite{PhysRevB.53.1490,efetov_supersymmetry_1996}, both in the diffusive regime $l\ll L\ll\xi$ as well as in the localised regime $L\gg\xi$. This convergence of models
indicates a large degree of universality, which we describe next \shortcite{RevModPhys.69.731}.

In the diffusive regime $l\ll L\ll\xi$ of a system with $N\gg 1$ channels, the density of transmission eigenvalues becomes independent of the symmetry class and is given by
\begin{align}
\rho(T)&=\frac{Nl}{2L}\frac{1}{T\sqrt{1-T}}\quad \mbox{for }T_-<T<1,
\nonumber \\
 T_- &\sim 4\exp(-2L/l).
\end{align}
Including the next-order corrections, the ensemble-averaged dimensionless conductance is
$\overline g=Nl/L+\frac{1}{3}(1-2/\beta)$ and the variance is $\mathrm{var}\,g=2/(15 \beta)$. Furthermore, $\overline{T(1-T)}=\overline{T}/3$ so that  the shot-noise Fano factor is $f=1/3$.

To capture the universal aspects of Anderson localization for $L\gg \xi$, it is useful to recall that the transfer matrix \eqref{eq:transfer}
of a composed system follows by multiplication of the transfer matrices  of the components, $M=\prod_{l=1}^L M_l$. These aspects are therefore linked to products of random matrices \shortcite{crisanti_products_1993,ThisIssueComtetPreprint}. For $L\to\infty$, the eigenvalues $x_n>1$ of $M^\dagger M$
display an exponential dependence, $\ln x_n\sim 2 L/\xi_n$ with Lyapunov exponents $\xi_n$. The scaled exponents $(\ln x_n)/L$  exhibit diminishing fluctuations, which are captured by a log-normal distribution for $x_n$.

These general features directly translate to the transmission eigenvalues $T_n=4/(x_n+2+1/x_n)$. The details follow from the DMPK equation, which recovers the log-normal behaviour of $\ln T_n$ in the localised regime. When ordered by magnitude, the transmission eigenvalues   fall into a pattern $1\gg T_1\gg T_2\gg\ldots\gg T_N$, with $\ln T_n\sim -2L(1+\beta n-\beta)/\xi$ and $\mathrm{var} \ln T \sim 4 L/\xi$. The dimensionless conductance is  dominated by the largest transmission eigenvalue, and also obeys a log-normal distribution.

These results indicate that the variance $\mathrm{var} \ln g=-2\overline {\ln g}$ in the localized regime is universal. This
relation can also be obtained in a diagrammatic approach, where it results from the random-phase approximation \shortcite{PhysRevB.22.3519}, and only breaks down when one approaches the band edges, while small corrections are observed near  spectral symmetry points in the clean system \shortcite{PhysRevB.67.100201}.
The strong universality of the log-normal distribution underpins qualitative descriptions based on renormalization arguments, which extend the considerations to higher dimensions \shortcite{PhysRevLett.42.673}.
For three dimensions, these arguments predict that localization sets in at a finite disorder strength, which is well supported by numerical investigations \shortcite{0034-4885-56-12-001,RevModPhys.80.1355}.
However, an accurate statistical description is still lacking.

\subsection{Thermalization and localization in many-body systems}

\begin{figure}[t]
\includegraphics[width=\columnwidth]{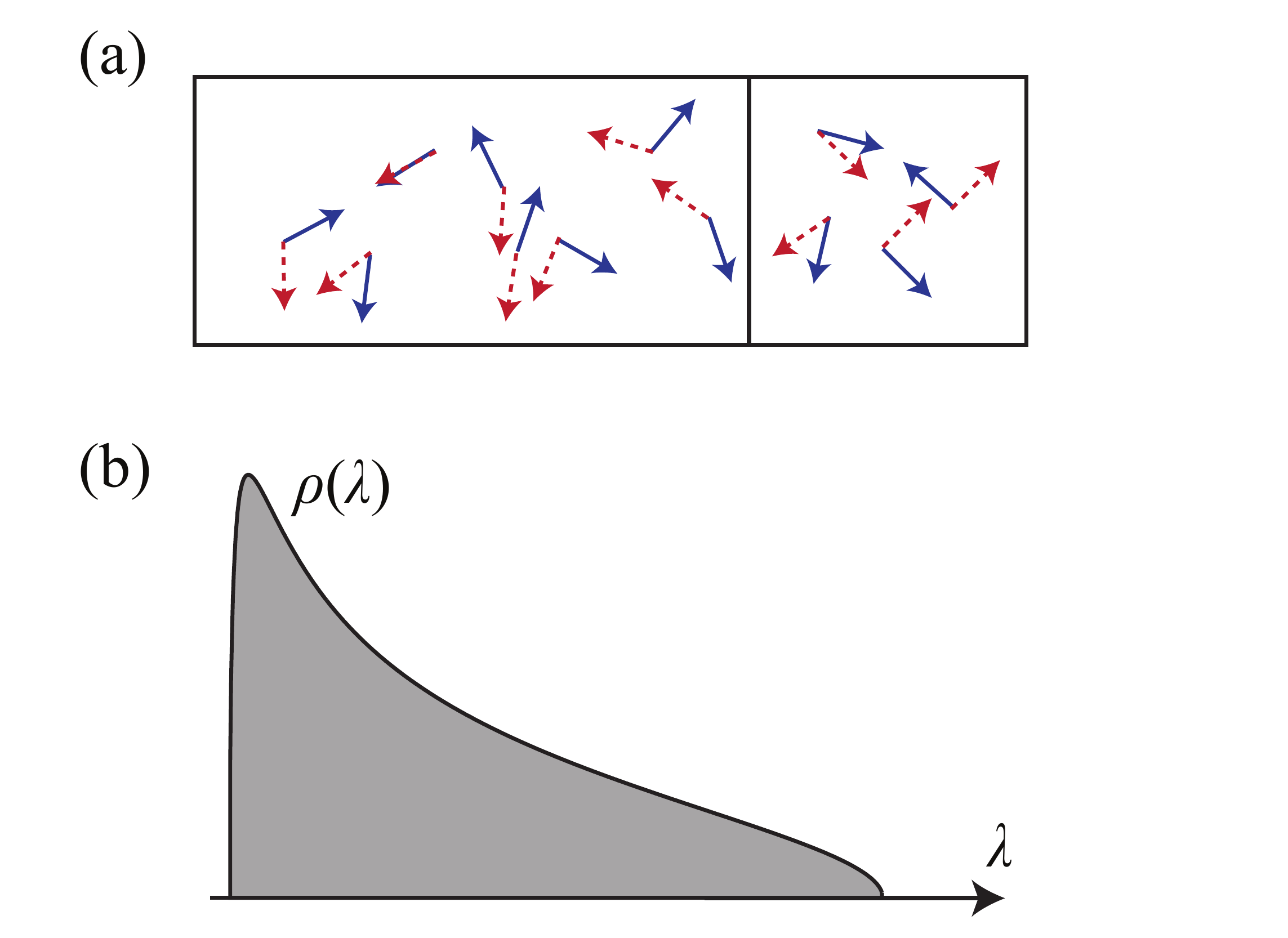}
\caption{(a) Bipartite entanglement concerns the quantum correlations between a subsystem and its complement. This information is captured in the eigenvalues $\lambda$ of the reduced density matrix. (b) Up to a small correction accounting for normalisation, the eigenvalues of a random reduced density matrix follow
the Marchenko-Pastur law \eqref{eq:mp} for the Wishart-Laguerre ensemble.}
\label{fig8}
\end{figure}

In a low-dimensional many-body system, the localizing properties of disorder can be overcome by interactions. The paradigm is provided by thermal energy fluctuations that can liberate a particle from a trapped state. Such processes are also facilitated by the fact that the many-body level spacing is much smaller than the single-particle level spacing---in fact, with increasing system size the number of available states proliferates exponentially, which can be characterised by an entropy. On the other hand, this proliferation also makes it harder to establish ergodic dynamics
\shortcite{PhysRev.109.1492,Basko20061126,altman_universal_2015,doi:10.1146/annurev-conmatphys-031214-014726}.
For the description of complex interacting systems it is therefore desirable to make contact with quantum statistical mechanics, where the posed questions tie to the concepts of ergodicity, entropy and entanglement \shortcite{peres_quantum_2002,gemmer_quantum_2010}.

Within the framework of quantum statistical mechanics, thermal equilibrium with a heat bath at temperature $T$ is described by the canonical ensemble with density matrix $\rho=Z^{-1}\exp(-H/T)$, where $Z=\mathrm{tr}\exp(-H/T)$ is the  partition function. The additional exchange of particles leads to the grandcanonical ensemble with density matrix $\rho=Z^{-1}\exp((\mu N-H)/T)$, where $\mu$ is the chemical potential and $N$ the fluctuating particle number. In this thermodynamic setting, the associated entropy $\mathcal{S}=-\mathrm{tr}\, \hat\rho \ln \hat\rho$  is an extensive quantity, which scales linearly with the volume, $\mathcal{S}\propto V_S=O(L^d)$ for a system of size $L$ in $d$ dimensions. This implies that an exponential number $\sim \exp(c V_S)$ of states are populated.
Deviations from these predictions occur when one departs from equilibrium. This includes systems in which thermalization is inhibited, with the most notable example found in glasses.

Intriguingly, quantum statistical mechanics also covers the case of closed systems with a fixed energy and particle number, which allows us to focus on the intrinsic quantum-mechanical properties. These systems are described by the microcanonical ensemble, where the density matrix $\rho =M^{-1}\sum_{|E_n-E|<\delta E/2}|\psi_n\rangle\langle\psi_n|$ gives equal weight to $M\gg 1$ eigenstates residing in a classically small energy window $\delta E$ around a fixed energy $E$. The expectation that the microcanonical entropy $S=\ln M$ is extensive indicates again that this  involves an exponential number of available states.

The applicability of this description is intimately related to the question of thermalization in closed system, which in turn reveals whether the internal dynamics are ergodic \shortcite{PhysRevA.43.2046,srednicki_chaos_1994}.
These links become apparent when we ask whether the microcanonical ensemble provides a good description of individual time-dependent quantum states.
More precisely, we form a generic superposition of the states within the energy window and ask whether the time-averaged expectation values of some well-behaved, preselected observables $\hat A_n$ agree  with their ensemble averages.
As it turns out, a good agreement occurs when the matrix elements of the observables in the basis of participating eigenstates are sufficiently random. The ensuing self-averaging leads to an approximate state-independence of the expectation values---a phenomenon known as eigenstate thermalization \shortcite{srednicki_chaos_1994,RevModPhys.83.863,doi:10.1146/annurev-conmatphys-031214-014726}.
Deviations from these predictions then serve as an efficient tool to detect inefficient coupling within the system.

In a useful picture, the state of a system becomes mixed because it is entangled with the environment.
Given a pure state $|\psi\rangle=\sum_{sb}x_{sb}|s\rangle\otimes |b\rangle\in \mathcal{H}_S\otimes \mathcal{H}_B$,  with $s$ labeling basis states of the system and $b$ labeling basis states of the environment, we define the reduced density matrix of the system as $\hat\rho_S=\sum_{bss'} x_{sb}x^*_{s'b}|s\rangle\langle s'|$ \shortcite{peres_quantum_2002}. For an observable $\hat A=\hat A_S\otimes 1$ that only depends on the state of the system, the expectation value follows from  $\mathcal{E}_\psi(A)={\rm tr}\,(\hat\rho_S\hat A_S)$. The information loss from ignoring the environment can be quantified by the entanglement entropy
\begin{equation}
\label{eq:bs}
\mathcal{S}_S=-\mathrm{tr}\, \hat\rho_S \ln \hat\rho_S=-\sum_k \lambda_k\ln \lambda_k,
\end{equation}
where $\lambda_k$ are the positive eigenvalues of $\rho_S$.
As indicated, this entropy measures the entanglement between the system and the environment; it vanishes when $\rho_S$ describes a pure state, which requires that $|\psi\rangle=|\psi_S\rangle\otimes|\psi_B\rangle$ is separable.

To apply these concepts to the microcanonical setting of a closed system, we select a subsystem with Hilbert space dimension $M$ and consider the remainder of (still finite) dimension $M'$ as the environment, where for convenience we assume $M'\geq M$. For each normalised eigenstate
$|\psi_n\rangle=\sum_{sb}x_{sb}|s\rangle\otimes |b\rangle$, we consider the coefficients $x_{sb}$ as the  elements of an $M\times M'$-dimensional matrix $x$, so that in this basis the reduced density matrix takes the form $\rho_S=xx^\dagger$, while $\rho_B=x^\dagger x$. Both are hermitian, positive semidefinite matrices normalised to $\mathrm{tr}\,\rho_S=\mathrm{tr}\,\rho_B=1$.
The bipartite entanglement entropy follows from Eq.~\eqref{eq:bs}. As we started out with a pure state for the total system we have $\mathcal{S}_S=\mathcal{S}_B$.
Indeed, the  non-vanishing eigenvalues of $\rho_S$ and $\rho_B$ are identical, so that we can pair each eigenstate $|\psi_{k,S}\rangle$ of $\rho_S$ with an eigenstate $|\psi_{k,B}\rangle$ of $\rho_B$. This determines the Schmidt decomposition $|\psi_n\rangle=\sum_k\sqrt{\lambda_k}|\psi_{k,S}\rangle\otimes|\psi_{k,B}\rangle$, which reconstructs the underlying pure eigenstate.

The bipartite entanglement entropy is a useful characteristics if the interactions in the system are local,
and plays a central role in a broad range of physical situations, including quantum information
\shortcite{nielsen2010quantum}
 critical phenomena
\shortcite{calabrese_entanglement_2009},
and quantum gravity \shortcite{1751-8121-42-50-504008}.
In the ground state of a many-body system with local interactions, the entanglement entropy is found to be small, scaling with the surface area $S_S\propto A_S=O(L^{d-1})$ instead of the volume $V_S$ of the subsystem.
This is termed an area law of entanglement. At phase transitions, the entanglement entropy in the ground state increases, and in 1D is often found to display a logarithmic dependence  $S_S\sim (c/3)\ln(L)$, where $c$ can be interpreted as the central charge in a conformal field theory \shortcite{calabrese_entanglement_2009}.

These considerations can be naturally informed by random-matrix theory, now applied directly to the structure of the eigenstates at a fixed energy. The simplest case arises when we assume that the system displays eigenstate thermalization. This can be modelled by a random reduced density matrix  $\rho_S=XX^\dagger/Z$, $Z=\mathrm{tr}\,XX^\dagger$,  where $X$ is distributed according to the Gaussian distribution \eqref{eq:px}
\shortcite{page_average_1993}. The density matrix can then be interpreted as a Wishart matrix with posterior normalization \shortcite{0305-4470-34-35-335,RMTHandbookMajumdar}. The joint probability density of the eigenvalues $\lambda_n$ follows directly by constraining the Wishart-Laguerre ensemble \eqref{eq:pwish} to a normalised trace,
\begin{align}
P(\{\lambda_n\})\propto &\,\delta\left(1-\sum_k \lambda_k\right)\prod_{n<m}|\lambda_n-\lambda_m|^\beta
\nonumber \\ & {}\times
\prod_k \lambda_k^{\beta(1+M'-M)/2 -1}.
\label{eq:zs}
\end{align}

For $1\ll M\leq M'$, the trace $\sum_k \lambda_k$ is self-averaging. The eigenvalue density then approaches the Marchenko-Pastur law \eqref{eq:mp} with $\overline\lambda=1/M$,
\begin{equation}
\rho(\lambda)
=
\frac{1}{2\pi\lambda}\sqrt{4 MM'-(M+M'-MM'\lambda)^2}
\end{equation}
for $(\sqrt{M'}- \sqrt{M})^2<MM'\lambda<(\sqrt{M'}+ \sqrt{M})^2$.
The average entanglement entropy follows as  \shortcite{page_average_1993}
\begin{equation}
\overline{\mathcal{S}_B} = -\int
d\lambda\,\rho(\lambda)\lambda \ln \lambda =
 \ln M - (M/2M'),
\end{equation}
independent of the ensemble. This result
signifies near-maximal entanglement. As the Hilbert space dimension of a many-body system grows exponentially with system size, the leading term corresponds to a volume law, $\mathcal{S}_S\propto V_S$, while the subleading term vanishes in the thermodynamic limit $M'\to\infty$.
For highly excited states in an ergodic system obeying eigenstate thermalization, we therefore recover the expected thermodynamic behaviour.

Deviations from the eigenstate thermalization conditions should reduce the entanglement entropy. The expectation is that one recovers an area law when the disorder is increased beyond a certain threshold, a phenomenon termed many-body localization \shortcite{PhysRev.109.1492,Basko20061126}. This transition is indeed confirmed in numerical studies, which also detect a significant reduction of the levels repulsion \shortcite{altman_universal_2015,doi:10.1146/annurev-conmatphys-031214-014726}.
As for the non-interacting case, a complete statistical description of this transition is still missing.

Beyond this  setting, many-body systems offer numerous deep applications of random-matrix theory.
For instance, the logarithmic scaling in critical one-dimensional systems can be recovered from group integrals over unitary, orthogonal or symplectic matrices equipped with the Haar measure \shortcite{PhysRevLett.94.050501}.
The ubiquitous appearance of such group integrals in field theories and other setting nicely leads us away from the theory of open quantum systems---see other notes in this issue---so we close here.

\section{Conclusions}
\label{chap:concl}

As these notes illustrate, the applications of random matrices to open quantum systems are very diverse. Indeed, one can reasonably expect that this setting provides natural physical applications for (almost?) any notable random-matrix result. After all, matrices appear naturally in quantum mechanics, while openness liberates us from some of the algebraic constraints otherwise encountered. The relevance comes from the richness of complex dynamics, which helps to
justify the approach for generic disorder \shortcite{efetov_supersymmetry_1996} or underlying classical chaos  \shortcite{haake_quantum_2010}, and confronts us with a large number of interesting questions about the physical behaviour.

This richness already appeared in the two pure settings covered here---elastic single-particle scattering, and purely interacting systems. The latter topic we only covered briefly, and both effects can of course be combined. This is the subject of much ongoing research---e.g., regarding many-body localization and the topological protection in interacting fermionic systems, to mention just two examples. Furthermore, by combining various effects, links can be established to many other areas that enjoy random-matrix applications,
as mentioned at various places in the text.
As an example we recall the case of photonic systems  with amplification and absorption, for which we can set up effectively non-hermitian descriptions of the wave dynamics. When driven to the laser threshold, these systems provide means to directly probe the poles and residues of the scattering matrix, which gives a physical meaning to the Petermann factor. We can also define new, genuinely non-hermitian symmetries, including the mentioned PT symmetry as well as non-hermitian variants of the chiral and charge-conjugation symmetry, which all provide interesting topological effects.
Such systems also display nonlinear phenomena, for which entirely new descriptions need to be developed.

It is of course  important to consider where the predictive power of random-matrix descriptions  may end. Take the design of small quantum devices. While their dedicated functionality is beyond the scope, random-matrix theory can still help to determine how well they may work---as is illustrated by our discussion of entanglement.
Our system may also be insufficiently ergodic. For instance, localization effects in low dimensions lead to the search for new ensembles, a search that has not been completed. More subtle effects can arise from ballistic dynamics. These are the short-time signatures of classically deterministic motion captured by the fractal Weyl law, and dynamical constraints as encountered in a classically mixed (partially regular and chaotic) phase space. Given some suitable questions, random-matrix theory can often still be adapted to such situations, and otherwise serves as a useful benchmark to quantify the system-specific behaviour.
In general,
deviations from random-matrix predictions can indicate exciting novel physics, leading to an endeavour that is nowhere near to end.

\acknowledgments

I gratefully acknowledge helpful remarks by Carlo Beenakker, Yan Fyodorov, Dmitry Savin and Jacobus Verbaarschot.

\appendix
\section{Eigenvalue densities of matrices with large dimensions}
\label{chap:meandos}

In the limit of large matrix dimensions $M$, eigenvalue distributions can be obtained very efficiently by applications of potential theory, which are based on the analogy of eigenvalues with fictitious particles in a Coulomb gas
\shortcite{dyson_statistical_1962-II,dyson_class_1972,RevModPhys.69.731,forrester2010log}. In the case of the Gaussian ensembles, the leading order can also be obtained from self-consistent equations for the Green function (or resolvent) $G$ \shortcite{pastur2011eigenvalue}. The latter approach links neatly to the theory of free probability \shortcite{janik_non-hermitian_1997,janik_correlations_1999}, %,ThisIssueGuionnet}
as we exploit in the following.

\subsection{Gaussian hermitian ensembles}
For the Gaussian ensembles of hermitian matrices $H$, we expand the Green function $G(E)=(E-H+i\varepsilon)^{-1}$
in a geometric series
\begin{equation}
G=E^{-1}\sum_{n=0}^\infty(HE^{-1})^n
\end{equation}
and average using Wick's theorem, but only retaining non-crossing contractions,
\begin{equation}
\overline G=E^{-1}+E^{-1}\dot H \overline {E^{-1}\sum_{n=0}^\infty(HE^{-1})^n}
\dot H \overline{E^{-1}\sum_{n=0}^\infty(HE^{-1})^n},
\end{equation}
where the dot denotes terms that remain to be contracted. Denoting the variance $\overline{|H_{lm}|^2}=\sigma^2$
this gives in leading order
$\overline G=E^{-1}+E^{-1}\sigma^2\,(\mathrm{tr}\,\overline G)\overline G$.
In terms of the trace $g=\mathrm{tr}\,\overline G$,
this leads to Pastur's equation
\begin{equation}
\label{eq:blue}
E=\sigma^2\,g+M/g.
\end{equation}
The solution $g=(1/2\sigma^2)\sqrt{E^2-4M\sigma^2}$ determines the density of states via
\begin{equation}
\rho(E)=-\lim_{\varepsilon\to0^+}\frac{1}{\pi}\,\mathrm{Im}\, g(E+i\varepsilon)=
\frac{2M}{\pi E_0^2}\sqrt{E_0^2-E^2}
\end{equation}
for $|E|<E_0$, where we identified
$\sigma^2=M\Delta^2/\pi^2= E_0^2/4M$. This is the semicircle law \eqref{eq:semi}.
In the language of free probability,
Eq.~\eqref{eq:blue} leads to the notion of a Blue function $B_r(z)=\sigma_r^2 g+M/g$,
where for later reference we equipped the variance with an index.

\subsection{Wishart-Laguerre ensembles}
For the Wishart-Laguerre ensembles, we analogously write
\begin{equation}
G(\lambda)=(\lambda-X^\dagger X)^{-1}=\lambda^{-1}\sum_{n=0}^\infty(X^\dagger X\lambda^{-1})^n
\end{equation}
and express the non-crossing contractions as
\begin{align}
\overline G&=\lambda^{-1}+\dot X^\dagger \overline{(\lambda-X X^\dagger )^{-1}} \dot X \overline G
\nonumber \\ &
=\lambda^{-1}+\sigma^2 \overline{\mathrm{tr}(\lambda-X X^\dagger )^{-1}}  \overline G
\end{align}
with  $\overline{|X_{lm}|^2}=\sigma^2$.
As $XX^\dagger$ differs from $X^\dagger X$ by $M'-M$ vanishing eigenvalues,
we obtain
\begin{equation}
\label{eq:gwishart}
g=\lambda^{-1}M+\sigma^2 ( \lambda^{-1}(M'-M) +g) g,
\end{equation}
where again $g=\mathrm{tr}\,\overline G$.
The solution
\begin{align}
g=&\frac{1}{2\sigma^2}-\frac{M'-M}{2\lambda}
\nonumber \\ & {}\pm
\frac{1}{2\lambda\sigma^2}\sqrt{\lambda^2-2\lambda \sigma^2(M+M')+\sigma^4(M-M')^2}
\end{align}
 gives the Marchenko-Pastur law \eqref{eq:mp} via $\rho(\lambda)=-\frac{1}{\pi}\,\mathrm{Im}\, g$.

Note that in the chiral ensembles with Hamiltonian
\eqref{eq:chiralh}, the eigenvalues $E_n^2$ can be obtained from a Wishart matrix $AA^\dagger$; the joint distributions \eqref{eq:pwish}  and \eqref{eq:pchiral} are thus related by a change of variables $\lambda_n\propto E_n^2$, and so are the densities
\eqref{eq:mp} and \eqref{eq:rhochiral}.

\subsection{Jacobi ensembles}
For the Jacobi ensembles, we base the considerations on the matrix $(X^\dagger X)^{-1}Y^\dagger Y$, whose eigenvalues $\lambda_n$ determine the eigenvalues of the MANOVA matrix $(X^\dagger X+Y^\dagger Y)^{-1}X^{\dagger} X$ by $T_n=1/(1+\lambda_n)$. Consider the Green function
\begin{align}
G&=\left(\begin{array}{cc} \lambda & Y \\ Y^\dagger & XX^\dagger-\lambda' \end{array}\right)=\left(\begin{array}{cc}G_{11} & G_{12} \\ G_{21} & G_{22} \end{array}\right),
\nonumber\\
 g&=\left(\begin{array}{cc}\mathrm{tr}\, G_{11} & \mathrm{tr}\,G_{12} \\ \mathrm{tr}\,G_{21} & \mathrm{tr}\,G_{22} \end{array}\right).
\end{align}
This has matrix elements
\begin{align}
G_{11}&=[\lambda-Y(X^\dagger X-\lambda')^{-1}Y^\dagger]^{-1},
\nonumber\\
G_{22}&=(X^\dagger X-\lambda'-Y^\dagger Y/\lambda)^{-1},
\end{align}
with traces
\begin{equation}
g_{11}=(M_y-M)/\lambda+g_0,
\quad
\mathrm{tr}\,X G_{22}X^\dagger=\lambda'g_{22}+\lambda g_0,
\end{equation}
where
\begin{equation}
g_0=\mathrm{tr}[\lambda-(X^\dagger X-\lambda')^{-1}Y^\dagger Y]^{-1}.
\end{equation}
The  eigenvalue density can then be obtained from $\rho(\lambda)=-\pi^{-1}\mathrm{Im}\,\overline{g_0}|_{\lambda'=0}$.

The non-crossing contractions give the relation
\begin{align}
\overline{G}=&\left(\begin{array}{cc} 1/\lambda & 0 \\ 0 & -1/\lambda' \end{array}\right)
+\sigma^2
\left(\begin{array}{cc} \mathrm{tr}\,\overline{G_{22}}/\lambda & 0 \\ 0 & -\mathrm{tr}\,\overline{G_{11}}/\lambda' \end{array}\right)\overline{G}
\nonumber\\& {}
+(\sigma^2/\lambda')
\left(\begin{array}{cc} 0 & 0 \\ 0 & M_x-\mathrm{tr}\,\overline{XG_{22}X^\dagger} \end{array}\right)\overline{G},
\end{align}
while on average $\overline {G_{12}}=\overline {G_{21}}=0$.
For $\lambda'\to 0$, we therefore have
\begin{align}
\overline{g_{11}}&=M_y/\lambda+\sigma^2 \overline{g_{11}}\,\overline{g_{22}}/\lambda
,
\nonumber\\
0&=M+\sigma^2[(1+\lambda)\overline{g_{11}}+M-M_x-M_y]\overline{g_{22}},
\end{align}
which determines
\begin{align}
\overline{g_{0}}=&
\frac{(M+M_x)\lambda+M-M_y}{2\pi\lambda(1+\lambda)}.
\nonumber \\ &{}
-\frac{\sqrt{[(M-M_x)\lambda+M-M_y]^2-4M_xM_y \lambda}}{2\pi\lambda(1+\lambda)}.
%\frac{(M+M_x)\lambda+M-M_y-\sqrt{[(M-M_x)\lambda+M-M_y]^2-4M_xM_y \lambda}}{2\pi\lambda(1+\lambda)}.
\end{align}
Denoting $c_x=M_x/M$, $c_y=M_y/M$,
the eigenvalue density is thus given by
\begin{equation}
\rho(\lambda)=\frac{M(c_x-1)\sqrt{(\lambda-\lambda_-)(\lambda_+-\lambda)}}{2\pi\lambda(1+\lambda)},
\end{equation}
where
\begin{align}
\lambda_\pm&=\left(\frac{\sqrt{c_xc_y}\pm\sqrt{c_x+c_y-1}}{c_x-1}\right)^2
\nonumber \\ &=
\left(\frac{c_y-1}{\sqrt{c_xc_y}\mp\sqrt{c_x+c_y-1}}\right)^2
\end{align}
determines the range where the density is finite.
In terms of the variables $T_n$, the density is then given by Eq.~\eqref{eq:rhot2}.

\subsection{Ginibre ensembles}
As an example of non-hermitian matrix ensembles we consider the complex Ginibre ensemble, defined by \eqref{eq:px} with $\beta=2$.
The Green function has now to be extended to the block form
\shortcite{janik_correlations_1999}
\begin{align}
G(z,z^*)&=\left(\begin{array}{cc}z-X & i\lambda \\ i\lambda & z^*-X^\dagger \end{array}\right)^{-1}=\left(\begin{array}{cc}G_{11} & G_{12} \\ G_{21} & G_{22} \end{array}\right),
\nonumber \\
g(z,z^*)&=\lim_{\lambda\to 0^+}\left(\begin{array}{cc}\mathrm{tr}\, G_{11} & \mathrm{tr}\,G_{12} \\ \mathrm{tr}\,G_{21} & \mathrm{tr}\,G_{22} \end{array}\right),
\end{align}
which delivers the density of complex eigenvalues $z_m$ via \begin{equation}
\frac{1}{\pi}\,\frac{\partial g_{11}}{\partial z^*}=\rho(z)=\sum_m\delta(z-z_m),
\end{equation}
while the Petermann factors $K_m$ are encoded in
\begin{equation}
\label{eq:oz}
-\frac{1}{\pi}g_{12}g_{21}=O(z)=\sum_m K_m\delta(z-z_m).
\end{equation}

We denote again $\overline{|X_{lm}|^2}=\sigma^2$ and employ the expansion
\begin{align}
&G=\mathfrak{Z}^{-1}\sum_{n=0}^\infty (\mathfrak{X}\mathfrak{Z}^{-1})^n,\quad \mathfrak{Z}=
\left(\begin{array}{cc}z & i\lambda \\ i\lambda & z^* \end{array}\right),
\nonumber \\ &
\mathfrak{X}= \left(\begin{array}{cc}X & 0 \\ 0 & X^\dagger\end{array}\right),
\end{align}
followed by the non-crossing approximation
\begin{align}
\overline{G}&=\mathfrak{Z}^{-1}+ \mathfrak{Z}^{-1}\dot{\mathfrak{X}} \overline{\mathfrak{Z}^{-1}\sum_{n=0}^\infty (\mathfrak{X}\mathfrak{Z}^{-1})^n } \dot{\mathfrak{X}}
\overline{\mathfrak{Z}^{-1}\sum_{n=0}^\infty (\mathfrak{X}\mathfrak{Z}^{-1})^n }
\nonumber\\
&=\mathfrak{Z}^{-1}+\mathfrak{Z}^{-1}\sigma^2\left(\begin{array}{cc}0 &\overline{g_{12}} \\ \overline{g_{21}} & 0 \end{array}\right) \overline{G},
\end{align}
where the dot denotes elements to be Wick-contracted.
The trace gives
\begin{equation}
\mathfrak{Z}=M/\overline{g}+\frac{\sigma^2}{2}(\overline{g}+\tilde g),\quad \tilde g=
\left(\begin{array}{cc}1 &0 \\ 0 & -1 \end{array}\right) \overline{g} \left(\begin{array}{cc}-1 &0 \\ 0 & 1 \end{array}\right).
\end{equation}
This agrees with the rules from free probability, according to which the Blue functions
of the real and imaginary parts are $B_r(A)=\sigma_r^2 A+M/A$, $B_i(A)= \sigma_i^2\tilde A+M/A$,
while the composition law reads $\mathfrak{Z}=B_r(g)+B_i(g)-M/g$;
here  $\sigma_r^2=\sigma_i^2=\sigma^2/2$.

Let us set $\sigma^2=1/M$.
For $|z|<1$ the solution is then given by
\begin{equation}
\overline{g}=M\left(\begin{array}{cc}z^* &\sqrt{|z^2|-1} \\ \sqrt{|z^2|-1} & z \end{array}\right).
\end{equation}
From this we recover Ginibre's circular law $\rho(z)=\frac{M}{\pi}\Theta(1-|z|)$, while $O(z)=M^2(1-|z|^2)/\pi$. According to Eq.~\eqref{eq:oz}, the ratio $O(z)/\rho(z) = M(1-|z|^2)\sim \overline{K_m}|_{z_m=z}$ gives the average Petermann factor within the support of the spectrum.

\bibliography{schomerus}

\end{document}